\documentclass[fleqn,usenatbib]{mnras}

\usepackage[T1]{fontenc}
\usepackage{graphicx}
\usepackage[cmex10]{amsmath}
\usepackage{array}
\usepackage{fixltx2e}
\usepackage{url}
\usepackage{amsfonts}
\usepackage{lmodern}
\usepackage{algpseudocode}
\usepackage{algorithmicx}
\usepackage{algorithm}
\usepackage{xfrac}
\usepackage{float}
\usepackage{multirow}
\usepackage{balance}
\usepackage{tikz}
\usetikzlibrary{arrows,shapes,automata}
\tikzstyle{every picture}+=[remember picture]

\DeclareMathOperator{\prox}{prox}

\DeclareMathOperator{\soft}{\op{S}}
\DeclareMathOperator{\identity}{\op{I}}

\DeclareMathOperator{\proj}{\op{P}}
\DeclareMathOperator*{\argmin}{argmin}

\newcommand{\ds}{\displaystyle}
\newcommand{\hquad}{\;\:}

\newcommand{\bs}{\boldsymbol}
\newcommand{\bm}[1]{\boldsymbol{\mathsf{#1}}}
\newcommand{\bb}{\mathbb}
\newcommand{\mc}{\mathcal}
\newcommand{\diff}{\mathrm{d}\hspace{-0.1ex}}
\newcommand{\acp}[1]{(\uppercase{#1})}
\newcommand{\ac}[1]{\uppercase{#1}}
\newcommand{\alg}[1]{\textsc{#1}}
\newcommand{\op}[1]{\boldsymbol{\mc{#1}}}

\algblock[Block]{Block}{EndBlock}
\algblockdefx[Block]{Block}{EndBlock}%
[1]{{#1}}%
[1]{{#1}}

\newcommand{\Given}[1]{\State{\bf given} {#1}}
\newcommand{\Func}[2]{\Block{{\bf function} \textsc{#1}}$\big($#2$\big)$}
\newcommand{\EndFunc}[1]{\EndBlock{{\bf return} #1}}
\newcommand{\FuncCall}[2]{\textsc{#1}$\big($#2$\big)$}
\newcommand{\RepeatFor}[1]{\Repeat {\bf~for} {#1}}

\newcommand{\ParForD}[2]{\Block{{#1} {\bf distribute} {#2} {\bf and do in parallel}}}
\newcommand{\EndParForD}[1]{\EndBlock{\bf end and gather} {#1}}

\newcommand{\ParFor}[1]{\Block{{#1} \bf do~in~parallel}}
\newcommand{\EndParFor}{\EndBlock{\bf end}}

\newcommand{\Set}[1]{\Block{{#1} \bf set}}
\newcommand{\EndSet}{\EndBlock{\bf end}}

\newcommand{\ignore}[1]{\color{magenta}\color{black}}

\newcommand{\bc}{\color{black}}
\newcommand{\ec}{\color{black}}

\newtheorem{define}{Definition}
\newtheorem{property}{Property}

\newcommand{\rnode}[2]{\tikz[baseline]{\node[minimum size = 3.5ex, anchor=base, inner sep=1mm,fill=#1,shape=rectangle,rounded corners]{#2};}}
\newcommand{\rnoden}[3]{\tikz[baseline]{\node[minimum size = 3.5ex, anchor=base, inner sep=1mm,fill=#1,shape=rectangle,rounded corners](#3){#2};}}

\title[Scalable splitting algorithms for SKA]{Scalable splitting algorithms for big-data interferometric imaging in the SKA era}

\author[A. Onose, R. Carrillo et al.]{
Alexandru Onose$^{1}$\thanks{E-mail: a.onose@hw.ac.uk}\textsuperscript{\thanks{The authors have contributed equally to the work herein.}},
Rafael E. Carrillo$^{2}${\color{blue} \textsuperscript{\footnotemark[2]}},
Audrey Repetti$^{1}$,
Jason D. McEwen$^{3}$,
\newauthor{
Jean-Philippe Thiran$^{2}$,
Jean-Christophe Pesquet$^{4}$ and
Yves Wiaux$^{1}$
}
\\
$^{1}$Institute of Sensors, Signals and Systems, Heriot-Watt University, Edinburgh EH14 4AS, United Kingdom\\
$^{2}$Signal Processing Laboratory (LTS5), Ecole Polytechnique F\'{e}d\'{e}rale de Lausanne, Lausanne CH-1015, Switzerland\\
$^{3}$Mullard Space Science Laboratory,  University College London, Surrey RH5 6NT, United Kingdom \\
$^{4}$Laboratoire d'Informatique Gaspard Monge, Universit\'{e} Paris-Est, Marne la Vall\'{e}e F-77454, France
}

\date{Accepted XXX. Received YYY; in original form ZZZ}

\pubyear{2016}

\begin{document}
\label{firstpage}
\pagerange{\pageref{firstpage}--\pageref{lastpage}}
\maketitle

\begin{abstract}

\bc
In the context of next generation radio telescopes, like the Square Kilometre Array, the efficient processing of large-scale datasets is extremely important.
Convex optimisation tasks under the compressive sensing framework have recently emerged and provide both enhanced image reconstruction quality and scalability to increasingly larger data sets.
We focus herein mainly on scalability and propose two new convex optimisation algorithmic structures able to solve the convex optimisation tasks arising in radio-interferometric imaging.
They rely on proximal splitting and forward-backward iterations and can be seen, by analogy with the \alg{clean} major-minor cycle, as running sophisticated \alg{clean}-like iterations in parallel in multiple data, prior, and image spaces.
Both methods support any convex regularisation function, in particular the well studied $\ell_1$ priors promoting image sparsity in an adequate domain.
\ec
Tailored for big-data, \bc they \ec employ parallel and distributed computations to achieve scalability, in terms of memory and computational requirements.
One of them also exploits randomisation, over data blocks at each iteration, offering further flexibility. 
We present simulation results showing the feasibility of the proposed methods as well as their advantages compared to state-of-the-art \bc algorithmic solvers. 
Our Matlab code is available online on GitHub. \ec
\end{abstract}


\section{Introduction}

Radio-interferometry \acp{ri} allows the observation of radio emissions with great sensitivity and angular resolution. The technique has been extensively investigated and provides valuable data driving many research directions in astronomy, cosmology or astrophysics \bc \citep{thompson01}\ec.
Next-generation radio telescopes, such as the LOw Frequency ARray (LOFAR) \citep{vanHaarlem2013} and the future Square Kilometre Array (SKA) \citep{Dewdney2009}, are envisaged to produce giga-pixel images and achieve a dynamic range of six or seven orders of magnitude.
This will be an improvement over current instruments by around two orders of magnitude, in terms of both resolution and sensitivity.
The amount of data acquired will be massive and the methods solving the inverse problems associated with the image reconstruction need to be fast and to scale well with the number of measurements.
Such challenges provided motivation for vigorous research to reformulate imaging and calibration techniques for \ac{ri}~\citep{Wijnholds2014}.

The construction of the first phase of SKA is scheduled to start in 2018. It will consist of two subsystems: a low frequency aperture array, the SKA1-low, operating in the 50-350 MHz frequency range and containing approximately $131,000$ antenna elements; a mid frequency array of reflector dishes, the SKA1-mid, operating above 350 MHz, consisting of 197 dishes~\citep{Dewdney2009,Broekema2015}. 
Both subsystems are planned to operate on the order of $65,000$ frequency bands.
Data rate estimates in this first phase are around five terabits per second for each subsystem~\citep{Broekema2015} and will present a great challenge for the infrastructure and signal processing.
The celebrated \alg{clean} algorithm \citep{hogbom74} and its variants do not scale well given the large dimension of the problem.
They rely on \emph{local} greedy iterative procedures and are slow compared to modern convex optimisation techniques, which are guaranteed to converge towards a \emph{global} optimal solution.
Moreover, they are not designed for large-scale parallelisation or distributed computing \citep{Carrillo2014}.

In the past few years, sparse models and convex optimisation techniques have been applied to RI imaging, showing the potential to outperform state-of-the-art imaging algorithms in the field \citep{wiaux09, rau09,li11,carrillo12, Carrillo2013, Carrillo2014,Garsden2015}. 
These methods typically solve the imaging problem by minimising an objective function defined as a sum of a data term, dependent on the measured visibilities, and several regularisation terms, usually promoting sparsity and positivity.
\ignore{Recently, convex optimisation algorithms have been specifically tailored for large-scale problems.
They employ advanced first-order methods and randomisation, as well as parallel and distributed schemes that are essential for scalability.
This is attractive for the development of new fast and scalable imaging methods able to work with the huge amounts of data the SKA will produce. 
The ultimate goal is to markedly reduce the computational, storage, and communication bottlenecks of traditional methods.}
\bc Scalable algorithms, specifically tailored for large-scale problems using parallel and distributed schemes, \ec are just now beginning to gain attention in the context of imaging~\citep{Carrillo2014, Ferrari2014} and calibration~\citep{Yatawatta2015} for next-generation radio telescopes. 

\bc In this context, \ec proximal splitting methods are very popular due to their ability to decompose the original problem into several simpler, easier to solve, sub-problems, each one associated with one term of the objective function \citep{combettes11}.
Another class of algorithms currently gaining traction for large-scale problems in optimisation is based on primal-dual (PD) methods~\citep{Komodakis2015}.
Such methods efficiently split the \bc optimisation problem \ec and, at the same time, maintain a highly parallelisable structure by solving concomitantly for a \emph{dual} formulation of the original problem.
Building on such tools, the simultaneous direction method of multipliers \acp{sdmm} was recently proposed in the context of RI imaging by \cite{Carrillo2014}.
It achieves the complete \emph{splitting} of the functions defining the minimisation task.
In the big-data context, \ac{sdmm} scales well with the number of \bc measurements\ec, however, an expensive matrix inversion is necessary when updating the solution, which limits the suitability of the method for the recovery of very large images.

\bc The scope of this article is to propose two new algorithmic structures for RI imaging.
We study their computational performance and parallelisation capabilities by solving the sparsity averaging optimisation problem proposed in the \ac{sara} algorithm \citep{carrillo12}, previously shown to outperform the standard \alg{clean} methods.
The application of the two algorithms is not limited to the \ac{sara} prior, any other convex prior functions being supported.
We assume a known model for the measured data such that there is no need for calibration.
We use \ac{sdmm}, solving the same minimisation problem, to compare the computational burden and parallelisation possibilities.
Theoretical results ensure convergence, all algorithms reaching the same solution.
We also showcase the reconstruction performance of the two algorithms coupled with the \ac{sara} prior in comparison with \ac{cs-clean} \citep{Schwab1984} and \ac{moresane} \citep{Dabbech2015}.
\ec

The first algorithmic solver is a sub-iterative version of the well-known alternating direction method of multipliers (ADMM).
The second is based on the PD method and uses forward-backward \bc\acp{fb} \ec iterations, typically alternating between gradient (forward) steps and projection (backward) steps.
\bc Such steps can be seen as interlaced \alg{clean}-like updates. \ec
Both \bc algorithms \ec are highly parallelisable and allow for an efficient distributed implementation.
ADMM however offers only partial splitting of the objective function \bc leading \ec to a sub-iterative algorithmic structure.
The PD method offers the \emph{full splitting} for both operators and functions. \bc It \ec does not need sub-iterations or any matrix inversion.
Additionally, it \bc can attain increased  \ec scalability by using \emph{randomised updates}.
\bc It works by selecting only a fraction of the visibilities at each iteration, thus \ec achieving great flexibility in terms of memory requirements and computational load per iteration, at the cost of requiring more iterations to converge.
Our simulations suggest no significant increase in the total computation cost.

The remainder of this article is organised as follows. Section~\ref{sec-ri} introduces the RI imaging problem and describes the state-of-the-art image reconstruction techniques used in radio astronomy. In Section~\ref{sec-conv-opt} we review some of the main tools from convex optimisation needed for RI imaging. Section~\ref{sec-large-scale} formulates the optimisation problem for RI imaging given the large-scale data scenario
\bc and presents \ec the proposed algorithms, ADMM and PD, respectively. We discuss implementation details and their computational complexity in Section~\ref{sec-imp}. Numerical experiments evaluating the performance of the algorithms are reported in Section~\ref{sec-results}. Finally, we briefly present the main contributions and envisaged future research directions in Section~\ref{sec-conc}.

\section{Radio-interferometric imaging}
\label{sec-ri}
Radio-interferometric data, the visibilities, are produced by an array of antenna pairs that measure radio emissions from a given area of the sky.
The \bc projected \ec baseline components, in units of the wavelength of observation, are commonly denoted $(u,v,w)$, where $w$ identifies the component in the line of sight and $\bs{u}=(u,v)$  the components in the orthogonal plane.
The sky brightness distribution $x$ is described in the same coordinate system, with components $l$, $m$, $n$ and with $\bs{l} = (l,m)$ and $n(\bs{l})=\sqrt{1 - l^2 - m^2},~l^2 + m^2 \leq 1$.
The general measurement equation for non-polarised monochromatic \ac{ri} imaging can be stated as
\begin{equation}
	y(\bs{u}) = \int D(\bs{l},\bs{u}) x(\bs{l}) e^{-2 i \pi \bs{u} \cdot \bs{l}} \diff^2 \bs{l},
	\label{measurement-eq}
\end{equation}
with $D (\bs{l}, \bs{u}) = \sfrac{1}{n(\bs{l})} \bar{D} (\bs{l},\bs{u})$ quantifying all the DDEs.
\bc Some dominant DDEs can be modelled analytically, like the
$w$ component which is expressed \ec as $\bar{D}_w(\bs{l},\bs{u}) = e^{-2i \pi w(n(\bs{l})-1)}$.
At high dynamic ranges however, unknown DDEs, related to the primary beam or ionospheric effects, also affect the measurements introducing the need for calibration.
\bc Here we work in the absence of DDEs.\ec

The recovery of $x$ from the visibilities relies on algorithms solving a discretised version of the inverse problem (\ref{measurement-eq}).
We denote by $\bs{x} \in \bb{R}^N$ the intensity \bc image \ec of which we take $M$ visibility measurements $\bs{y} \in \bb{C}^M$.
The measurement model is defined by
\begin{equation}
	\bs{y} = \bm{\Phi} \bs{x} + \bs{n},
	\label{inverse-problem}
\end{equation}
where the \emph{measurement operator} $\bm{\Phi} \in \bb{C}^{M \times N}$ is a linear map from the image domain to the visibility space and $\bs{y}$ denotes the vector of measured visibilities corrupted by the additive noise $\bs{n}$.
Due to limitations in the visibility sampling scheme, equation (\ref{inverse-problem}) defines an ill-posed inverse problem.
Furthermore, the large number of the data points, $M \gg N$, introduces additional challenges related to the computational and memory requirements for finding the solution.
\bc In what follows, we assume the operator $\bm{\Phi}$ to be known is advance such that no calibration step is needed to estimate it.\ec

Due to the highly iterative nature of the reconstruction algorithms, a fast implementation of all operators involved in the image reconstruction is essential, for both regularisation and data terms.
To this purpose, the measurement operator is modelled as the product between a matrix $\bm{G} \in \bb{C}^{M \times n_{\rm{o}}N}$ 
and an $n_{\rm{o}}$-oversampled Fourier operator,
\begin{equation}
	\bm{\Phi} = \bm{G} \bm{F} \bm{Z}.
\end{equation}
The matrix $\bm{Z} \in \bb{R}^{n_{\rm{o}}N \times N}$ accounts for the oversampling and the scaling of the image to pre-compensate for possible imperfections in the interpolation \citep{Fessler2003}.
In the absence of DDEs, $\bm{G}$ only contains compact support kernels that enable the computation of the continuous Fourier samples from the discrete Fourier coefficients provided by $\bm{F}$. 
\bc Alternatively, seen as a transform from the $u$--$v$ space to the discrete Fourier space, $\bm{G}^\dagger$, the adjoint operator of $\bm{G}$, \emph{grids} the continuous measurements onto a uniformly sampled Fourier space associated with the oversampled discrete Fourier coefficients provided by $\bm{F}\bm{Z}$. \ec
This representation of the measurement operator enables a fast implementation thanks to the use of the fast Fourier transform for $\bm{F}$ and to the fact that the \bc convolution kernels used \ec are in general modelled with compact support in the Fourier domain, which leads to a sparse matrix $\bm{G}$\footnote{
\bc Assuming pre-calibrated data in \ec the presence of DDEs, the line of $\bm{G}$ associated with frequency $\bs{u}$, is explicitly given by the convolution of the discrete Fourier transform of $D(\bs{l},~\bs{u})$, centred on $\bs{u}$, with the associated gridding kernel.
\bc This maintains the sparse structure of $\bm{G}$, since the DDEs are generally modelled with compact support in the Fourier domain. \ec
\bc A non-sparse $\bm{G}$ drastically increases the computational requirements for the implementation of the measurement operator.
However, it is generally transparent to the algorithms since they do not rely on the sparsity structure explicitly. This is the case for all the algorithmic structures discussed herein.\ec
}.

\subsection{Classical imaging algorithms}
Various methods have been proposed for solving the inverse problem defined by (\ref{inverse-problem}).
The standard imaging algorithms belong to the \alg{clean} family and perform a greedy non-linear deconvolution based on local iterative beam removal \citep{hogbom74, schwarz78, Schwab1984, thompson01}.
A sparsity prior on the \bc solution is implicitly introduced since the method reconstructs the image pixel by pixel. Thus, \alg{clean} is very similar to \ec the matching pursuit (MP) algorithm \citep{mallat93}.
It may also be seen as a regularised gradient descent method.
It minimises the residual norm $\| \bs{y} - \bm{\Phi} \bs{x} \|_2^2$ via a gradient descent subject to an implicit sparsity constraint on $\bs{x}$.
An update of the solution takes the following form
\begin{equation}
	\bs{x}^{(t)} = \bs{x}^{(t-1)} + \op{T} \Big( \bm{\Phi}^\dagger \big( \bs{y} - \bm{\Phi} \bs{x}^{(t-1)} \big) \Big),
\end{equation}
where $\bm{\Phi}^\dagger$ is the adjoint of the linear operator $\bm{\Phi}$.
In the astronomy community, the computation of the residual image $\bm{\Phi}^\dagger \big( \bs{y} - \bm{\Phi} \bs{x}^{(t-1)} \big)$, which represents a gradient step of the residual norm, is being referred to as the \emph{major} cycle while the deconvolution performed by the operator $\op{T}$ is named the \emph{minor} cycle.
\bc All proposed versions of \alg{clean} use variations of these major and minor cycles \citep{rau09}. \ec
\bc \alg{clean} \ec builds the solution image iteratively by searching for atoms associated with the largest magnitude pixel from the residual image.
A \emph{loop gain factor} controls how aggressive is the update step, by only allowing a fraction of the chosen atoms to be used.

Multiple improvements of \alg{clean} have been suggested. In the multi-scale version \citep{cornwell08b} the sparsity model is augmented through a multi-scale decomposition.
An adaptive scale variant was proposed by \cite{bhatnagar04} and can be seen as \ac{mp} with over-complete dictionaries since it models the image as a superposition of atoms over a redundant dictionary.
Another class of solvers, the maximum entropy method \citep{Ables74, Gull78, cornwell85} solves a regularised global optimisation problem through a general entropy prior.
In practice however, \alg{clean} and its variants have been preferred even though they are slow and require empirically chosen configuration parameters.
Furthermore, these methods also lack the scalability required for working with huge, \ac{ska}-like data.

\subsection{Compressed sensing in radio-interferometry}

Imaging algorithms based on convex optimisation and using sparsity-aware models have also been proposed, especially under the theoretical framework of compressed sensing \acp{cs}, reporting superior reconstruction quality with respect to \alg{clean} and its multi-scale versions. CS proposes both the optimisation of the acquisition framework, going beyond the traditional Nyquist sampling paradigm, and the use of non-linear iterative algorithms for signal reconstruction, regularising the ill-posed inverse problem through a low dimensional signal model \citep{donoho06,candes06}. The key premise in CS is that the underlying signal has a sparse representation, $\bs{x}=\bm{\Psi} \bs{\alpha}$ with $\bs{\alpha}\in \bb{C}^{D}$ containing only a few nonzero elements \citep{fornasier11}, in a dictionary $\bm{\Psi} \in \bb{C}^{N \times D}$, \textit{e.g.} a \bc collection of wavelet bases or, more generally, \ec an over-complete frame.

The first study of CS applied to RI was done by \citet{wiaux09}, who demonstrated the versatility of convex optimisation methods and their superiority relative to standard interferometric imaging techniques. A CS approach was developed by \citet{wiaux10a} to recover the signal induced by cosmic strings in the cosmic microwave background. \citet{mcewen11a} generalised the CS imaging techniques to wide field-of-view observations. Non-coplanar effects and the optimisation of the acquisition process, were studied by \citet{wiaux09b} and \citet{wolz13}. All the aforementioned works solve a synthesis-based problem defined by
\begin{equation}
	\min_{\bs{\alpha}} \| \bs{\alpha} \|_1 \quad \rm{subject~to} \quad \|\bs{y} - \bm{\Phi} \bm{\Psi} \bs{\alpha}\|_2  \leq \epsilon,
	\label{synthesis-l1-problem}
\end{equation}
where $\epsilon$ is a bound on the $\ell_2$ norm of the noise $\bs{n}$. Synthesis-based problems recover the image representation $\bs{\alpha}$ with the final image obtained from the synthesis relation $\bs{x}=\bm{\Psi} \bs{\alpha}$.
\bc Here, the best model for the sparsity, the non-convex $\ell_0$ norm, is replaced with its closest convex relaxation, the $\ell_1$ norm, to allow the use of efficient convex optimisation solvers.
Re-weighting schemes are generally employed to approximate the $\ell_0$ norm from its $\ell_1$ relaxation \citep{Candes2008, Daubechies2010}. \ec
Imaging approaches based on unconstrained versions of \eqref{synthesis-l1-problem} have also been studied~\citep{wenger10,li11,Hardy2013,Garsden2015}.
For example, \citet{Garsden2015} applied a synthesis-based reconstruction method to LOFAR data.

As opposed to synthesis-based problems, analysis-based approaches recover the signal itself, solving
\begin{equation}
	\min_{\bs{x}} \| \bm{\Psi}^\dagger \bs{x} \|_1 \quad \rm{subject~to} \quad \|\bs{y} - \bm{\Phi} \bs{x}\|_2 \leq \epsilon.
	\label{analysis-l1-problem}
\end{equation}
The sparsity averaging reweighed analysis \acp{sara}, based on the analysis approach and an average sparsity model, was introduced by \citet{carrillo12}. \citet{Carrillo2014} proposed a \bc scalable \ec algorithm, based on \ac{sdmm}, to solve \eqref{analysis-l1-problem}. For such large-scale problems, the use of sparsity operators $\bm{\Psi}$ that allow for a fast implementation is fundamental. Hybrid analysis-by-synthesis greedy approaches have also been proposed by \citet{Dabbech2015}.

\bc To provide an analogy between \alg{clean} and the \ac{fb} iterations employed herein, we can consider one of the most basic approaches, the unconstrained version of the minimisation problem (\ref{analysis-l1-problem}), namely $\min_{\bs{x}} \| \bm{\Psi}^\dagger \bs{x} \|_1 + \beta \|\bs{y} - \bm{\Phi} \bs{x}\|^2_2$ with $\beta$ a free parameter.
To solve it, modern approaches using \ac{fb} iterations perform a gradient step together with a soft-thresholding operation in the given basis $\bm{\Psi}^\dagger$ \citep{Combettes2007b}.
This \ac{fb} iterative structure is conceptually extremely close to the major-minor cycle structure of \alg{clean}.
At a given iteration, the forward (gradient) step consists in doing a step in the opposite direction to the gradient of the $\ell_2$ norm of the residual.
It is essentially equivalent to a major cycle of \alg{clean}. 
The backward (soft-thresholding) step consists in decreasing the absolute values of all the coefficients of $\bm{\Psi}^\dagger\bs{x}$ that are above a certain threshold by the threshold value, and setting to zero those below the threshold.
This step is very similar to the minor cycle of \alg{clean}, with the soft-threshold value being an analogous to the loop gain factor. 
The soft-thresholding intuitively works by removing small and insignificant coefficients, globally, on all signal locations simultaneously
while \alg{clean} iteratively builds up the signal by picking up parts of the most important coefficients, a local procedure, until the residuals become negligible.
Thus, \alg{clean} can be intuitively understood as a very specific version of the \ac{fb} algorithm.
As will be discussed in Section \ref{sec-large-scale}, from the perspective of \alg{clean}, the algorithms presented herein can be viewed as being composed of complex \alg{clean}-like \ac{fb} steps performed in parallel in multiple data, prior and image spaces.
\ec

\section{Convex optimisation}
\label{sec-conv-opt}
\bc Optimisation techniques play a central role in solving the large-scale inverse problem (\ref{inverse-problem}) from \ac{ri}. \ec
Some of the main \bc methods \ec from convex optimisation \citep{Bauschke2011} \ignore{,  fundamental for solving large-scale inverse problems such as the \ac{ri} problem (\ref{inverse-problem}),} are presented in what follows.

Proximal splitting techniques are very attractive due to their flexibility \bc and ability to produce scalable algorithmic structures. \ec
Examples of proximal splitting algorithms include the Douglas-Rachford method \citep{Combettes2007, Bot2013}, the projected gradient approach \citep{Calamai1987}, the iterative thresholding algorithm \citep{Daubechies2004, Beck2009}, the alternating direction method of multipliers \citep{Boyd2011} or the simultaneous direction method of multipliers \citep{Setzer2010}.
All \bc proximal \ec splitting methods solve optimisation problems like 
\begin{equation}
	\min_{\bs{z}} g_1(\bs{z}) + \cdots + g_n(\bs{z}),
	\label{splitting-example}
\end{equation}
with $g_i$, $i \in \{1,\ldots,n\}$, proper, lower-semicontinuous, convex functions.
No assumptions are required about the smoothness, each non-differentiable function being incorporated into the minimisation through its proximity operator \bc (\ref{proximity-operator}). \ec
Constrained problems are reformulated to fit (\ref{splitting-example}) through the use of the \emph{indicator function} \bc (\ref{indicator-function}) \ec
of the convex set $\mc{C}$ defined by the constraints.
As a general framework, proximal splitting methods minimise (\ref{splitting-example}) iteratively
by handling each function $g_i$, possibly non smooth, through its proximity operator.
A good review of the main proximal splitting algorithms and some of their applications to signal and image processing is presented by \cite{combettes11}.

Primal-dual methods \citep{Komodakis2015} introduce another framework over the proximal splitting approaches and are able to achieve \emph{full splitting}.
All the operators involved, not only the gradient or proximity operators, but also the linear operators, can be used separately.
Due to this, no inversion of operators is required, which gives important computational advantages when compared to other splitting schemes \citep{Combettes2012}.
The methods solve optimisation tasks of the form
\begin{equation}
	\min_{\bs{z}}  g_1(\bs{z})+ g_2(\bm{L}\bs{z}),
	\label{primal-problem}
\end{equation}
with $g_1$ and $g_2$ proper, lower semicontinuous convex functions and $\bm{L}$ a linear operator.
They are easily extended to problems, similar to (\ref{splitting-example}), involving multiple functions.
The minimisation (\ref{primal-problem}), usually referred to as the \emph{primal problem}, accepts a \emph{dual problem} \citep{Bauschke2011},
\begin{equation}
	\min_{\bs{v}}  g_1^*(-\bm{L}^\dagger\bs{v}) + g_2^*(\bs{v}),
	\label{dual-problem}
\end{equation}
where $\bm{L}^\dagger$ is the adjoint of the linear operator $\bm{L}$ and $g_2^*$ is the Legendre-Fenchel \emph{conjugate function} of $g_2$, defined \bc in (\ref{f-conj}). \ec
Under our assumptions for $g_1$ and $g_2$ and, if a solution to (\ref{primal-problem}) exists, 
\bc efficient algorithms for solving together the primal and dual problems can be devised \citep{Condat2013, Vu2013, Combettes2012}. \ec
\ignore{
The goal of this class of methods is to find a Kuhn-Tucker point $(\hat{\bs{z}}, \hat{\bs{v}})$ that satisfies 
\begin{equation}
	-\bm{L}^\dagger \hat{\bs{v}} \in \partial g_1(\hat{\bs{z}}), \qquad \bm{L} \hat{\bs{z}} \in \partial g_2^*(\hat{\bs{v}}),
	\label{kk-condition}
\end{equation}
which ensures that $\hat{\bs{z}}$ and $\hat{\bs{v}}$ are solutions to the primal and dual problems, respectively.}
Such \ac{pd} approaches are able to produce highly scalable algorithms that are well suited for solving inverse problems similar to (\ref{inverse-problem}).
They are flexible and offer a broad class of methods ranging from distributed computing to randomised or block coordinate approaches \citep{Pesquet2014, Combettes2015}.

Augmented Lagrangian \acp{al} methods \citep{Bertsekas1982} have been traditionally used for solving constrained optimisation problems through an equivalent unconstrained minimisation.
In our context, the methods can be applied for finding the solution to a constrained optimisation task equivalent to (\ref{primal-problem}),%
\begin{equation}
	\min_{\bs{z}, \bs{r}}  g_1(\bs{z}) + g_2(\bs{r}),~~\rm{subject~to~} \bs{r} = \bm{L}\bs{z},
	\label{admm-constained-opt}
\end{equation}
by the introduction of the slack variable $\bs{r}$.
The solution is found by searching for a saddle point \bc of the augmented Lagrange function associated with (\ref{admm-constained-opt}), \ec
\begin{equation}
	\max_{\bs{s}} \min_{\bs{z}, \bs{r}} g_1(\bs{z}) + g_2(\bs{r}) + \frac{ \bs{s}^\dagger}{\mu} \left( \bm{L} \bs{z} - \bs{r} \right) + \frac{1}{2\mu} \| \bm{L} \bs{z} - \bs{r} \|_2^2.
	\label{saddle-point-admm}
\end{equation}
The vector $\bs{s}$ and parameter $\mu$, correspond to the Lagrange multipliers.
No explicit assumption is required on the smoothness of the functions $g_1$ and $g_2$.
Several algorithms working in this framework have been proposed.
The alternating direction method of multipliers \acp{admm} \citep{Boyd2011, Yang2011} is directly applicable to the minimisation (\ref{admm-constained-opt}). 
A generalisation of the method, solving (\ref{splitting-example}), is the simultaneous direction method of multipliers \acp{sdmm}\citep{Setzer2010}. It finds the solution to an extended augmented Lagrangian, defined for multiple functions $g_i$.
Both methods can also be characterised from the \ac{PD} perspective \citep{Boyd2011, Komodakis2015}.
Algorithmically, they split the minimisation step by alternating between the minimisation over each of the primal variables \ignore{ of interests}, $\bs{z}$ and $\bs{r}$, followed by a maximisation with respect to the multipliers $\bs{s}$, performed via  a gradient ascent\ignore{step}.

\section{Large-scale optimisation}
\label{sec-large-scale}
The next generation telescopes will be able to produce a huge amount of visibility data.
\ignore{The \ac{ska}, when fully built, is expected to contain on the order of one million antennas and will generate exabytes of data per day.}
To this regard, there is much interest in the development of fast and well performing reconstruction algorithms \citep{Carrillo2014, mcewen11a}.
Highly scalable algorithms, distributed or parallelised, are just now beginning to gather traction \citep{Carrillo2014, Ferrari2014}.
Given their flexibility and parallelisation capabilities, the \ac{pd}  and \ac{al} algorithmic frameworks are prime candidates for solving the inverse problems from \ac{ri}.

\subsection{Convex optimisation algorithms for radio-interferometry}

Under the \ac{cs} paradigm, we can redefine the inverse problem as the estimation of the image $\bs{x} \in \bb{R}^N$ given the measurements $\bs{y} \in \bb{C}^M$ under the constraint that the image is sparse in an over-complete dictionary $\bm{\Psi}$.
Since the solution of interests is an intensity image, we also require $\bs{x}$ to be real and positive.
The analysis formulation (\ref{analysis-l1-problem}) is more tractable \bc since it generally produces a simpler optimisation
problem when over-complete dictionaries are used \citep{Elad2007}.
Additionally, the constrained formulation \ec offers an easy way of defining the minimisation given accurate \bc noise estimates.\ec

Thus, we state the reconstruction task as the convex minimisation problem \citep{Carrillo2013, Carrillo2014}
\begin{equation}
	\min_{\bs{x}} f(\bs{x}) + l(\bm{\Psi}^\dagger \bs{x}) + h(\bm{\Phi}\bs{x})
	\label{basic-min-problem}
\end{equation}
with the functions involved including all the aforementioned constraints,
\begin{equation}
	\begin{aligned}
		l\phantom{(\bs{z})}  & = \| \cdot \|_1, 		        && \\
		f\phantom{(\bs{z})}  & = \iota_{\mc{C}},        & \mc{C} & = \bb{R}^N_+, \\
		h(\bs{z}) & = \iota_{\mc{B}}  (\bs{z}),       & \mc{B} & = \{ \bs{z} \in \bb{C}^M: \| \bs{z} - \bs{y} \|_2 \leq \epsilon \}.
	\end{aligned}
	\label{basic-min-problem-function-def}
\end{equation}
The function $f$ introduces the reality and positivity requirement for the recovered solution, $l$ represents the sparsity prior in the given dictionary $\bm{\Psi}$ and $h$ is the term that ensures data fidelity constraining the residual to be situated in an $\ell_2$ ball defined by the noise level $\epsilon$.

We set the operator $\bm{\Psi} \in \bb{C}^{N \times n_{\rm{b}}N}$ to be a collection of $n_{\rm{b}}$ sparsity inducing bases \citep{Carrillo2014}. The \ac{sara} wavelet bases \citep{carrillo12} are a good candidate but problem (\ref{basic-min-problem}) is not restricted to them.
A re-weighted $\ell_1$ approach \citep{Candes2008} may also be used by implicitly imposing weights on the operator $\bm{\Psi}$ but it is not specifically dealt with herein \bc since it does not change the algorithmic structure. \ec
This would serve to approximate the $\ell_0$ pseudo norm, $\|\bm{\Psi}^\dagger \bs{x}\|_0$, by iteratively re-solving the same problem as in (\ref{basic-min-problem}) with refined weights based on the inverse of the solution coefficients from the previous re-weighted problem.

An efficient parallel implementation can be achieved from (\ref{inverse-problem}) by splitting of the data into multiple blocks 
\begin{equation}
    		\bs{y} = \begin{bmatrix}
                		\bs{y}_1 \\
                		\vdots \\
                		\bs{y}_{n_{\rm{d}}}
                	\end{bmatrix}, \qquad
                	\bm{\Phi} = \begin{bmatrix}
                		\bm{\Phi}_1 \\
                		\vdots \\
                		\bm{\Phi}_{n_{\rm{d}}}
                	\end{bmatrix}
                	= \begin{bmatrix}
                		\bm{G}_1 \bm{M}_1\\
                		\vdots \\
                		\bm{G}_{n_{\rm{d}}} \bm{M}_{n_{\rm{d}}} 
                	\end{bmatrix}  \bm{F}\bm{Z}.
        	\label{data-split}
\end{equation}
Since $\bm{G}_j \in \bb{C}^{M_j \times n_{\rm{o}}N_j}$ is composed of compact support kernels, the matrices $\bm{M}_j \in \bb{R}^{n_{\rm{o}}N_j \times n_{\rm{o}} N}$ can be introduced to select only the parts of the discrete Fourier plane involved in computations for block $j$, masking everything else.
The selected, $n_{\rm{o}}N_j$, $N_j \leq N$, frequency points are directly linked to the continuous $u$--$v$ coordinates associated with each of the visibility measurements from block $\bs{y}_j$.
Thus, for a \emph{compact grouping} of the visibilities in the $u$--$v$ space, each block only deals with a limited frequency interval.
These frequency ranges are not disjoint since a discrete frequency point is generally used for multiple visibilities due to the interpolation kernels and DDEs modelled through the operator $\bm{G}_j$.
Since both have a compact support in frequency domain, without any loss of generality, we consider for each block $j$ an overlap of $n_{\rm{v}}$ such points.

We rewrite (\ref{inverse-problem}) for each data block as%
\begin{equation}%
\bs{y}_j = \bm{\Phi}_j \bs{x} + \bs{n}_j,
\end{equation}
with $\bs{n}_j$ being the noise associated with the measurements $\bs{y}_j$.
\bc Thus\ec, we can redefine the minimisation problem (\ref{basic-min-problem}) as%
\begin{equation}%
	\min_{\bs{x}} f(\bs{x}) + \sum_{i=1}^{n_{\rm{b}}} l_i(\bm{\Psi}^\dagger_i\bs{x}) + \sum_{j=1}^{n_{\rm{d}}} h_j(\bm{\Phi}_j\bs{x})
	\label{split-min-problem}
\end{equation}
where, similarly to (\ref{basic-min-problem-function-def}), we have
\begin{equation}
	\begin{aligned}
		&l_i\phantom{(\bs{z})}~ = \| \cdot \|_1,\\
		&h_j(\bs{z}) = \iota_{\mc{B}_j}  (\bs{z}), \hquad \mc{B}_j = \{ \bs{z} \in \bb{C}^{M_j}: \| \bs{z} - \bs{y}_j \|_2 \leq \epsilon_j \}.
	\end{aligned}
	\label{split-function-definition}
\end{equation}
\bc Here, $\epsilon_j$ represents the bound on the noise for each block. \ec
For the sparsity priors, the $\ell_1$ norm is additively separable and the splitting of the bases used,
\begin{equation}
            	\bm{\Psi} = \begin{bmatrix}
            		\bm{\Psi}_1 & \ldots & \bm{\Psi}_{n_{\rm{b}}}
            	\end{bmatrix},
        	\label{basis-split}
\end{equation}
with $\bm{\Psi}_i \in \bb{C}^{N \times N}$ for $i \in \{1, \ldots, n_{\rm{b}}\}$, is immediate.
The new formulation involving the $\ell_1$ terms remains equivalent to the original one.
Note that there are no restrictions on the number of blocks $\bm{\Psi}$ is split into.
However, a different splitting strategy may not allow for the use of fast algorithms for the computation of the operator.

Hereafter we focus on the block minimisation problem defined in (\ref{split-min-problem}) and we describe two main algorithmic structures for finding the solution.
The first class of methods uses a proximal \ac{admm} and details the preliminary work of \cite{Carrillo2015}.
The second is based on the \ac{pd} framework and introduces to \ac{ri}, a new algorithm able to achieve the full splitting previously mentioned.
These methods have a much lighter computational burden than the \ac{sdmm} solver previously proposed by \cite{Carrillo2014}.
They are still able to achieve a similar level of parallelism, either through an efficient implementation in the case of \ac{admm} or, in the case of \ac{pd}, by making use of the inherent parallelisable structure of the algorithm.
The main bottleneck of \ac{sdmm}, which the proposed algorithms avoid, is the need to compute the solution of a linear system of equations, at each iteration.
Such operation can be prohibitively slow for the large \ac{ri} data sets and makes the method less attractive.
The structure of \ac{sdmm} is presented \bc in Appendix \ref{sdmm}, Algorithm \ref{alg-sdmm}. \ec
For its complete description in the \ac{ri} context we direct the reader to \cite{Carrillo2014}, the following presentation being focused on the \ac{admm} and \ac{pd} algorithms.

\subsection{\bc Dual forward-backward based alternating direction method of multipliers \ec}
\label{sec-admm}

The \ac{admm} is only applicable to the minimisation of a sum of two functions and does not exhibit any intrinsic parallelisation structure.
However, by rewriting the minimisation problem from (\ref{split-min-problem}) as
\begin{equation}
	\min_{\bs{x}}  \bar{f}(\bs{x})+ \bar{h}(\bm{\Phi}\bs{x}),
	\label{split-min-problem-admm}
\end{equation}
an efficient parallel implementation may be achieved.
We define the two functions involved in as
\bc
\begin{equation}
	\bar{f}(\bs{x}) = f(\bs{x}) + \sum_{i=1}^{n_{\rm{b}}} l_i(\bm{\Psi}^\dagger_i\bs{x}), \quad
	\bar{h}(\bm{\Phi}\bs{x}) = \sum_{j=1}^{n_{\rm{d}}} h_j(\bm{\Phi}_j\bs{x}).
\end{equation}
\ec
Furthermore, since $\bar{h}$ is a sum of indicator functions $\iota_{\mc{B}_j}(\bm{\Phi}_j\bs{x})$, we can redefine it as \bc
	$\bar{h}(\bm{\Phi}\bs{x}) = \iota_{\bar{\mc{B}}} (\bm{\Phi}\bs{x})$,
with $\bar{\mc{B}}=\mc{B}_1 \times \mc{B}_2  \times \ldots  \times \mc{B}_{n_{\rm{d}}}$. \ec

\ac{admm} iteratively searches for the solution to an augmented Lagrangian function similar to~(\ref{saddle-point-admm}).
The computations are performed in a serial fashion and explicit parallelisation may only be introduced inside each of its three algorithmic steps.
Thus, at each iteration, \ac{admm} alternates between the minimisation
\begin{equation}
	\min_{\bs{x}} \mu \bar{f}(\bs{x}) + \frac{1}{2} \big\|\bm{\Phi} \bs{x} + \bs{s} - \bs{r} \big\|_2^2\\
	\label{admm-basic-min-steps-x}
\end{equation}
over the variable of interest $\bs{x}$ and the minimisation involving the slack variable $\bs{r}$,
\begin{equation}
	\min_{\bs{r}} \mu \bar{h}(\bs{r}) + \frac{1}{2} \big\| \bs{r} - \bm{\Phi} \bs{x} - \bs{s} \big\|_2^2.
	\label{admm-basic-min-steps-r}
\end{equation}
These are followed by a gradient ascent with a step $\varrho$ performed for the Lagrange multiplier variable $\bs{s}$.
Given the definition of the function $\bar{h}(\bs{r})$, the minimisation involving $\bs{r}$ can be split into $n_{\rm{d}}$ independent sub-problems
\begin{equation}
	\min_{\bs{r}_j} \mu \bar{h}_j(\bs{r}_j)  +  \frac{1}{2} \big\| \bs{r}_j - \bm{\Phi}_j \bs{x} - \bs{s}_j \big\|_2^2, \quad j \in \{1, \ldots, n_{\rm{d}}\}.
	\label{admm-basic-min-slack-var-split}
\end{equation}
This minimisation amounts to computing the proximity operator of $\mu \bar{h}_j$ at $\bm{\Phi}_j \bs{x} + \bs{s}_j$, which, given the definition of the function $\bar{h}_j$, reduces to a projection operation.
The method imposes that every $\bs{r}_j$ approaches $\bm{\Phi}_j\bs{x}$ while $\bs{x}$ converges towards the solution.
The convergence speed is governed by the Lagrange multiplier $\mu$ and by the ascent step $\varrho$ associated with the maximisation over the Lagrange multiplier variable $\bs{s}$.

\begin{figure}
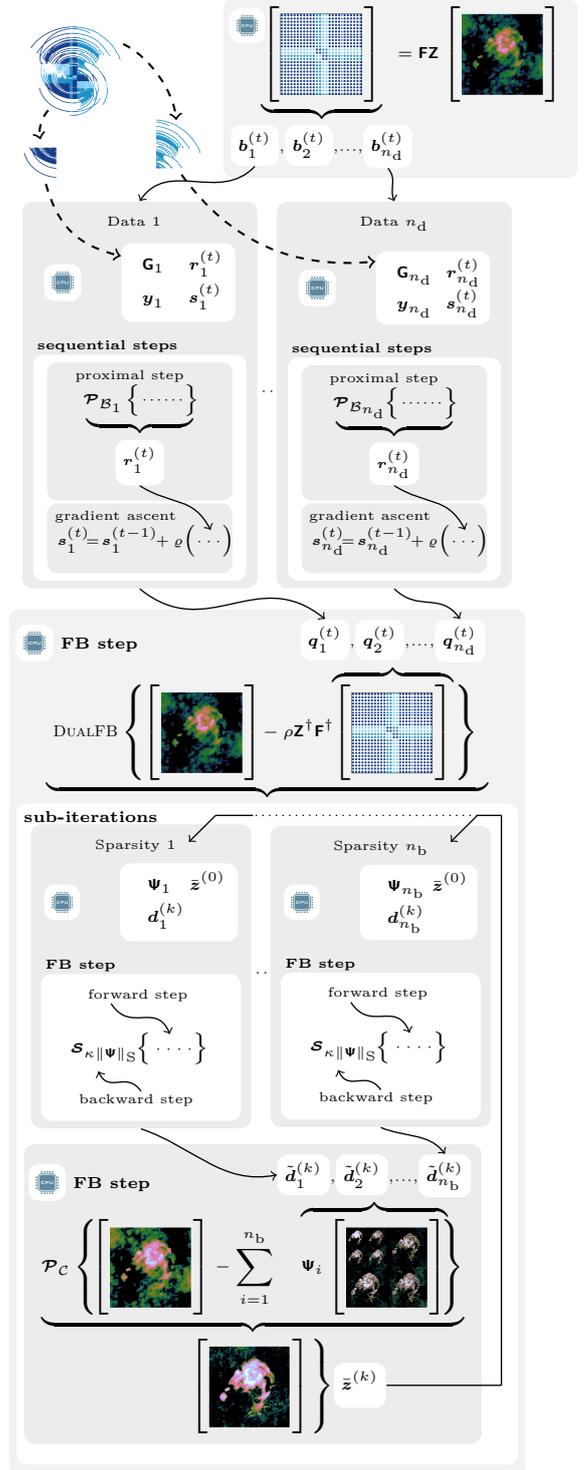

    	\hspace{-5pt}
	\begin{minipage}{0.98\linewidth}
		\bc
		\include{algo-admm}
		\ec
    	\end{minipage}
	\vspace{15pt}
	\caption{\bc The diagram of the structure of \ac{admm}, detailed in Algorithm \ref{alg-admm}, showcasing the parallelism capabilities and overall computation flow. The algorithm performs in parallel proximal and gradient updates (similarly to the \ac{clean} performing major-minor cycle) for all data fidelity terms. Its structure is sub-iterative and enforces sparsity and positivity through the dual \ac{fb} algorithm. These updates, performed in parallel for each sparsity basis, can be again seen as analogous to \alg{clean}. Thus, the whole algorithm can be seen as composed of interlaced \alg{clean}-like proximal splitting and \ac{fb} updates running in parallel in multiple data, prior, and image spaces. \ec
	}
	\label{algo-fig-admm}
\end{figure}

\bc A proximal version of \ec \ac{admm} deals with the non-smooth functions from (\ref{admm-basic-min-steps-x}) and (\ref{admm-basic-min-slack-var-split}) by approximating the solution via proximal splitting.
Algorithm~\ref{alg-admm} presents the details.
\bc In Figure~\ref{algo-fig-admm} we present a diagram of the algorithm to provide further insight into its parallelisation and distribution capabilities. It can also be used to understand the algorithm from the \alg{clean} perspective, performing \ac{fb} \alg{clean}-like updates in multiple data, prior and image spaces. \ec
Data fidelity is enforced through the slack variables $\bs{r}^{(t)}_j$, by minimising (\ref{admm-basic-min-slack-var-split}) and thus constraining the residual to belong to the $\ell_2$ balls $\mc{B}_j$.
This accepts a closed form solution and, for each ball $j$, represents the projection,
\begin{equation}
	\proj_{\mc{B}_j}(\bs{z})  \overset{\Delta}{=} \left\{ 
	\begin{aligned}
		\epsilon_j \frac{\bs{z} - \bs{y}_j}{\|\bs{z} - \bs{y}_j\|_2} + \bs{y_j} & \qquad \|\bs{z} - \bs{y}_j\|_2 > \epsilon_j\\
		\bs{z} \qquad \qquad & \qquad \|\bs{z} - \bs{y}_j\|_2 \leq \epsilon_j
	\end{aligned}\right.
	\label{proj-L2}
\end{equation}
onto the feasible regions defined by it.
Given the structure of the function $\bar{h}$, this is implemented in parallel with distributed computations and presented in Algorithm \ref{alg-admm}, step $8$, together with the update of the Lagrange variables $\bs{s}_j^{(t)}$, in step $9$.
The variables $\bs{b}_j^{(t)} \in \bb{C}^{ n_{\rm{o}}N_j}$, computed in steps $3$ to $6$, are required in the computations and need to be transmitted to the different processing nodes.
The nodes compute the solution updates $\bs{q}^{(t)}_j \in \bb{C}^{ n_{\rm{o}}N_j}$ in step $10$ after which they are centralised and used to revise the previous solution estimate $\bs{x}^{(t-1)}$ and to compute $\bs{x}^{(t)}$.
Thus, by carefully defining the minimisation problem, a high degree of parallelism is achieved.
Note that this step can easily incorporate all types of weighting of the data specific to \ac{ri}.

For our specific problem, the minimisation over $\bs{x}$ from (\ref{admm-basic-min-steps-x}) does not accept a closed form solution.
We approximate it by using a \ac{fb} step.
The \emph{forward} step corresponds to a gradient step and the \emph{backward} step is an implicit sub-gradient-like step performed through the proximity operator.
Thus, in step $12$, the solution is updated using the descent step $\rho$, in the direction of the gradient of the smooth part.
This is followed by the iterative dual \ac{fb} \citep{Combettes2011} updates necessary to approximate the proximity operator to the non smooth $\bar{f}$.
Algorithm~\ref{alg-admm}, function \alg{DualFB}, details the required sub-iterations.
In steps $23$ and $20$, the method alternates between, projections onto the convex set $\mc{C}$, which, component wise, are defined as
\begin{equation}
	\Big( \proj_{\mc{C}}(\bs{z}) \Big)_k  \overset{\Delta}{=} \left\{ 
	\begin{array}{cl}
		\Re(z_k) & \qquad \Re(z_k) > 0 \\
		0 & \qquad \Re(z_k) \leq 0
	\end{array}\right. \quad \forall k,
	\label{proj-plus}
\end{equation}
and the application of the proximity operator to the sparsity prior functions $l_i$, which is the component wise soft-thresholding operator
\begin{equation}
	\Big( \soft_{\alpha}(\bs{z}) \Big)_k \overset{\Delta}{=} \left\{ 
	\begin{array}{cl}
		\ds \frac{z_k\{ | z_k | - \alpha \}_{+}}{| z_k |} & \qquad | z_k | > 0\\
		\ds 0 & \qquad | z_k | = 0\\
	\end{array}\right. \quad \forall k,
	\label{prox-L1}
\end{equation}
with threshold $\alpha$.
The soft threshold resulting for the algorithm is $\eta \rho \mu$.
However, since $\mu$ is a free parameter, we re-parameterise the operation to use the soft threshold $\kappa \|\bm{\Psi}\|_{\rm{S}}$, with $\kappa$ as a new scale-free parameter, independent of the operator $\bm{\Psi}$ used. Here, we denote by $\|\bm{\Psi}\|_{\rm{S}}$ the operator norm of the sparsifying transform.
The operator $\{\cdot\}_{+}$ from (\ref{prox-L1}) sets the negative values to $0$.
The parameter $\eta$ serves as an update step for the sub-problem.
In step $20$, we have additionally used the Moreau decomposition (\ref{moreau-decomposition}) to replace the proximity operator of the conjugates $l_i^*$ with that of the functions $l_i$,
with  $\identity$ denoting the identity operator.
The computations involving each basis $\bm{\Psi}^\dagger_i$ are to be performed in parallel, locally.
Distributed processing is problematic here due to the large size of the image $\bar{\bs{z}}^{(k)}$ that would need to be transmitted.

\begin{algorithm}[t]
\caption{Dual forward-backward \ac{admm}.}
\label{alg-admm}

\begin{algorithmic}[1]
\small
\Given{$\bs{x}^{(0)}, \bs{r}_j^{(0)}, \bs{s}_j^{(0)}, \bs{q}_j^{(0)}, \kappa, \rho, \varrho$}
\RepeatFor{$t=1,\ldots$}

\State $\ds \tilde{\bs{b}}^{(t)} = \bm{F}\bm{Z} \bs{x}^{(t-1)}$
\Set{$\forall j \in \{1, \ldots, n_{\rm{d}}\}$}
	\State $\ds \bs{b}_j^{(t)} = \bm{M}_j \tilde{\bs{b}}^{(t)}$
\EndSet
\ParForD{$\forall j \in \{1, \ldots, n_{\rm{d}}\}$}{$\bs{b}^{(t)}_j$}
	\State $\bs{r}_j^{(t)} = \proj_{\mc{B}_j} \Big( \bm{G}_j \bs{b}^{(t)}_j + \bs{s}_j^{(t-1)} \Big)$
	\State $\bs{s}_j^{(t)} = \bs{s}_j^{(t-1)} + \varrho \big( \bm{G}_j \bs{b}^{(t)}_j - \bs{r}_j^{(t)} \big)$
	\State $\bs{q}_j^{(t)} = \bm{G}^\dagger_j \bigg( \bm{G}_j \bs{b}^{(t)}_j + \bs{r}_j^{(t)} - \bs{s}_j^{(t)} \bigg)$
\EndParForD{$\bs{q}_j^{(t)}$}
\State $\ds \tilde{\bs{x}}^{(t)} = \bs{x}^{(t-1)} - \rho \bm{Z}^\dagger \bm{F}^\dagger \sum_{j=1}^{n_{\rm{d}}} \bm{M}_j^\dagger \bs{q}_j^{(t)}$
\State $\bs{x}^{(t)} =$ \FuncCall{DualFB}{$\tilde{\bs{x}}^{(t)}, \kappa$}
\Until {\bf convergence \normalfont}
\vspace{5px}
\Func{DualFB}{$\bs{z}, \kappa$}
\Given{$\bs{d}^{(0)}_i, \eta$}
\State $\ds \bar{\bs{z}}^{(0)} = \proj_{\mc{C}}  \big( \bs{z} \big)$
\RepeatFor{$k=1,\ldots$}
	\ParFor{$\forall i \in \{1, \ldots, n_{\rm{b}}\}$}
		\State $\bs{d}^{(k)}_i = \frac{1}{\eta} \Bigg(\! \identity - \soft_{\kappa \|\bm{\Psi}\|_{\rm{S}}} \!\!\Bigg)\! \Big( \eta \bs{d}^{(k-1)}_i + \bm{\Psi}_i^\dagger \bar{\bs{z}}^{(k-1)} \Big)$
		\State $\bc \tilde{\bs{d}}^{(k)}_i = \bm{\Psi}_i \bs{d}^{(k)}_i \ec $
	\EndParFor
	\vspace{-5px}
	\State $\ds \bar{\bs{z}}^{(k)} = \proj_{\mc{C}}  \bigg(\bs{z} -  \sum_{i=1}^{n_{\rm{b}}}  \bc \tilde{\bs{d}}^{(k)}_i \ec \bigg)$
\Until {\bf convergence \normalfont}
\EndFunc{$\bar{\bs{z}}^{(k)}$}
\end{algorithmic}
\end{algorithm}

\subsection{\bc Primal-dual algorithms with randomisation \ec}
\label{sec-pdr}

\begin{figure}
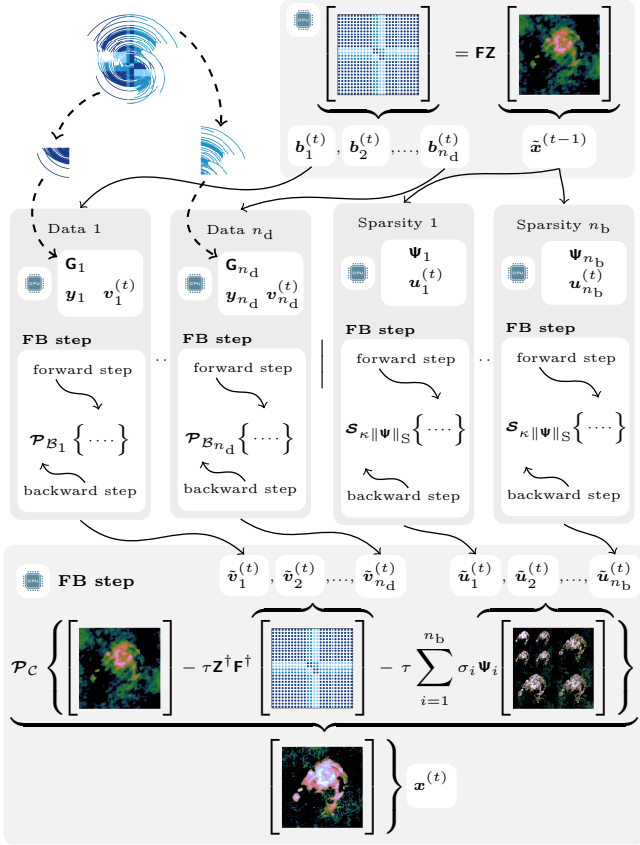


    	\hspace{-5pt}
	\begin{minipage}{0.98\linewidth}
		\bc
		\include{algo-pd}
		\ec
    	\end{minipage}
	\vspace{8pt}
	\caption{\bc The diagram of structure of \ac{pd}, detailed in Algorithm \ref{alg-primal-dual}, showcasing the parallelism capabilities and overall computation flow.
In contrast with \ac{admm}, the \ac{pd} algorithm is able to perform all updates on the dual variables $\bs{v}_i^{(t)}$ and $\bs{u}_j^{(t)}$ using \ac{fb} iterations and in parallel.
The update of the primal variable $\bs{x}^{(t)}$ is also a \ac{fb} step.
Viewed though the perspective of the intuitive similarity between a \ac{fb} iteration and \alg{clean}, this translates to performing \alg{clean}-like iterations in parallel in multiple data, prior, and image spaces.\ec
	}
	\label{algo-fig-pd}
\end{figure}

\begin{algorithm}[t]
\caption{Randomised forward-backward \ac{pd}.}
\label{alg-primal-dual}

\begin{algorithmic}[1]
\small

\Given{$\bs{x}^{(0)}, \bc \tilde{ \ec \bs{x} }^{\ec (0)} \ec, \bs{u}_i^{(0)}, \bs{v}_j^{(0)}, \tilde{\bs{u}}^{(0)}_i, \tilde{\bs{v}}^{(0)}_j, \kappa, \tau, \sigma_i, \varsigma_j, \lambda$}

\RepeatFor{$t=1,\ldots$}
\State {\bf generate sets} $\mc{P} \subset \{1,\ldots, n_{\rm{b}}\}$ {\bf and} $\mc{D} \subset \{1,\ldots, n_{\rm{d}}\}$

\State $\ds \tilde{\bs{b}}^{(t)} = \bm{F}\bm{Z} \tilde{\bs{x}}^{(t\bc -1 \ec)}$
\Set{$\forall j \in \mc{D}$}
	\State $\ds \bs{b}_j^{(t)} = \bm{M}_j \tilde{\bs{b}}^{(t)}$
\EndSet
\Block{\bf run simultaneously}
\ParForD {$\forall j \in \mc{D}$}{$\bs{b}_j^{(t)}$}
	\State $\ds \bar{\bs{v}}_j^{(t)} = \Bigg( \identity - \proj_{\mc{B}_j} \Bigg) \Big( \bs{v}_j^{(t-1)} + \bm{G}_j \bs{b}^{(t)}_j \Big)$
	\State $\ds \bs{v}_j^{(t)} = \bs{v}_j^{(t-1)} + \lambda \left( \bar{\bs{v}}_j^{(t)} - \bs{v}_j^{(t-1)} \right)$
	\State $\ds \tilde{\bs{v}}^{(t)}_j = \bm{G}_j^\dagger \bc \bs{v}^{(t)}_j \ec$
\EndParForD{$\tilde{\bs{v}}^{(t)}_j$}
\Set{$\forall j \in \{1, \ldots n_{\rm{d}}\} \setminus \mc{D}$}
	\State $\ds \bs{v}^{(t)}_j = \bs{v}^{(t-1)}_j$
	\State $\ds \tilde{\bs{v}}^{(t)}_j = \tilde{\bs{v}}^{(t-1)}_j$
\EndSet
\ParFor{$\forall i \in \mc{P}$}
	\State $\ds \bar{\bs{u}}_i^{(t)} = \Bigg( \identity - \soft_{\kappa \|\bm{\Psi}\|_{\rm{S}}} \Bigg) \Big( \bs{u}_i^{(t-1)} + \bm{\Psi}_i^\dagger  \tilde{\bs{x}}^{(t\bc - 1\ec)} \Big)$
	\State $\ds \bs{u}_i^{(t)} = \bs{u}_i^{(t-1)} + \lambda \left( \bar{\bs{u}}^{(t)} - \bs{u}_i^{(t-1)} \right)$
	\State $\ds \tilde{\bs{u}}^{(t)}_i = \bm{\Psi}_i \bs{u}^{(t)}_i$
\EndParFor
\Set{$\forall i \in \{1, \ldots n_{\rm{b}}\} \setminus \mc{P}$}
	\State $\ds \bs{u}^{(t)}_i = \bs{u}^{(t-1)}_i$
	\State $\ds \tilde{\bs{u}}^{(t)}_i = \tilde{\bs{u}}^{(t-1)}_i$
\EndSet
\EndBlock{\bf end}
\vspace{-5px}
\State $\ds \bar{\bs{x}}^{(t)} \! = \proj_{\mc{C}} \! \Bigg(\! \bs{x}^{(t-1)}  - \tau  \Big(\!  \bm{Z}^\dagger \bm{F}^\dagger \! \sum_{j=1}^{n_{\rm{d}}} \! \varsigma_j  \bm{M}_j^\dagger \tilde{\bs{v}}_j^{(t)} \! +\! \sum_{i=1}^{n_{\rm{b}}} \! \sigma_i \tilde{\bs{u}}_i^{(t)} \Big)\!\!\Bigg)$
\State $\ds \bs{x}^{(t)} \!= \bs{x}^{(t-1)} + \lambda \left( \bar{\bs{x}}^{(t)} - \bs{x}^{(t-1)} \right)$
\State $\ds \bc \tilde{\bs{x}}^{(t)} = 2\bar{\bs{x}}^{(t)} - \bs{x}^{(t-1)} \ec$
\Until {\bf convergence}
\end{algorithmic}
\end{algorithm}

The main advantage that makes the \ac{pd} algorithms attractive for solving inverse problems is their flexibility and scalability.
They are able to deal with both differentiable and non-differentiable functions and are applicable to a broad range of minimisation tasks.
The inherent parallelisation on the level of splitting the functions gives a direct approach for solving (\ref{split-min-problem}).
Another important aspect is given by the use of randomisation, allowing the update for a given component function to be performed less often and thus lowering the computational cost per iteration.
Block coordinate computations are also supported but are not explicitly used herein.

We define the minimisation task to be solved using \ac{pd} methods, similarly to~(\ref{split-min-problem}), as
\begin{equation}
	\min_{\bs{x}} f(\bs{x}) + \gamma \sum_{i=1}^{n_{\rm{b}}} l_i(\bm{\Psi}^\dagger_i\bs{x}) + \sum_{j=1}^{n_{\rm{d}}} h_j(\bm{\Phi}_j\bs{x}),
	\label{split-min-problem-gamma}
\end{equation}
where  $\gamma$ is an additional tuning parameter.
Note that the minimisation problem does not change, regardless of the value $\gamma$ takes due to the use of the indicator functions in $f$ and $h_j$ which are invariant to scaling.
This fits under the framework introduced by \cite{Condat2013, Vu2013, Pesquet2014} and we devise a \ac{pd} algorithm towards finding the solution.
The method iteratively alternates between solving the primal problem (\ref{split-min-problem-gamma}) and the dual problem,%
\begin{equation}%
	\begin{aligned}%
		\min_{\substack{\bs{u}_i \\ \bs{v}_j}} f^* \Bigg(-\sum_{i=1}^{n_{\rm{b}}} \bm{\Psi}_i\bs{u}_i - & \sum_{j=1}^{n_{\rm{d}}} \bm{\Phi}^\dagger_j \bs{v}_j \Bigg) \\
		& + \frac{1}{\gamma} \sum_{i=1}^{n_{\rm{b}}} l_i^*(\bs{u}_i) + \sum_{j=1}^{n_{\rm{d}}} h_j^*(\bs{v}_j),
	\end{aligned}
	\label{split-min-dual-problem}
\end{equation}
essentially converging towards a Kuhn-Tucker point.
This produces the algorithmic structure of Algorithm \ref{alg-primal-dual} where additionally we have used the Moreau decomposition (\ref{moreau-decomposition}) to rewrite the proximal operations and replace the function conjugates.
\bc A diagram of the structure is presented in Figure~\ref{algo-fig-pd} further exemplifying the conceptual analogy between the \ac{pd} algorithm and \alg{clean}. \ec
\bc The algorithm \ec allows the full split of the operations and performs all the updates on the dual variables in parallel.
The update of the primal variable, the image of interest $\bs{x}^{(t)}$, requires the contribution of all dual variables $\bs{v}^{(t)}_i$ and $\bs{u}^{(t)}_j$.
The algorithm uses the update steps $\tau$, $\sigma_i$ and $\varsigma_j$ to iteratively revise the solution and allows for a relaxation with the factor $\lambda$.
\ac{fb} iterations, consisting of a gradient descent step coupled with a proximal update, are used to update both the primal and the dual variables.
\bc These \ac{fb} updates can be seen as \alg{clean}-like steps performed in the multiple signal spaces associated with the primal and the dual variables. \ec
In the deterministic case, the \emph{active sets} $\mc{P}$ and $\mc{D}$ are fixed such that all the dual variables are used.
The randomisation capabilities of the algorithm are presented later on, given a probabilistic construction of the active sets.

When applied in conjunction with the functions from (\ref{split-function-definition}),
the \emph{primal update} from step $28$ is performed through the projection (\ref{proj-plus}) onto the positive orthant defined by $\mc{C}$.
The dual variables are updated in steps $10$ and $19$ using the proximity operators for $h_j$ and $l_i$, which become the projection onto an $\ell_2$ ball $\mc{B}_j$ defined by (\ref{proj-L2}) 
and the component wise soft-thresholding operator (\ref{prox-L1}).
We use the Moreau decomposition (\ref{moreau-decomposition}) to replace the proximity operator of the conjugate functions $l_i^*$ and $h_j^*$ with that of the function $l_i$ and $h_j$, respectively. The identity operator is denoted by $\op{I}$.
Step $19$ also contains a re-parametrisation similar to the one performed for \ac{admm}.
We replace the implicit algorithmic soft-threshold size $\sfrac{\gamma}{\sigma_i}$ with $\kappa \|\bm{\Psi}\|_{\rm{S}}$ by appropriately choosing the free parameter $\gamma$. This ensures that we are left with the scale-free parameter $\kappa$ independent to the operator $\bm{\Psi}$.
Steps $11$, $20$ and $29$ represent the relaxation of the application of the updates.
To make use of the parallelisation, the application of the operators $\bm{G}_j^\dagger$ and $\bm{\Psi}_i$ is also performed in parallel, in steps $12$ and $21$.
Note that the splitting of the operators is presented in (\ref{data-split}), more specifically $\bm{\Phi}_j = \bm{G}_j \bm{M}_j \bm{F}\bm{Z}$, $\forall j \in \{1, \ldots, n_{\rm{d}}\}$.
\bc These 
operations are given in steps $4$ to $7$.\ec

The computation of the dual variables $\bs{u}_i^{(t)}$ associated with the sparsity priors requires the current solution estimate.
This solution estimate is then revised with the updates $\tilde{\bs{u}}_i^{(t)}$ computed from the dual variables. 
Both $\bs{x}^{(t)}$ and $\tilde{\bs{u}}_i^{(t)}$ are of size $N$ and their communication might not be desirable in a loosely distributed system.
In such case all computations involving $\bs{u}_i^{(t)}$ can be performed in parallel but not in a distributed fashion.
The dual variables $\bs{v}_j^{(t)}$, associated with the data fidelity functions, should be computed over a distributed computing network. 
They only require the communication of the updates $\bs{b}^{(t)}_j \in \bb{C}^{ n_{\rm{o}}N_j}$ and dual updates $\tilde{\bs{v}}_j^{(t)} \in \bb{C}^{ n_{\rm{o}}N_j}$ which remains feasible.


The main challenge associated with the inverse problem defined by (\ref{inverse-problem}) is linked with the dimensionality of the data.
The large data size is a limiting factor not only from the computational perspective but also from that of memory availability.
A randomisation of the computations following the same \ac{pd} framework \citep{Pesquet2014} is much more flexible at balancing memory and computational requirements.
By selectively deciding which data fidelity and sparsity prior functions are active at each iterations, full control over the memory requirements and computational cost per iteration can be achieved.
In Algorithm \ref{alg-primal-dual}, this is controlled by changing the sets $\mc{P}$, containing the active sparsity prior dual variables, and $\mc{D}$, which governs the selection of the data fidelity dual variables.
At each iteration, each dual variable has a given probability of being selected, $p_{\mc{P}_i}$ for the sparsity prior, and  $p_{\mc{D}_j}$ for the data fidelity, respectively.
These probabilities are independent of each other.
Note that the algorithm has \emph{inertia} still performing the primal updates using all dual variables even though some dual variables remain unchanged.

\section{Implementation details and computational complexity}
\label{sec-imp}

\begin{table*}

  	\caption{Complexity of \ac{admm} (top) and \ac{pd} (bottom) algorithms for one iteration. Each node has its computational load listed. 
The \ac{admm} algorithm iterates $n_{\bar{f}}$ times over steps $18$ to $23$. The serial nature of its structure can be observed, the nodes not operating simultaneously.
The \ac{pd} methods alternate between updating the primal and the dual variables. All dual variables are computed in parallel. The visibility data are assumed to be split into compact blocks composed of an equal number of visibilities in the $u$--$v$ space.}
  	\label{complexity-table}
  	\centering
	\small
  	\begin{tabular}{@{\hspace{15pt}}l@{\hspace{2pt}}l@{\hspace{3pt}}lcc@{\hspace{22pt}}}
	\hline
	\multicolumn{3}{c}{Algorithm~\ref{alg-admm}\bc~\acp{admm}\ec} & central node & $n_{\rm{d}}$ data fidelity nodes \\ \hline {\phantom{$\bigg(\bigg.$}}
	& steps & $3$-$6$ &  $\mc{O}\big(n_{\rm{o}}N \log n_{\rm{o}}N \big)$  & --- \\{\phantom{$\bigg(\bigg.$}}
	&steps & $8$-$10$ & --- & $\qquad \mc{O}\bigg(2\frac{n_{\rm{s}}n_{\rm{o}}}{n_{\rm{d}}}M N_j + \bc M_j \ec\!\bigg) \quad$ \\{\phantom{$\bigg(\bigg.$}}
	& step & $12$ &  $\: \hquad \qquad \qquad \mc{O}\big(n_{\rm{o}}N \log n_{\rm{o}}N \big) + \mc{O}\big(n_{\rm{o}}N + n_{\rm{d}} n_{\rm{v}}\big) \hquad \qquad $  & --- \\{\phantom{$\bigg(\bigg.$}}
	$n_{\bar{f}} \times $ & steps & $17$-$22$ &  $\mc{O}\big(2n_{\rm{b}}N\big)$  & --- \\\hline 
  	\end{tabular}%
	\vspace{8.5pt}\hfill%
	\begin{tabular}{l@{\hspace{2pt}}lcc}
	\hline
	&  Algorithm~\ref{alg-primal-dual}\bc~\acp{pd}\ec & central node & $n_{\rm{d}}$ data fidelity nodes \\ \hline {\phantom{$\bigg(\bigg.$}}
	& steps  $4$-$8$ &  $\mc{O}\big(n_{\rm{o}}N \log n_{\rm{o}}N \big)$ & --- \\ {\phantom{$\bigg(\bigg.$}}
	& steps $10$-$27$ &  $p_{\mc{P}_i}  \mc{O}\big(2n_{\rm{b}}N\big)$ & $ \qquad p_{\mc{D}_j} \mc{O}\bigg( 2\frac{n_{\rm{s}}n_{\rm{o}}}{n_{\rm{d}}}M N_j + \bc M_j \ec \!\bigg) \quad$ \\ {\phantom{$\bigg(\bigg.$}} 
	& steps  $29$-$30$ &  $\qquad \quad ~ \mc{O}\big(n_{\rm{o}}N \log n_{\rm{o}}N \big) + \mc{O}\big((n_{\rm{b}} + n_{\rm{o}})N + n_{\rm{d}} n_{\rm{v}}\big) \qquad$ & --- \\ \hline 
  	\end{tabular}
\end{table*}

An efficient implementation of the \ac{admm} and the \ac{pd} algorithms takes advantage of the data split and of the implicit parallelisation from the definition of the minimisation problem.
For presentation simplicity, we consider the processing to be split between a \emph{central meta-node}, a single processing unit or possibly a collection of nodes, centralising the update on the desired solution $\bs{x}^{(t)}$ and performing the computations associated with the sparsity priors, and a number of \emph{data fidelity nodes} dealing with the constraints involving the balls $\mc{B}_j$.
The computation of the sparsity prior terms can be easily parallelised however, the distribution of the data can be too costly.
In this case, a shared memory architecture might be more appropriate than distributed processing.
For the data nodes, the communication cost is low and a distributed approach is feasible.
We have assumed these two different strategies for dealing with the different terms in the presentation of Algorithms \ref{alg-admm} and \ref{alg-primal-dual}.

Most of the operations to be performed are proportional with $N$ since the main variable of interest $\bs{x}^{(t)}$ is the image to be recovered.
The most demanding operation performed on $\bs{x}^{(t)}$ is the application of the oversampled Fourier operators.
When computed with a fast Fourier algorithm \acp{fft} \citep{Cooley1965}, the computational cost of the transforms $\bm{F}$ and $\bm{F}^\dagger$ applied to $n_{\rm{o}}$-oversampled data scales as $\mc{O}\left( n_{\rm{o}}N \log n_{\rm{o}}N \right)$.
It should be noted that the \ac{fft} implementation can be sped up by using multiple processing cores or nodes.
The wavelet operators $\bm{\Psi}$ and $\bm{\Psi}^\dagger$ are applied to the image  $\bs{x}^{(t)}$ as well. 
The Discrete Wavelet Transform \acp{dwt} can be performed with fast wavelet implementations using lifting schemes or filter banks \citep{Cohen1993, Daubechies1998, Mallat2008} and achieves a linear complexity of $\mc{O}(N)$ for compactly supported wavelets.
A distributed processing of the operations involved in the application of each sparsity basis $\bm{\Psi}_i$ may be used. 
However, this requires the communication of the current solution estimate, which might not be feasible.
We consider that these computations are performed locally, on the central meta-node.

For the data nodes, a manageable computational load and an efficient communication can be achieved by both algorithms by adopting a \emph{balanced} and \emph{compact} split of the data; splitting the data into blocks of similar size having a compact frequency range as proposed in (\ref{data-split}).
An overlap of size $n_{\rm{v}}$ between discrete frequency ranges is necessary for an efficient interpolation \citep{Fessler2003} to the uniform frequency grid which allows fast Fourier computations or to include DDEs \citep{wolz13}.
Besides this overlap, each block only deals with a limited frequency range reducing the communication performed.
In such case, the matrices $\bm{M}_j$ mask out the frequencies outside the range associated with the blocks $\bs{y}_j$.
Furthermore, the use of compact support interpolation kernels and DDEs with compact support in the Fourier domain makes $\bm{G}_j$ sparse, which lowers the computational load significantly.
We consider it has a generic sparsity percentage $n_{\rm{s}}$.

Details on the levels of parallelisation and the scaling to multiple nodes for both methods are presented below.
As mentioned earlier, the main computational difficulties arise from working with large images and data sets, thus making important the way the complexity of the algorithms scales with $N$ and $M$.
An overview of the complexity requirements is presented in Table~\ref{complexity-table}.

\subsection{Alternating direction method of multipliers}

The efficient implementation of \ac{admm} for the problem defined by (\ref{split-min-problem-admm}) offloads the data fidelity computations to the data nodes.
As can be seen from \bc Figure~\ref{algo-fig-admm} and \ec Table~\ref{complexity-table}, the basic structure of the algorithm is serial and the processing is just accelerated by parallelising each serial step.

The iterative updates follow the operations presented in Algorithm \ref{alg-admm}.
The central node computes an estimate $\tilde{\bs{x}}^{(t)}$ of the solution and iteratively updates it to enforce sparsity and positivity.
The update from step $12$ requires $\mc{O}\left( n_{\rm{o}}N \log n_{\rm{o}}N \right)$ operations for the computation of the oversampled \ac{fft}.
Given a compact partitioning of the matrix $\bm{G}$, the sum involving the updates $\bs{q}_j^{(t)}$ requires computations of the order 
$\mc{O}(n_{\rm{o}}N) + \mc{O}(n_{\rm{d}} n_{\rm{v}})$.
Note that it may be accelerated by using the data node network, however since generally $n_{\rm{v}}$ is not large, the gain remains small.
The computation of the Fourier coefficients from step $3$ also incurs a complexity $\mc{O}\left( n_{\rm{o}}N \log n_{\rm{o}}N \right)$.

For the approximation of the proximal operator of the function $\bar{f}$, the algorithm essentially remains serial and requires a number $n_{\bar{f}}$ of iterations.
In this case, the complexity of each update performed for the sparsity prior is dominated by the application of the operators $\bm{\Psi}$ and $\bm{\Psi}^\dagger$,  which, given an efficient implementation of the \ac{dwt} requires $\mc{O}(N)$ operations.
The updates $\bs{d}_i^{(k)}$ \bc and $\tilde{\bs{d}}_i^{(k)}$ from step $20$ and $21$ \ec may be computed in parallel. Given a serial processing \bc however \ec this would need $\mc{O}(n_{\rm{b}}N)$ computations. \bc Note that although in this case the complexity scales linearly with $N$, the scaling constants can make the computations to be of the same level as the \ac{fft}. \ec

The data fidelity nodes perform steps $8$ to $10$ in parallel using the Fourier coefficients $\bs{b}_j^{(t)}$ precomputed in step $5$.
The computations are heavier due to the linear operator $\bm{G}_j$.
As mentioned earlier, the operator has a very sparse structure.
This reduces the computation cost for applying $\bm{G}_j$ or $\bm{G}^\dagger_j$ to $\mc{O}(n_{\rm{s}}M_j n_{\rm{o}}N_j)$, where $n_{\rm{o}} N_j$ is the number of uniformly gridded, frequency points associated with each visibility block $\bs{y}_j$.
The remaining operations only involve vectors of size \bc $M_j$\ec. 
The overall resulting complexity per node is $\mc{O}(n_{\rm{s}}M_j n_{\rm{o}}N_j) + \mc{O}(\bc M_j \ec)$.
Under the assumption that the blocks contain an equal number of visibilities, this further reduces to 
$\mc{O}(\sfrac{ n_{\rm{s}}}{n_{\rm{d}}}M n_{\rm{o}}N_j) + \mc{O}(\bc M_j \ec)$
The communication required between the central and the data fidelity nodes is of order $n_{\rm{o}}N_j$, the size of frequency range of each data block.

\subsection{Primal-dual algorithm}

An implementation of the \ac{PD} algorithms benefits from the full split achieved by the methods which allows for the computation of all the dual variables to be completed in parallel.
The processing is performed in two synchronous alternating serial steps to update the primal and dual variables, respectively. 
Each step is however highly parallelisable.
The central node uses the current estimate of the solution $\bs{x}^{(t-1)}$ and distributes the oversampled Fourier transform coefficients $\bs{b}_j^{(t)}$ to the data fidelity nodes.
The data fidelity and central nodes compute simultaneously the dual variables and provide the updates $\tilde{\bs{v}}_j^{(t)}$ and $\tilde{\bs{u}}_i^{(t)}$ to be centralised and included in the next solution estimate on the central node.
Such a strategy requires at each step the propagation of variables of size $n_{\rm{o}}N_j$, between the central and data fidelity nodes.
As suggested in Algorithms \ref{alg-primal-dual}, the computation of the sparsity prior dual variables is also highly parallelisable.
However, the communication of the current image estimate is required, limiting the possibility to distribute the data due to its large size.
We leave the computation to be performed by the central node, without an explicit exploitation of the possible parallelism.

All dual variables can be computed simultaneously \bc as can be seen in Figure~\ref{algo-fig-pd}. \ec
The data fidelity nodes need to apply the linear operators $\bm{G}_j$ as in steps $10$ and $12$. 
Similarly to \ac{admm}, this incurs the heaviest computational burden.
Given the very sparse structure of the matrix $\bm{G}_j$ this accounts for a complexity of $\mc{O}(n_{\rm{s}}M_j n_{\rm{o}}N_j)$ with $n_{\rm{o}} N_j$ being the previously mentioned number of, uniformly gridded, frequency points for the visibilities $\bs{y}_j$.
The remaining operations only involve vectors of size \bc $M_j$ \ec and thus the overall resulting complexity is $\mc{O}(2n_{\rm{s}}M_j n_{\rm{o}}N_j) + \mc{O}(2\bc M_j\ec)$.
The wavelet decomposition from steps $19$ and $21$ achieves a linear complexity of $\mc{O}(N)$ for compactly supported wavelets.
The other operations from steps $19$ and $20$ are of order $\mc{O}(N)$ resulting in a load for the sparsity prior nodes that scales linearly with $N$.

In step $28$ of Algorithm \ref{alg-primal-dual}, the summing of the sparsity prior updates requires $\mc{O}(n_{\rm{b}}N)$ operations.
For the $\ell_2$ data fidelity terms, given a compact partitioning in frequency for the matrix $\bm{G}$, the computation requires $\mc{O}(n_{\rm{o}}N) + \mc{O}(n_{\rm{d}} n_{\rm{v}})$ operations.
The computational cost of the transforms $\bm{F}$ and  $\bm{F}^\dagger$, steps $4$ and $28$, scales as $\mc{O}\left( n_{\rm{o}}N \log n_{\rm{o}}N \right)$ since this requires the \ac{fft} computation of the $n_{\rm{o}}$-oversampled image.
The remaining operations, including the projection, are $\mc{O}(N)$, giving the complexity of the primal update step
$\mc{O} \left( n_{\rm{o}}N \log n_{\rm{o}}N \right) + \mc{O} \left( (n_{\rm{b}}+n_{\rm{o}})N\right) + \mc{O}(N) +  \mc{O} \left(n_{\rm{d}}n_{\rm{v}}\right)$.
We kept the terms separate to give insight on how the algorithms scales for different configurations.
\bc Similarly to \ec \ac{admm}, the sums may be performed over the network in a distributed fashion, further reducing the complexity and leaving the primal update step dominated by the Fourier computations.

The randomised primal-dual algorithm introduces an even more scalable implementation.
To achieve a low computational burden per data node, the number of nodes has to be very large in order to reduce the size of $M_j$ and $N_j$ for each block.
The randomised algorithms achieve greater flexibility by allowing some of the updates for the sparsity prior or data fidelity dual variables, to be skipped at the current iteration.
Given a limited computing infrastructure, by carefully choosing the probabilities we can ensure that data fit into memory and that all available nodes are processing parts of it.
The average computational burden per iteration is lowered proportionally to the probability of selection, $p_{\mc{P}_i}$ and  $p_{\mc{D}_j}$.
In practice this also produces an increase in the number of iterations needed to achieve convergence, requiring a balanced choice for the probabilities.

\subsection{Splitting the data}

As reported earlier, the modality in which the data are split can have a big impact in the scalability of the algorithms.
Ideally, each data node should process an identical number of visibilities for the computation to be spread evenly.
If the visibilities used by one node are however spread over the whole $u$--$v$ plane, their processing requires all the discrete Fourier points.
Due to this, a compact grouping in frequency domain is also important since it determines the size of the data to be communicated.
Ideally, the splitting should be performed taking into account the computing infrastructure and should balance the communication and computation loads which are linked to the size of $N_j$ and $M_j$.

\section{Simulations and results}
\label{sec-results}

\bc
We study the performance of the algorithms developed herein for different configuration parameters and compare the reconstruction performance against \ac{cs-clean} \citep{Schwab1984} and \ac{moresane} \citep{Dabbech2015}.
We denote the methods as follows: \ac{sdmm}, the method introduced by \cite{Carrillo2014}; \ac{admm}, the approach described in Algorithm \ref{alg-admm}; \ac{pd} and \ac{pd-r}, the algorithms presented in Algorithm \ref{alg-primal-dual} without and with randomisation, respectively;
MORESANE, the algorithm\footnote{\bc We have used the MORESANE implementation from \alg{ws-clean} \citep{Offringa2014}, \url{https://sourceforge.net/p/wsclean/wiki/Home/}.\ec} from \cite{Dabbech2015}; CS-CLEAN, the Cotton-Schwab \alg{clean} \citep{Schwab1984} algorithm\footnote{\bc We have used the CS-CLEAN implementation of \alg{LWImager} from Casacore, \url{https://github.com/casacore/}.\ec}.
For both MORESANE and CS-CLEAN we perform tests for three types of weighting: natural weighting denoted by -N, uniform weighting denoted by -U and Briggs weighting with the robustness parameter set to 1 denoted by -B.

The reconstruction performance is assessed in terms of the signal to noise ratio,
\begin{equation}
	{\rm SNR} = 20 \log_{10} \left( \frac{\|\bs{x}^{\bc \circ \ec} \|_2}{\|\bs{x}^{\bc \circ \ec} \! - \bs{x}^{\bc (t) \ec}\|_2}\right),
\end{equation}
where $\bs{x}^{\circ}$ is the original image and $\bs{x}^{(t)}$ is the reconstructed estimate of the original, averaged over $10$ simulations performed for different noise realisations.
For the tests involving the comparison with \ac{cs-clean} and \ac{moresane} on the VLA and SKA coverages we do not perform this averaging.
In the latter case we also report the dynamic range
\begin{equation}
	{\rm DR} = \frac{\sqrt{N}\|\bm{\Phi}\|_{\rm{S}}^2}{\|\bm{\Phi}^\dagger(\bs{y} - \bm{\Phi}\bs{x})\|_2} \max_{k, l} {x_{k, l}}
\end{equation}
obtained by all algorithms.

\subsection{Simulation setup}

\begin{figure}
	\centering
	\begin{minipage}{.49\linewidth}
		\centering
  		\includegraphics[trim={0px 0px 0px 0px}, clip, height=3.80cm]{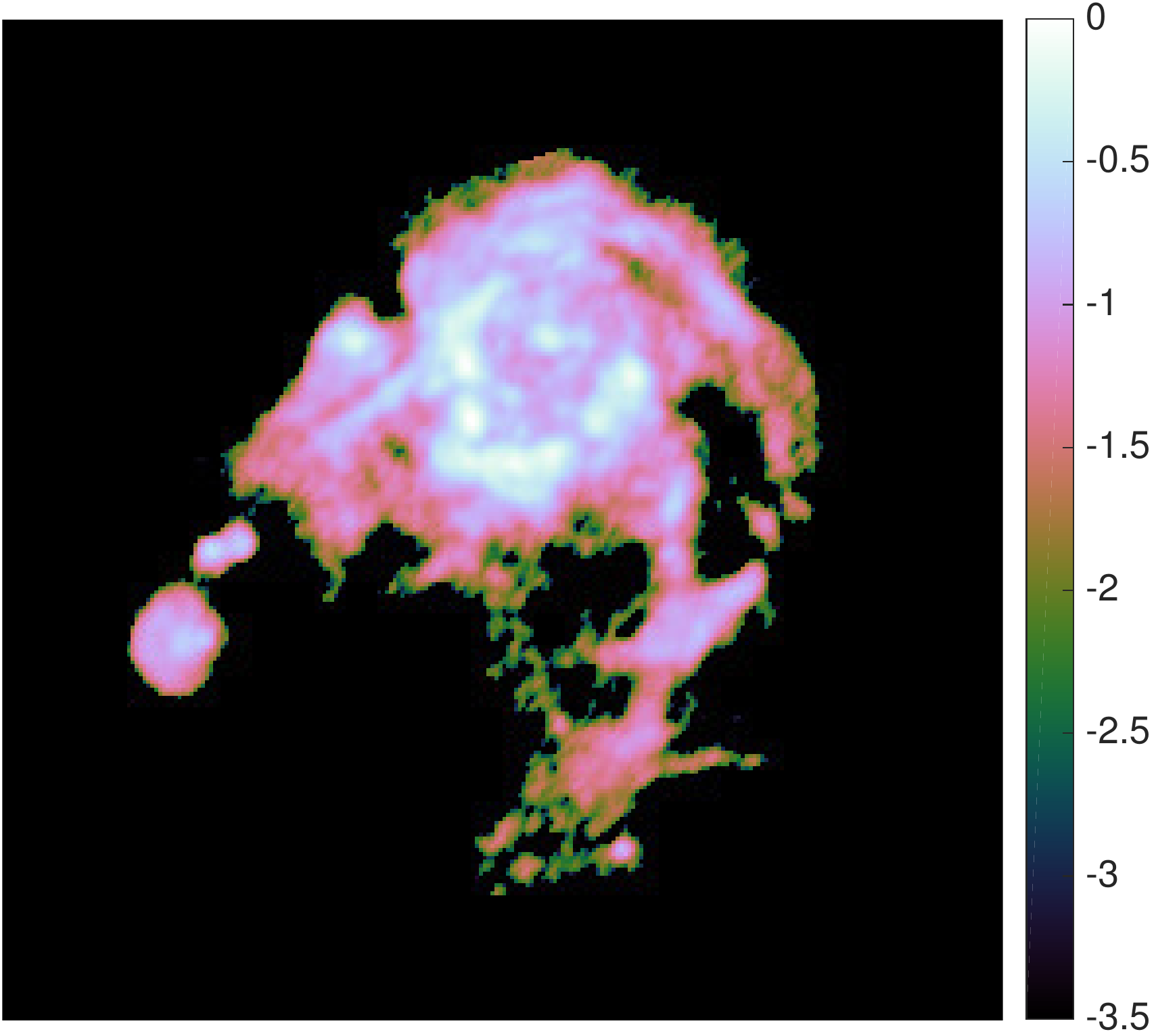}
	\end{minipage}
	\begin{minipage}{.49\linewidth}
		\centering
  		\includegraphics[trim={0px 0px 0px 0px}, clip, height=3.80cm]{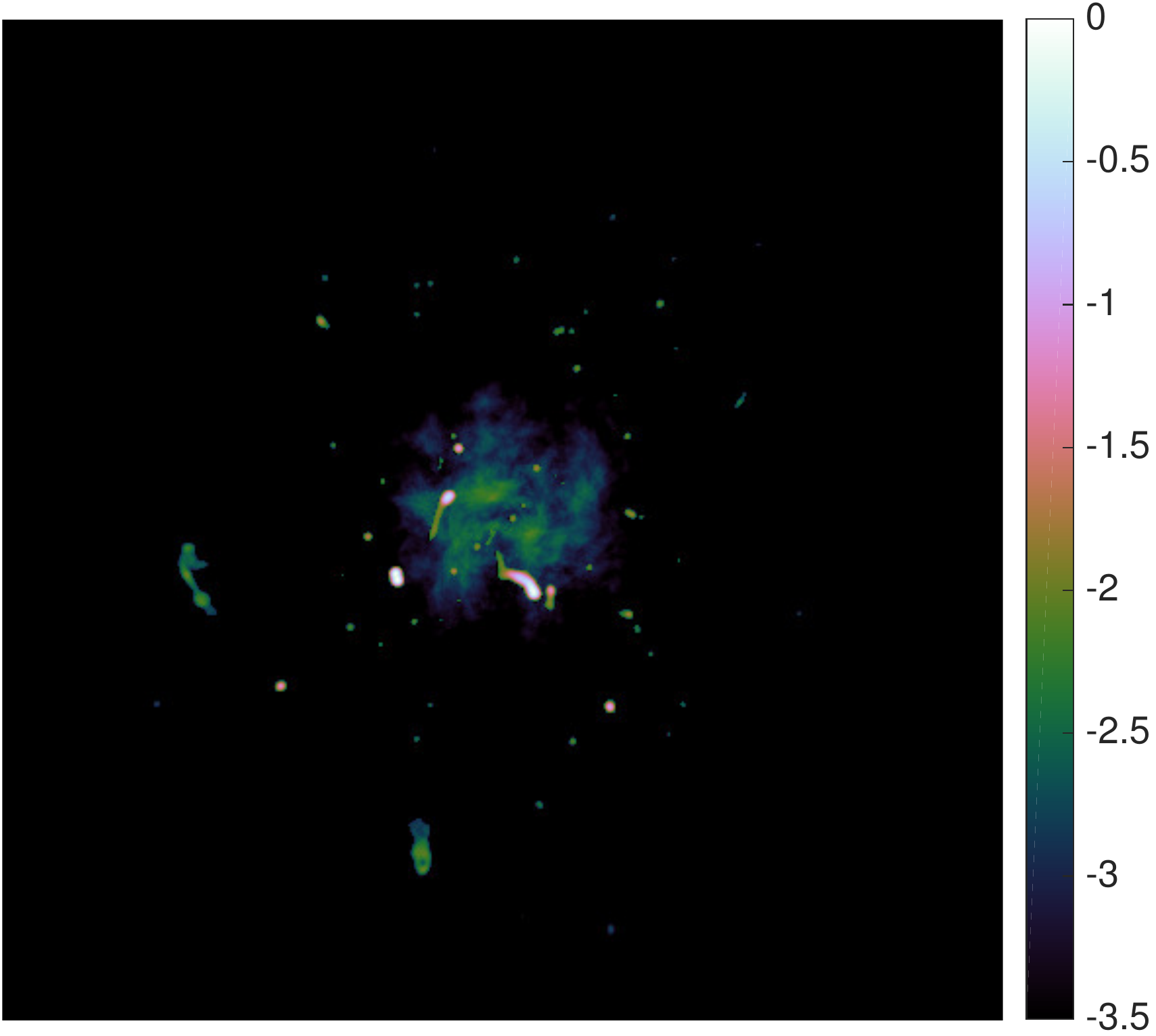}
	\end{minipage}

	\hspace{-6pt}
	\begin{minipage}{.98\linewidth}
  		\centering
  		\includegraphics[trim={0px 0px 0px 0px}, clip, height=3.81cm]{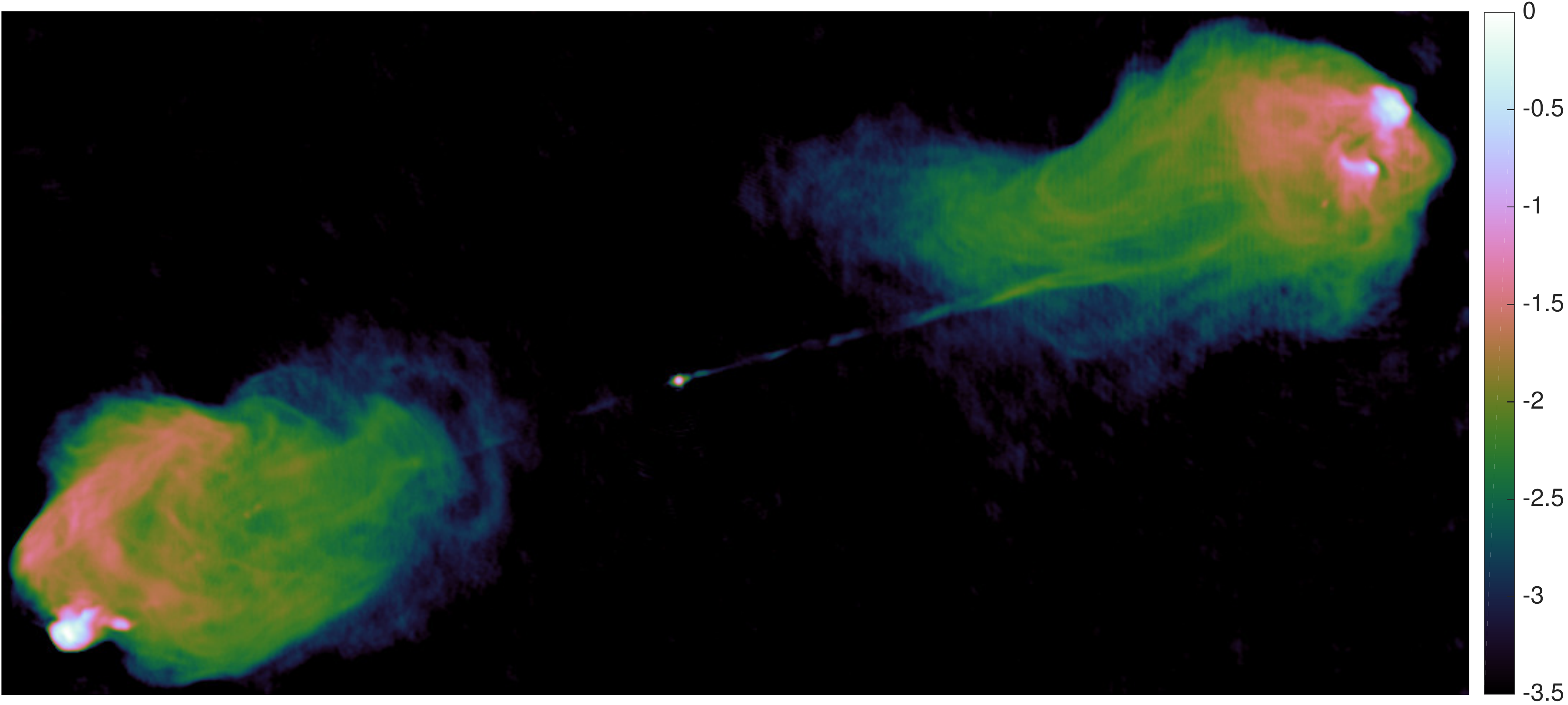}
	\end{minipage}

	\begin{minipage}{.49\linewidth}
	        \centering
  		\includegraphics[trim={0px 0px 0px 0px}, clip, height=3.80cm]{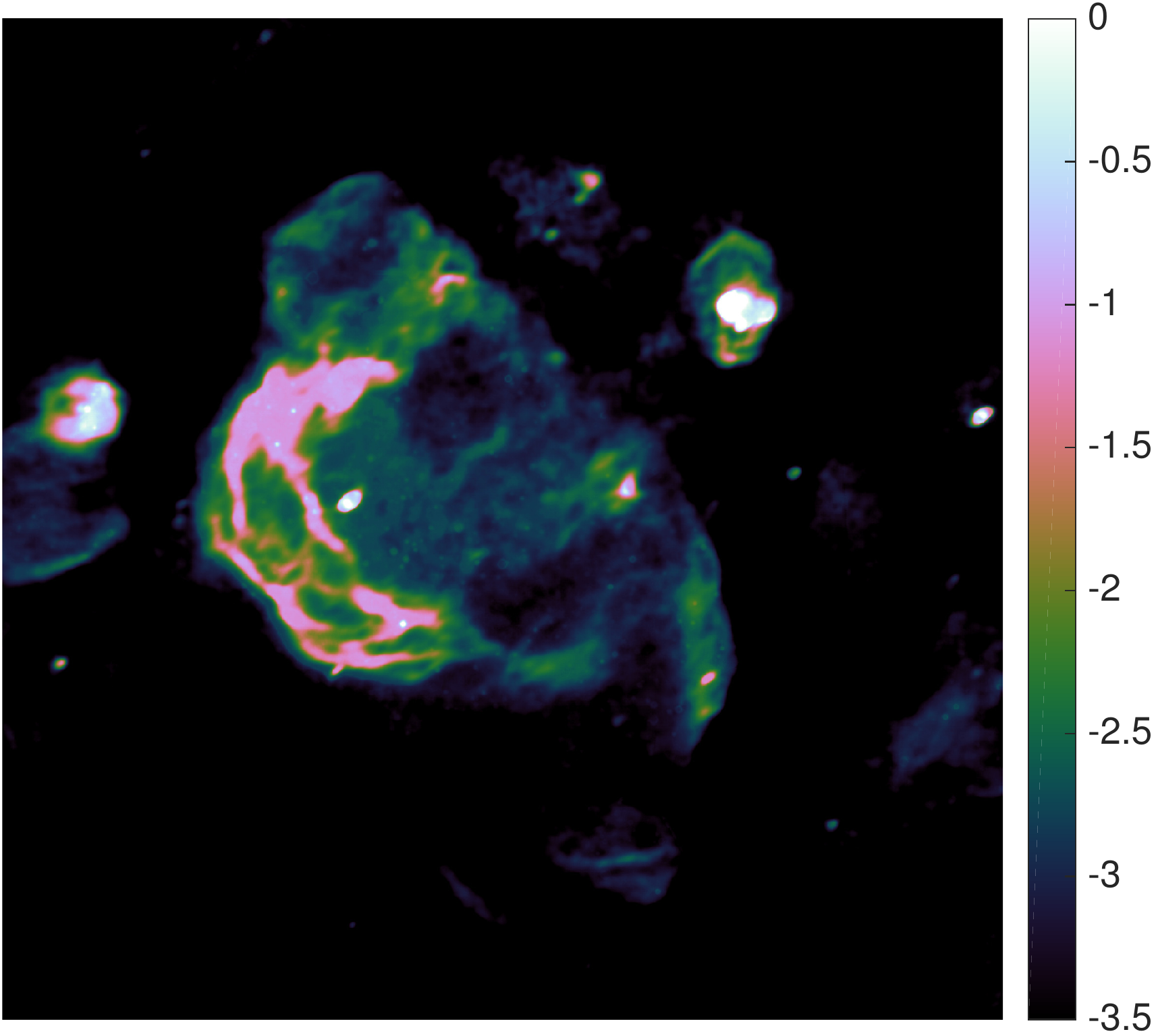}
	\end{minipage}
	
	\caption{\bc The test images, from left to right, top to bottom, a $256 \times 256$ image of the M31 galaxy, a $512 \times 512$ galaxy cluster image, a $477 \times 1025$ image of Cygnus A and a $1024 \times 1024$ image of the W28 supernova remnant, all shown in $\log_{10}$ scale.\ec}
	\label{fig-test-images}
\end{figure}

\begin{figure*}
	\centering
	\begin{minipage}{.33\linewidth}
	        \centering
  		\includegraphics[trim={0px 0px 0px 0px}, clip, height=5.72cm]{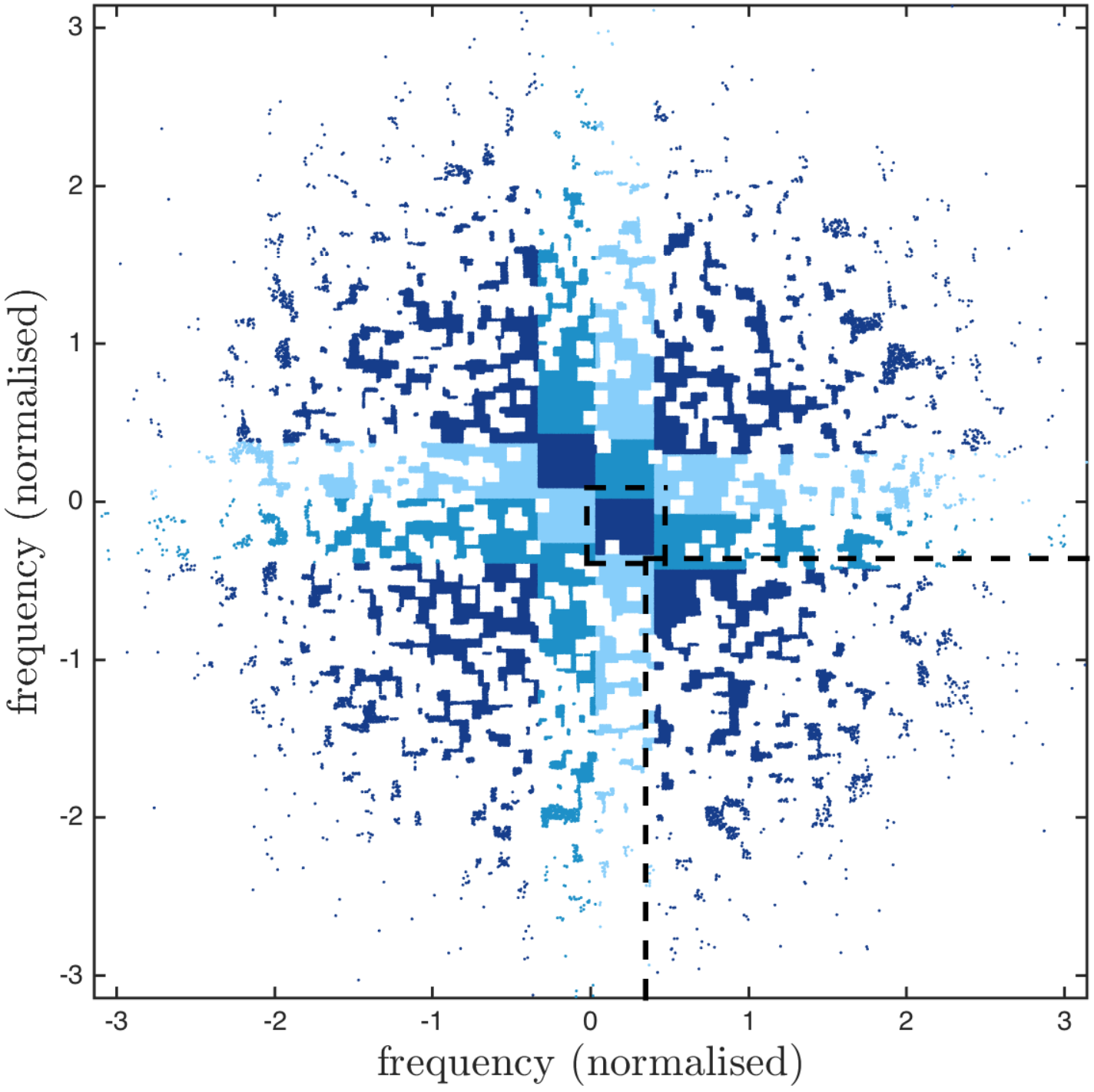}
	\end{minipage}
	\begin{minipage}{.33\linewidth}
	        \centering
  		\includegraphics[trim={0px 0px 0px 0px}, clip, height=5.72cm]{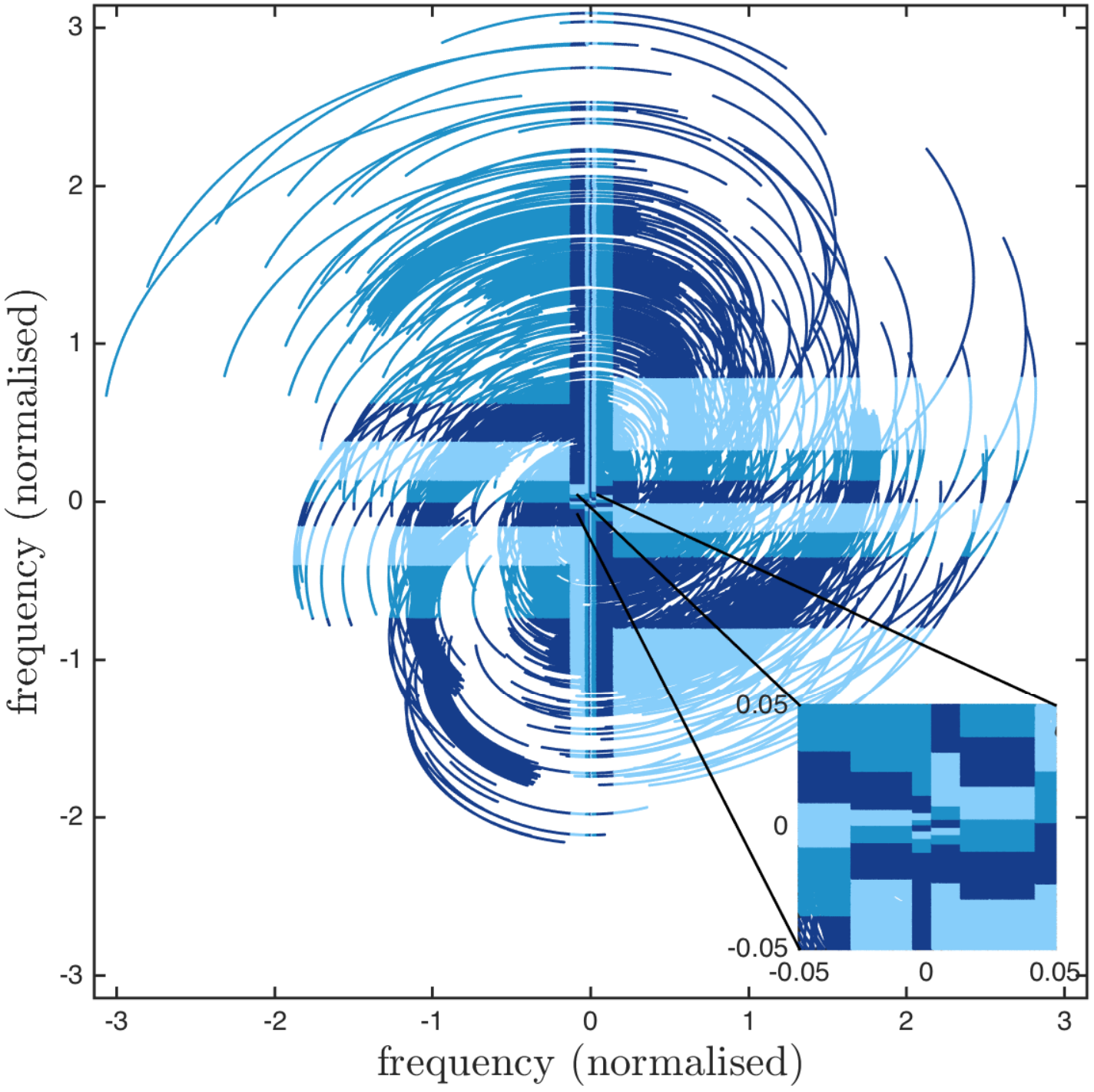}
	\end{minipage}
	\begin{minipage}{.33\linewidth}
	        \centering
  		\includegraphics[trim={0px 0px 0px 0px}, clip, height=5.72cm]{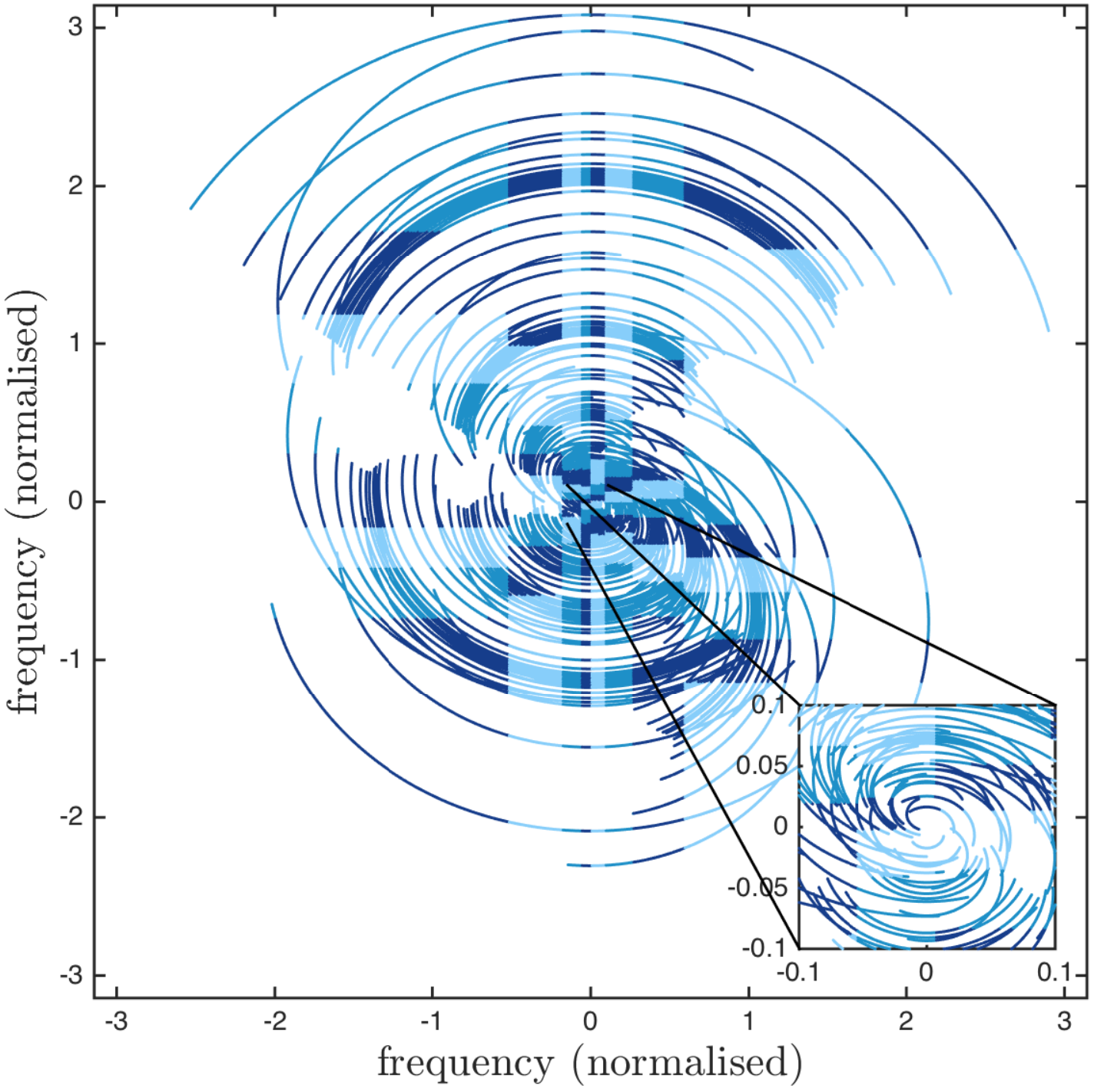}
	\end{minipage}
  	
	\caption{\bc (left) \ec An example of randomly generated coverage with the number of visibilities $M=\bc 655360 \ec$. \bc The visibilities are split into $16$ equal size blocks, marked with different colours, with compact $u$--$v$ grouping. \ec The dashed lines mark the parts of the discrete Fourier space involved in the computations associated with the central-bottom-right and the bottom-right blocks, respectively. \bc In this case, \ec the whole discrete frequency space \bc is considered to have \ec $512 \times 512$ points. \bc (centre) The SKA $u$--$v$ coverage for $5$ hours of observation corresponding to $M=5791800$. (right) The VLA $u$--$v$ coverage for $9$ hours of observations corresponding to $M=1788480$. The SKA and VLA data are split into $64$ blocks containing an equal number of visibilities.\ec}
	\label{fig-coverage-example}
\end{figure*}

In the first part of the simulations, we evaluate the influence of the different configuration parameters for \ac{pd}, \ac{pd-r}, and \ac{admm}.
Here, we also validate their performance against \ac{sdmm}, a previously proposed solver \citep{Carrillo2014} for the same optimisation task.
The test images, as shown in Figure~\ref{fig-test-images}, represent a small $256 \times 256$ image of the H\textsc{ii} region of the M31 galaxy, a $512 \times 512$ high dynamic range image of a galaxy cluster with faint extended emissions, and a $477 \times 1025$ image of the Cygnus A radio galaxy, respectively.
The galaxy cluster image was produced using the \alg{faraday} tool \citep{Murgia2004}.
We reconstruct them from simulated visibility data.
We use a $u$--$v$ coverage generated randomly through Gaussian sampling, with zero mean and variance of $0.25$ of the maximum frequency, creating a concentration of visibility data in the centre of the plane, for low frequencies.
We introduce holes in the coverage with an inverse Gaussian profile, placing the missing spectrum information predominantly in high frequency.
This generates very generic profiles and allows us to study the algorithm performance with a large number of different coverages.
A typical $u$--$v$ coverage is presented in Figure~\ref{fig-coverage-example}.

The second part of the simulations involves testing the algorithm reconstruction using simulated VLA and SKA coverages\footnote{\bc The SKA and VLA $u$--$v$ coverages are generated using the Casa and Casacore software package: \url{https://casa.nrao.edu/} and \url{https://github.com/casacore} \ec} corresponding to $5$ and $9$ hours of observations, respectively.
The coverages are presented in Figure~\ref{fig-coverage-example}.
For the tests we use an additional large $1024 \times 1024$ image, also presented in Figure~\ref{fig-test-images}, representing the W28 supernova remnant\footnote{\bc Image courtesy of NRAO/AUI and \cite{Brogan2006}\ec}.
We showcase the reconstruction quality and speed of convergence for PD and ADMM without performing any re-weighting\footnote{\bc Performing the re-weighting improves the reconstruction \citep{carrillo12, Carrillo2014} but falls outside the scope of this study.\ec} and compare the results with those produced by \ac{cs-clean} and \ac{moresane}.

In both cases, we have normalised the frequencies to the interval $\left[-\pi,~\pi \right]$.
The visibilities are corrupted by zero mean complex Gaussian noise producing a signal to noise level of $20~\rm{dB}$.
The bound $\epsilon_j$, for the ball $\mc{B}_j$ defined by (\ref{split-function-definition}), can be therefore estimated based on the noise variance $\sigma_{\chi}^2$ of the real and imaginary parts of the noise, the residual norm being distributed according to a $\chi^2$ distribution with $2M_j$ degrees of freedom.
Thus, we impose that the square of the global bound $\epsilon^2$ is 2 standard deviations above the mean of the $\chi^2$ distribution, $\epsilon^2 = \left(2M + 2\sqrt{4M}\right)\sigma_{\chi}^2$. The resulting block constraints must satisfy $\sum_{j=1}^{n_{\mathrm{d}}} \epsilon_j^2 = \epsilon^2$.
When all blocks have the same size, this results in $\epsilon_j^2 = \big(2M_j + \frac{2}{\sqrt{n_{\mathrm{d}}}}\sqrt{4M_j}\big)\sigma_{\chi}^2$.

We work with pre-calibrated measurements.
For simplicity we assume, without loss of generality, the absence of DDEs and a small field of view, the measurement operator reducing to a Fourier matrix sampled at the $M$ frequencies that characterise the visibility points.
We have used an oversampled Fourier transform $\bm{F}$ with $n_{\rm{o}}=4$ and a matrix $\bm{G}$ that performs an interpolation of the frequency data, linking the visibilities to the uniformly sampled frequency space. The $8 \times 8$ interpolation kernels \citep{Fessler2003} average nearby uniformly distributed frequency values to estimate the value at the frequencies associated with each visibility.
A scaling is also introduced in image space to pre-compensate for imperfections in the interpolation.
This allows for an efficient implementation of the operator.

To detail the behaviour of the algorithms, we vary the number of blocks $n_{\rm{d}}$ used for the data fidelity term. Tests are performed for $4$, $16$ and $64$ blocks.
In each case, the blocks are generated such that they have an equal number of visibility points, which cover a compact region in the $u$--$v$ space.
An example of the grouping for the $16$ blocks is overlaid on the randomly generated coverage from Figure~\ref{fig-coverage-example}.
The figure also contains, marked with dashed lines, an example of the discrete frequency points required to model the visibilities for two of the blocks, under our previous assumptions, for the M31 image.
The number of discrete frequency points required for each block would only grow slightly in the presence of DDEs due to their, possible larger, compact support. 
The overall structure from Figure~\ref{fig-coverage-example} would remain similar.
For the SKA and VLA coverages, the data are also split into blocks of equal size. The resulting block structure is also presented in Figure~\ref{fig-coverage-example}.
As sparsity prior, we use the \ac{sara} collection of wavelets \citep{carrillo12}, namely a concatenation of a Dirac basis with the first eight Daubechies wavelets.
We split the collection of bases into $n_{\rm{b}}=9$ individual basis.

\subsection{Choice of parameters}

The \ac{admm}, \ac{pd}, and \ac{pd-r} algorithms converge given that (\ref{convergence-req-admm}) and (\ref{convergence-req-pd}), respectively, are satisfied.
To ensure this we set for \ac{pd} $\sigma = \sfrac{1}{\| \bm{\Psi} \|_{\rm{S}}^2}$, $\varsigma = \sfrac{1}{\| \bm{\Phi} \|_{\rm{S}}^2}$ and $\tau=0.49$.
The relaxation parameter is set to 1.
For the \ac{admm} algorithm we set $\rho = \sfrac{1}{\| \bm{\Phi} \|_{\rm{S}}^2}$ and $\eta = \sfrac{1}{\| \bm{\Psi} \|_{\rm{S}}^2}$. The ascent step is set $\varrho = 0.9$.
The maximum number of sub-iterations is set to $n_{\bar{f}}=100$. 
We consider the convergence achieved, using a criterion similar to (\ref{stop-crit-rel-norm-variation}), when the relative solution variation for $\bar{\bs{z}}^{(k)}$ is below $10^{-3}$.
The norms of the operators are computed a priori using the power iterative method.
They act as a normalisation of the updates, enabling the algorithm to deal with different data or image scales.

We leave the normalised soft-threshold values $\kappa$ as a configuration parameter for both \ac{pd} and \ac{admm}.
\ac{sdmm} has a similar parameter $\kappa$.
It influences the convergence speed which is of interest since, given the scale of the problem, we want to minimise the computational burden which is inherently linked to the number of iterations performed.
We aim at providing a general prescription for this tuning parameter, similarly to the standard choices for the loop gain factor used by  \alg{clean}.
Intuitively, this soft-thresholding parameter can be seen as analogous to this factor, deciding how aggressive we are in enforcing the sparsity requirements.
The stopping parameter $\bar{\delta}$, essentially linked to the accuracy of the solution given a certain convergence speed, is also configurable.
For simplicity we also set equal probabilities for \ac{pd-r}, namely $p_{\mc{P}_i}=p_{\mc{P}}$, $\forall i$ and $p_{\mc{D}_j}=p_{\mc{D}}$, $\forall j$ and we show how the different choices affect the performance.
We choose to randomise only over the data fidelity terms since the \ac{sara} sparsity prior is light from the computational perspective \bc when compared to the data fidelity term, thus $p_{\mc{P}} = 1$ for all tests performed.
Different strategies for the choice of probabilities, with values different for each block, are also possible.
For example setting a higher probability for the blocks containing low frequency data will recover faster a coarse image.
The details are incorporated into the solution through the lower probability updates of the high frequency data.
An overview of all the parameters used for defining the optimisation task and for configuring both \ac{admm} and \ac{pd} algorithms is presented in Appendix \ref{sec:param-overview}, Table~\ref{param-table-opt} and Table~\ref{param-table-alg}, respectively.

We ran MORESANE with a $5$ major loops and a major loop gain $0.9$. The loop gain inside MORESANE was set to $0.1$.
We use the model image to compare against the other methods.
CS-CLEAN was run with two loop gain factors, $l_g=0.1$ and $l_g=0.001$. The results shown are the best of the two.
We compare against the model image convolved with a Gaussian kernel associated with the main beam. We scale the resulting image to be closest to the true model image in the least square sense. 
Additionally, we also present results with the main beam scaled by a factor $b$ chosen such that the best SNR is achieved.
This introduces a large advantage for \ac{cs-clean} when compared to the other algorithms.
To avoid edge artefacts, both MORESANE and CS-CLEAN were configured to produce a padded double sized image and only the centre was used for comparison.

For \ac{pd}, \ac{pd-r}, and \ac{admm}, the stopping criterion for the algorithms is composed of two criteria.
We consider the constraints satisfied when the global residual norm is in the vicinity of the bound $\epsilon$ of the global $\ell_2$ ball, namely below a threshold $\bar{\epsilon}$. This is equivalent to stopping if $\sum_{j=1}^{n_{\mathrm{d}}} \| \bs{y}_j - \bm{\Phi}_j\bs{x}^{(t)} \|^2_2 \leq \bar{\epsilon}^2$.
We set $\bar{\epsilon}^2 = \left(2M + 3\sqrt{4M}\right)\sigma_{\chi}^2$, namely $3$ standard deviations above the mean.
The second criterion relates to the relative variation of the solution, measured by%
\begin{equation}%
	\delta = \frac{\|\bs{x}^{(t)}-\bs{x}^{(t-1)}\|_2}{\|\bs{x}^{(t)}\|_2}.
	\label{stop-crit-rel-norm-variation}
\end{equation}
The iterations stop when the $\ell_2$ ball constraints are satisfied and when the relative change in the solution norm is small, $\delta \leq \bar{\delta}$.
The data fidelity requirements are explicitly enforced, ensuring that we are inside or very close to the feasible region. 
However, this does not guarantee the minimisation of the $\ell_1$ prior function.
The algorithms should run until the relative variation of the solution is small between iterations.
To better understand the behaviour of the algorithms, for most simulations we perform tests over a fixed number of iterations without applying the stopping conditions  above.

The stopping criterion for MORESANE and CS-CLEAN was set to be $3$ standard deviations above the noise mean. This level was seldom reached by CS-CLEAN after the deconvolution, the algorithm seeming to stop because of the accumulation of false detections leading to the increase of the residual between iterations.

\subsection{Results using random coverages}

\bc

\begin{figure}
	\centering
	\includegraphics[trim={0px 0 0px 0}, clip, width=0.99\linewidth]{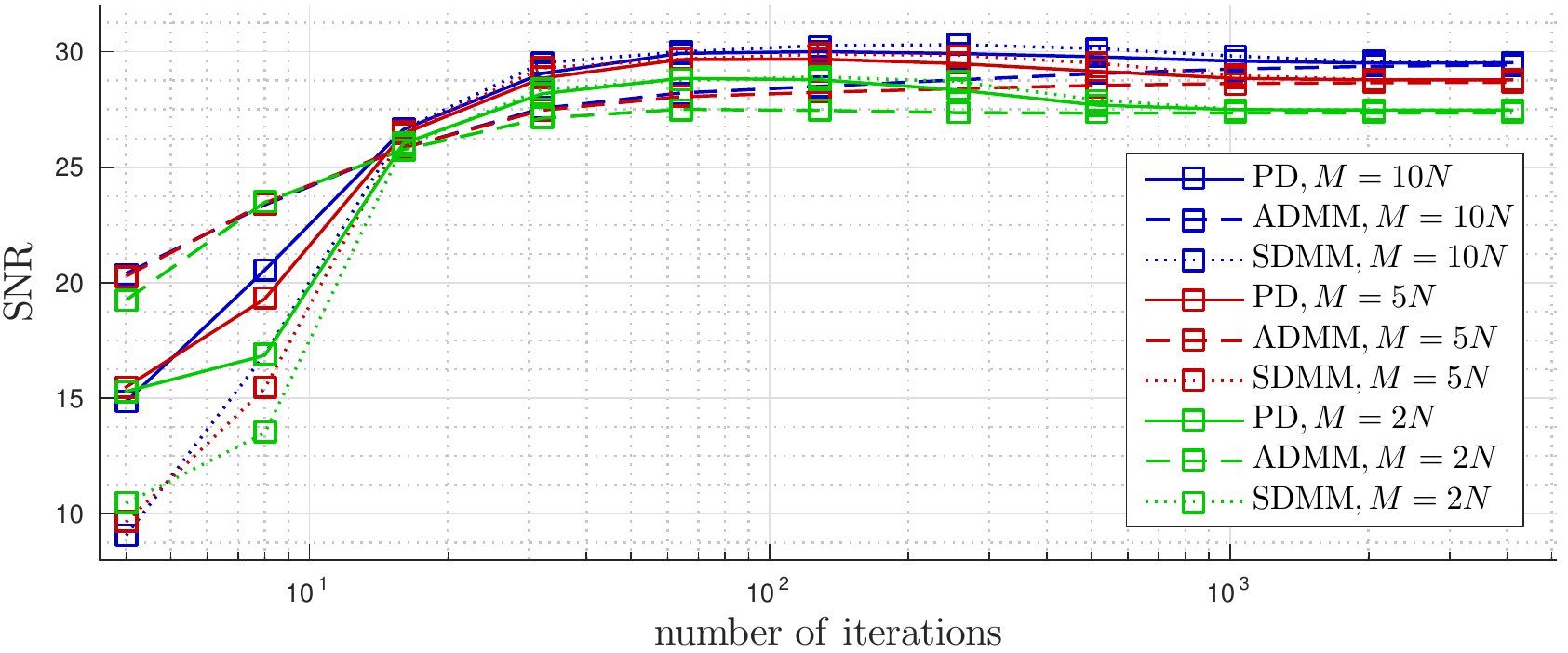}
	\caption{The evolution of the SNR for \ac{pd}, \ac{admm} and \ac{sdmm} as a function of the number of iterations for the M31 test image. The configuration parameter, $\kappa = 10^{-3}$, is the same for \ac{admm}, \ac{pd} and \ac{sdmm}. The number of visibilities $M$ used is $10N$, $5N$ and $2N$.  The input data are split into $4$ blocks.}
	\label{fig-m31-results-f-alg}
\end{figure}

\begin{figure}
	\centering
  	\includegraphics[trim={-8px 0 0px 0}, clip, width=0.99\linewidth]{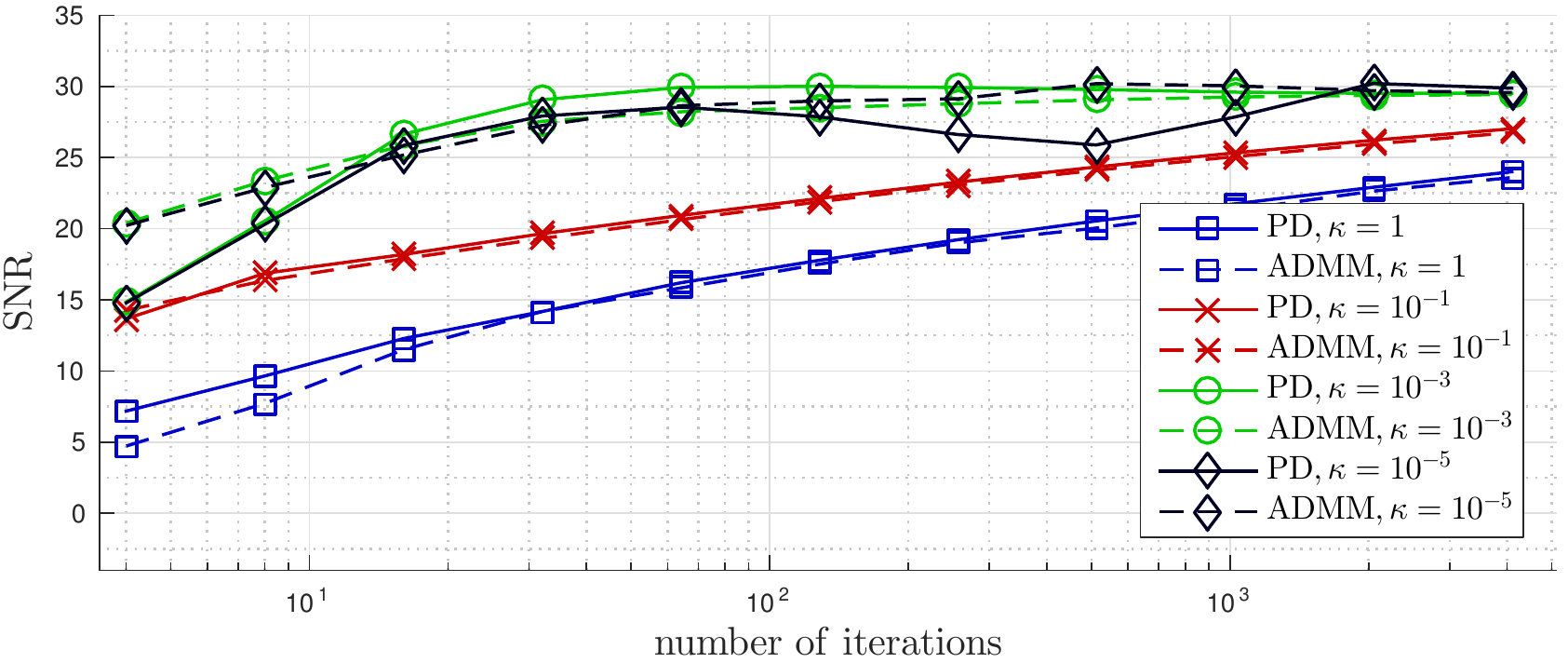}
	\includegraphics[trim={0px 0 0px 0}, clip, width=0.99\linewidth]{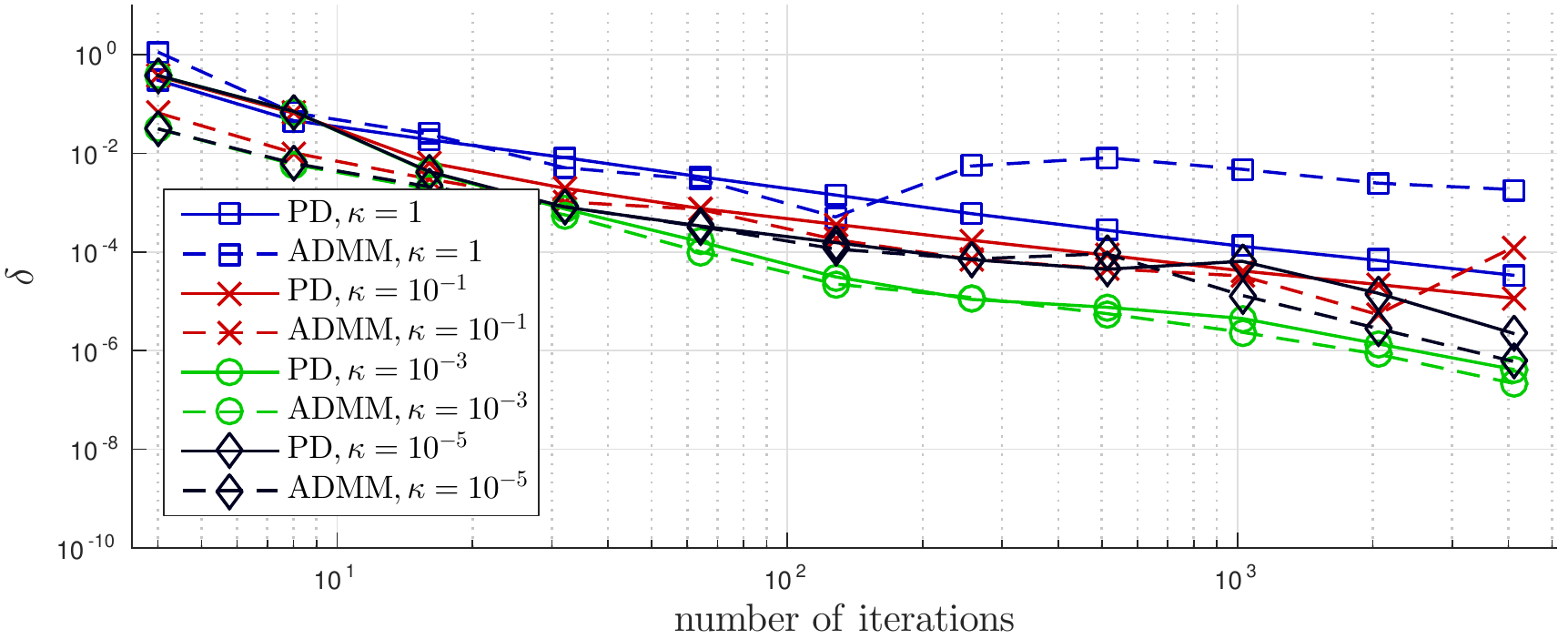}
	\caption{The reconstruction of the M31 image from $M=10N$ visibilities. The input data are split into $4$ blocks. (top) The evolution of the SNR for \ac{pd} and \ac{admm} as a function of the number of iterations for different values of the parameter $\kappa$. (bottom) The value of $\delta$ for both methods.}
	\label{fig-m31-10M-results-f-gamma}
\end{figure}

We begin by analysing the evolution of the $\rm SNR$ for the \ac{admm} and \ac{pd} algorithms in comparison with that produced by the previously proposed \ac{sdmm} solver.
Figure~\ref{fig-m31-results-f-alg} contains the $\rm SNR$ as a function of number of iterations for the three algorithms for the reconstruction of the M31 image from $M=10N$, $M=5N$ and $M=2N$ visibilities.
The two newly introduced algorithms have the same convergence rate as \ac{sdmm} but have a much lower computational burden per iteration, especially the \ac{pd} method.
In these tests, all three method use the parameter $\kappa = 10^{-3}$, suggested also by \cite{Carrillo2014}.
The reconstruction performance is comparable for the different test cases, the \ac{pd} and \ac{admm} obtaining the same reconstruction quality.
 Adding more data improves the reconstruction SNR by $2$-$3~\mathrm{dB}$ because the noise is better averaged.
 However, note that the $\rm SNR$ gain stagnates slightly when more visibility data are added mainly because the holes in the frequency plane are still not covered.
The problem remains very ill-posed with similar coverage.
In a realistic situation, adding more data will also fill the coverage more and the $\rm SNR$ improvement will be larger. Since all three algorithms explicitly solve the same minimisation problem, they should have similar behaviour for any other test case.

\begin{figure}
	\centering
  	\includegraphics[trim={-8px 0 0px 0}, clip, width=0.99\linewidth]{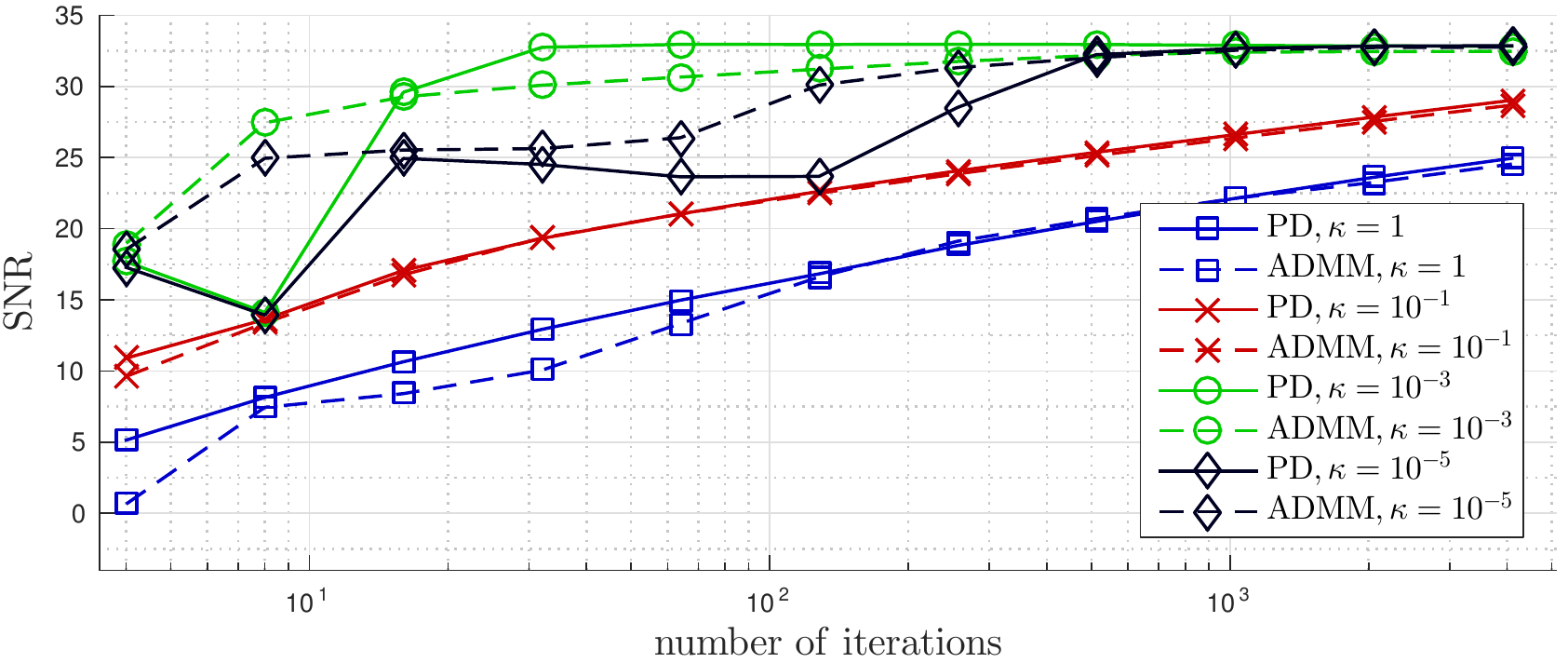}
	\includegraphics[trim={0px 0 0px 0}, clip, width=0.99\linewidth]{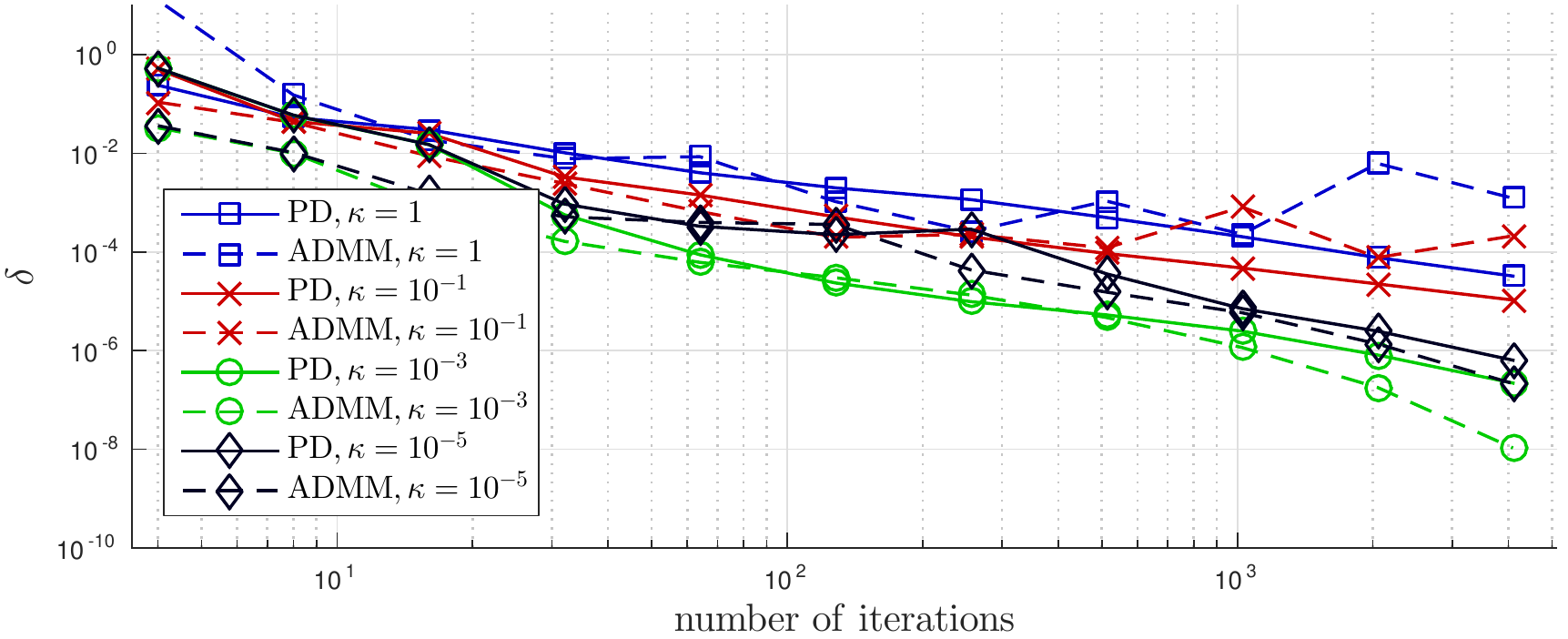}
	\caption{The reconstruction of the Cygnus A image from $M=N$ visibilities. The input data are split into $4$ blocks. (top) The evolution of the SNR for \ac{pd} and \ac{admm} as a function of the number of iterations for different values of the parameter $\kappa$. (bottom) The value of $\delta$ for both methods.}
	\label{fig-cyn-1M-results-f-gamma}
\end{figure}

\begin{figure}
	\centering
  	\includegraphics[trim={-8px 0 0px 0}, clip, width=0.99\linewidth]{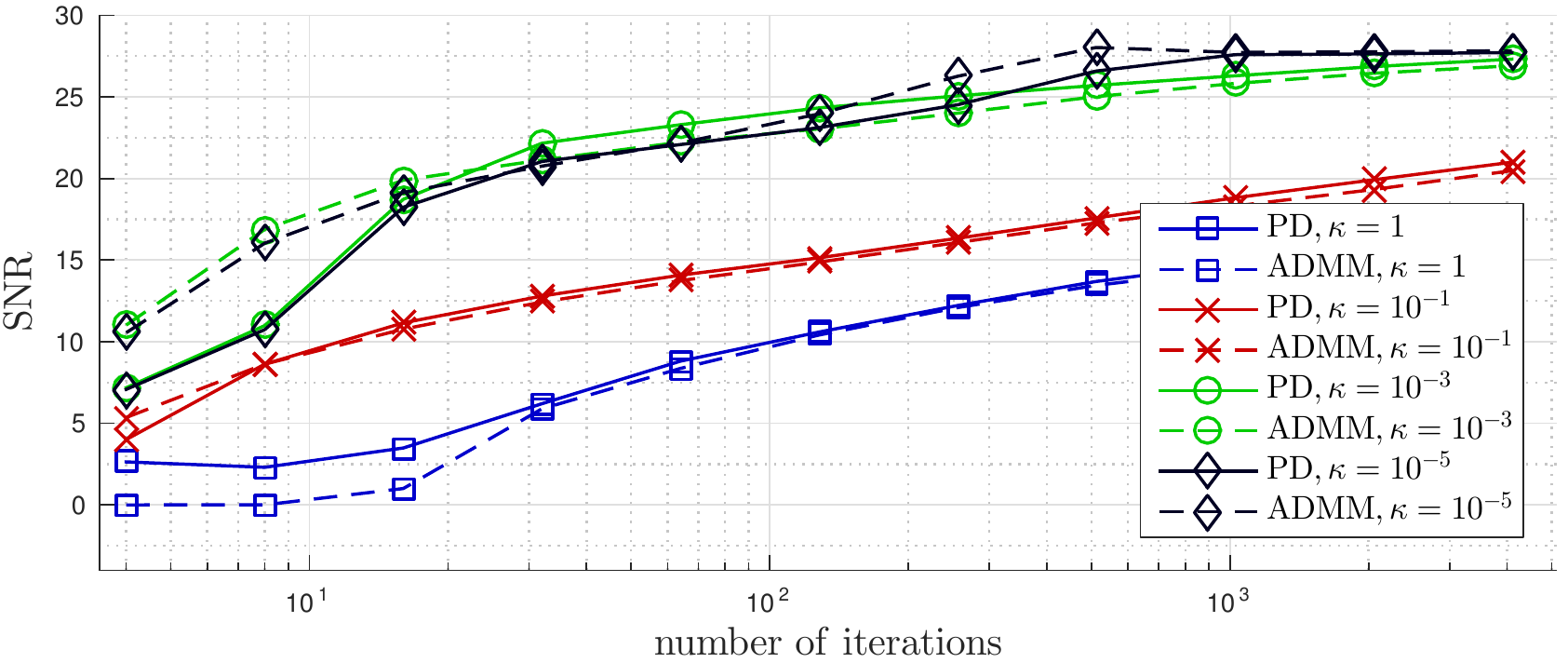}
	\includegraphics[trim={0px 0 0px 0}, clip, width=0.99\linewidth]{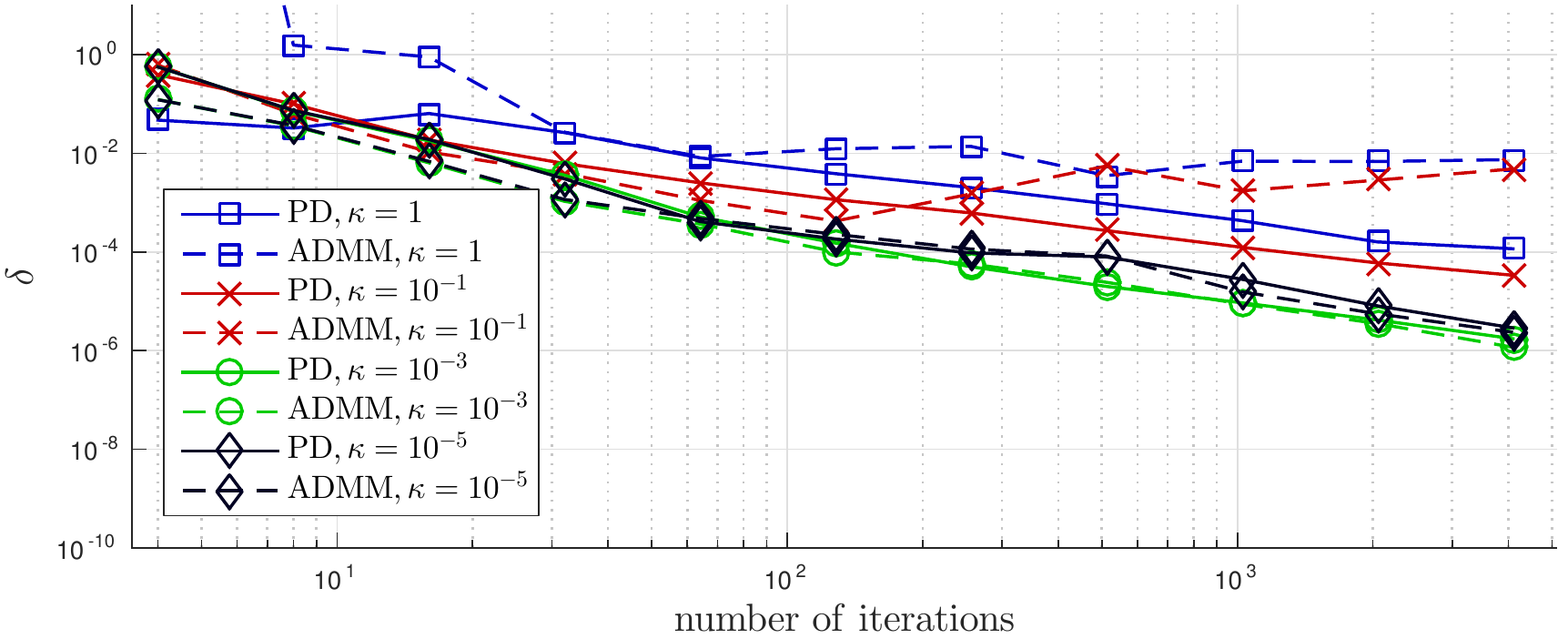}
	\caption{\bc The reconstruction of the galaxy cluster image from $M=2N$ visibilities. The input data are split into $4$ blocks. (top) The evolution of the SNR for \ac{pd} and \ac{admm} as a function of the number of iterations for different values of the parameter $\kappa$. (bottom) The value of $\delta$ for both methods.\ec}
	\label{fig-gc-2M-results-f-gamma}
\end{figure}

We continue by investigating the performance of the \ac{pd} and \ac{admm} algorithms as a function of the parameter $\kappa$ in Figures \ref{fig-m31-10M-results-f-gamma}, \ref{fig-cyn-1M-results-f-gamma} and \ref{fig-gc-2M-results-f-gamma} for the reconstruction of the M31, Cygnus A and galaxy cluster test images, respectively.
The parameter $\kappa$ serves as a normalised threshold and essentially governs the convergence speed.
The values $\kappa = 10^{-3}$ \bc to $\kappa = 10^{-5}$ generally produce good and consistent performance. This behaviour was also observed for similar tests, with smaller $M$.
Larger values for $\kappa$ reduce the convergence speed since they emphasise greatly the sparsity prior information at the expense of the data fidelity.
The smaller values place less weight on the sparsity prior and, after an initial fast convergence due to the data fidelity term, typically require more iterations to minimise the $\ell_1$ prior.
The average variation of the solution norm $\delta$ is also reported since the stopping criterion is based on it.
It links the convergence speed with the recovery performance.
For the galaxy cluster, the tests exhibits slower convergence speed when compared to the M31 and Cygnus A tests.
The values $\kappa=10^{-3}$ and $\kappa=10^{-5}$ produce similar behaviour.
It should be also noted that the variation of the solution decreases smoothly until convergence and that \ac{admm} shows a larger variability.

\begin{figure}
	\centering

  	\includegraphics[trim={0px 0px 0px 0}, clip, width=0.99\linewidth]{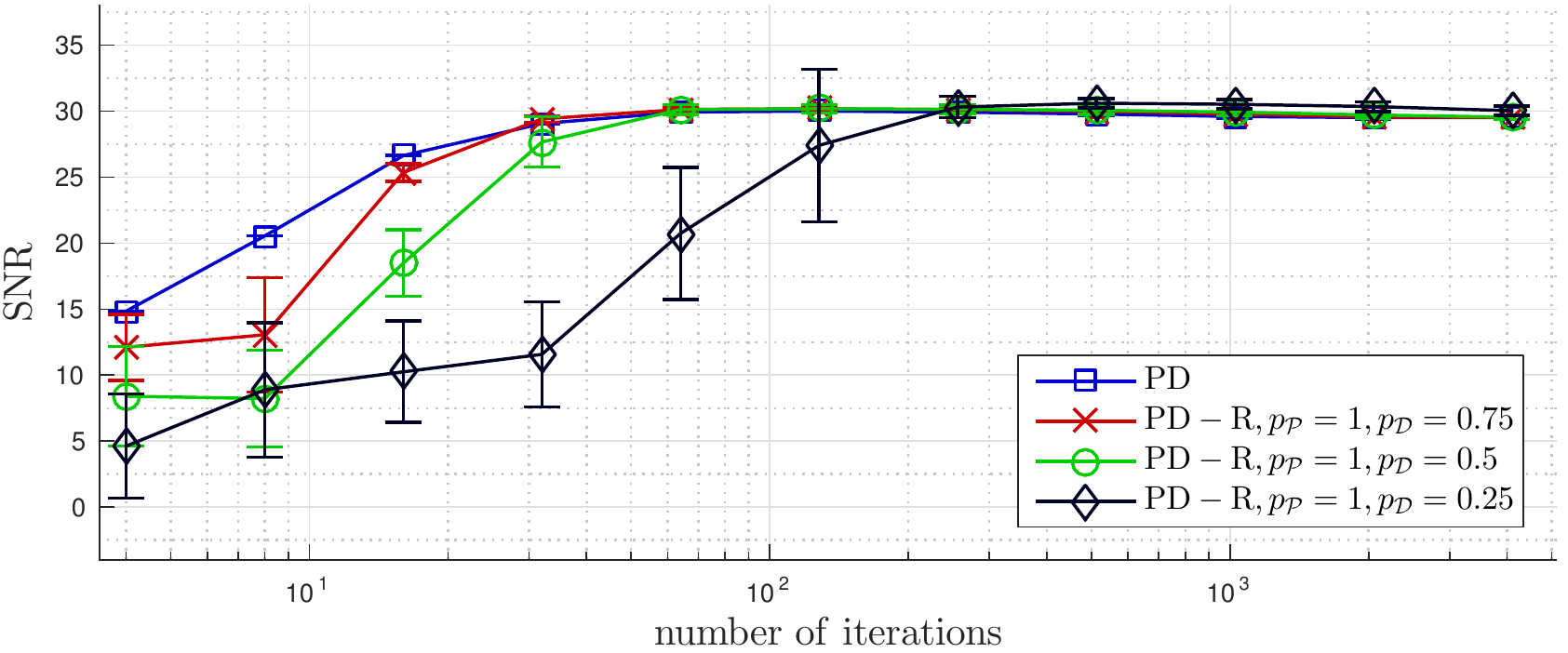}
  	\includegraphics[trim={0px 0px 0px 0}, clip, width=0.99\linewidth]{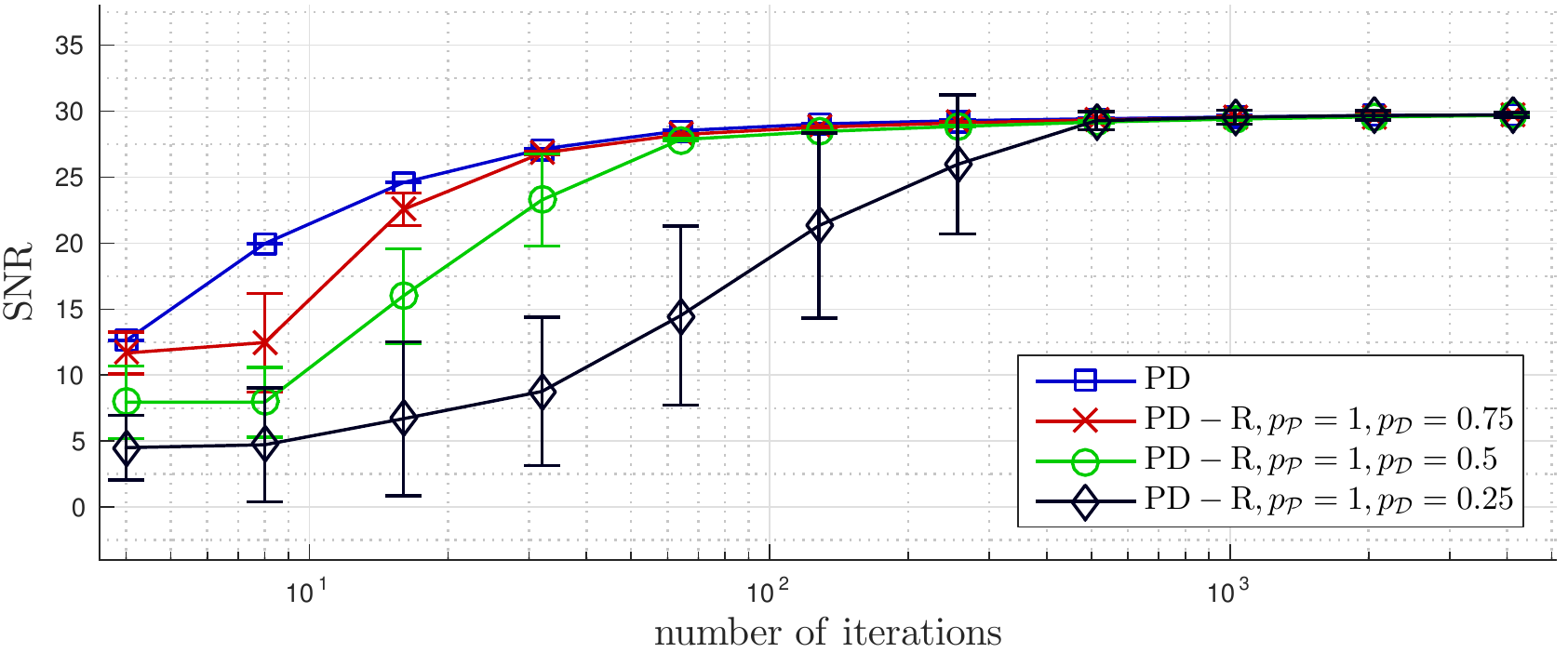}
  	\includegraphics[trim={0px 0px 0px 0}, clip, width=0.99\linewidth]{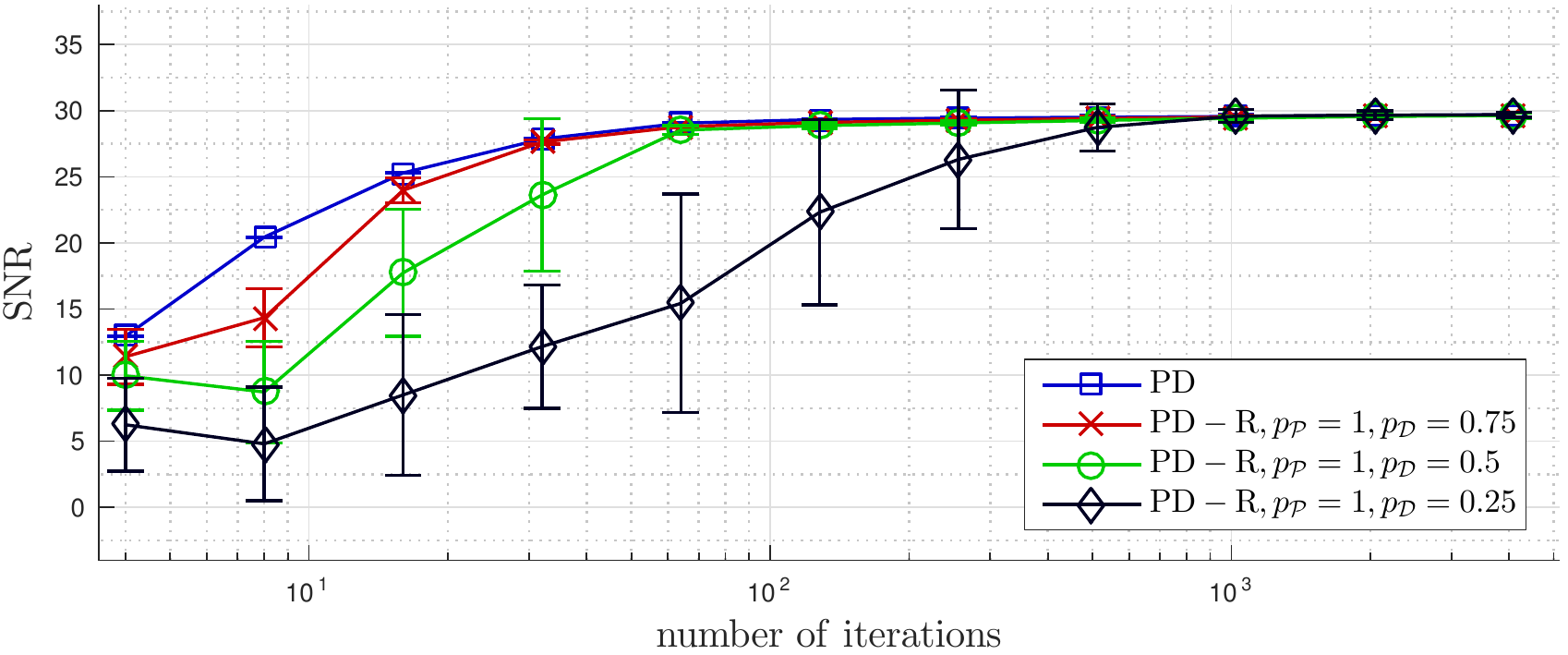}
	
	\caption{The SNR for the reconstruction of the M31 image from $M=10N$ visibilities for the \ac{pd}  and \ac{pd-r} algorithms with parameter $\kappa=10^{-3}$. 
	The algorithms split the input data into: (top) $4$ blocks, (middle) $16$ blocks, (bottom) $64$ blocks.}
	\label{fig-m31-10M-results-f-prob-L2}
\end{figure}

\begin{figure}
	\centering

  	\includegraphics[trim={0px 0px 0px 0}, clip, width=0.99\linewidth]{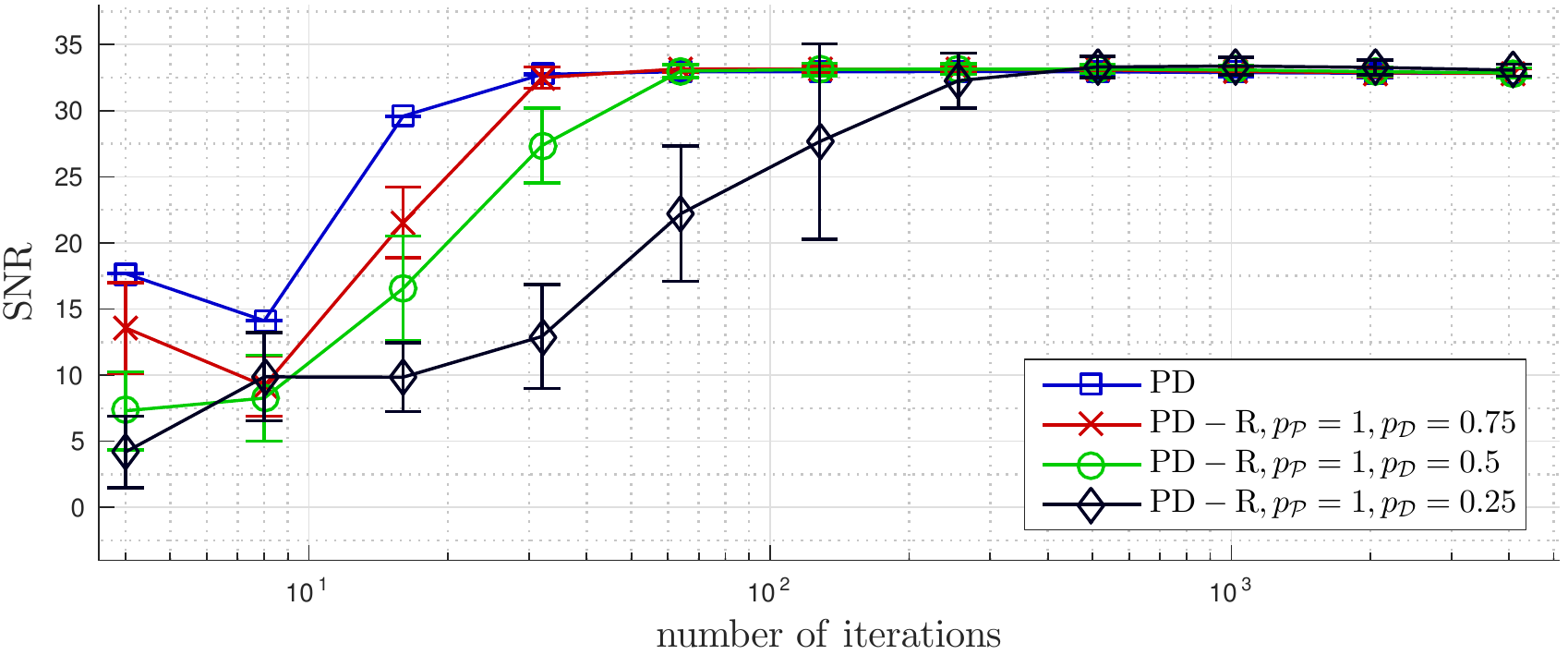}
  	\includegraphics[trim={0px 0px 0px 0}, clip, width=0.99\linewidth]{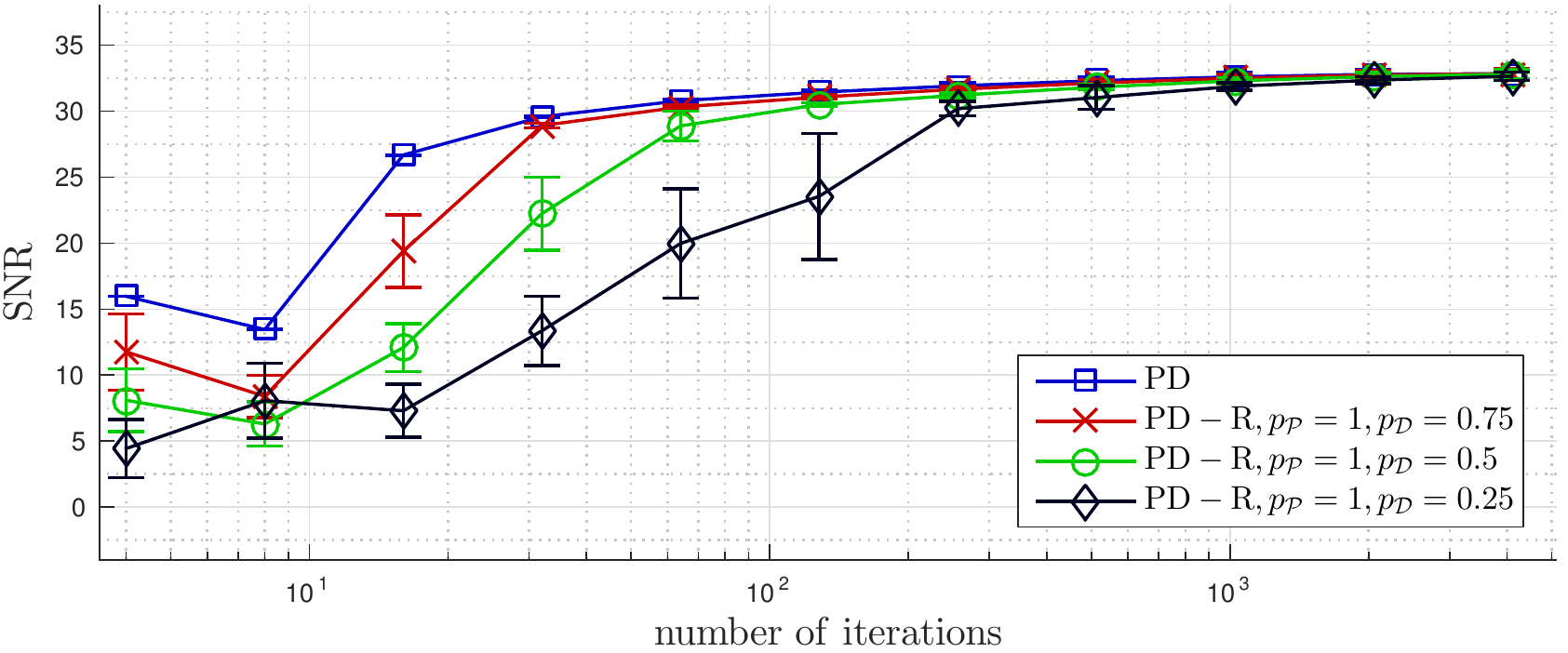}
  	\includegraphics[trim={0px 0px 0px 0}, clip, width=0.99\linewidth]{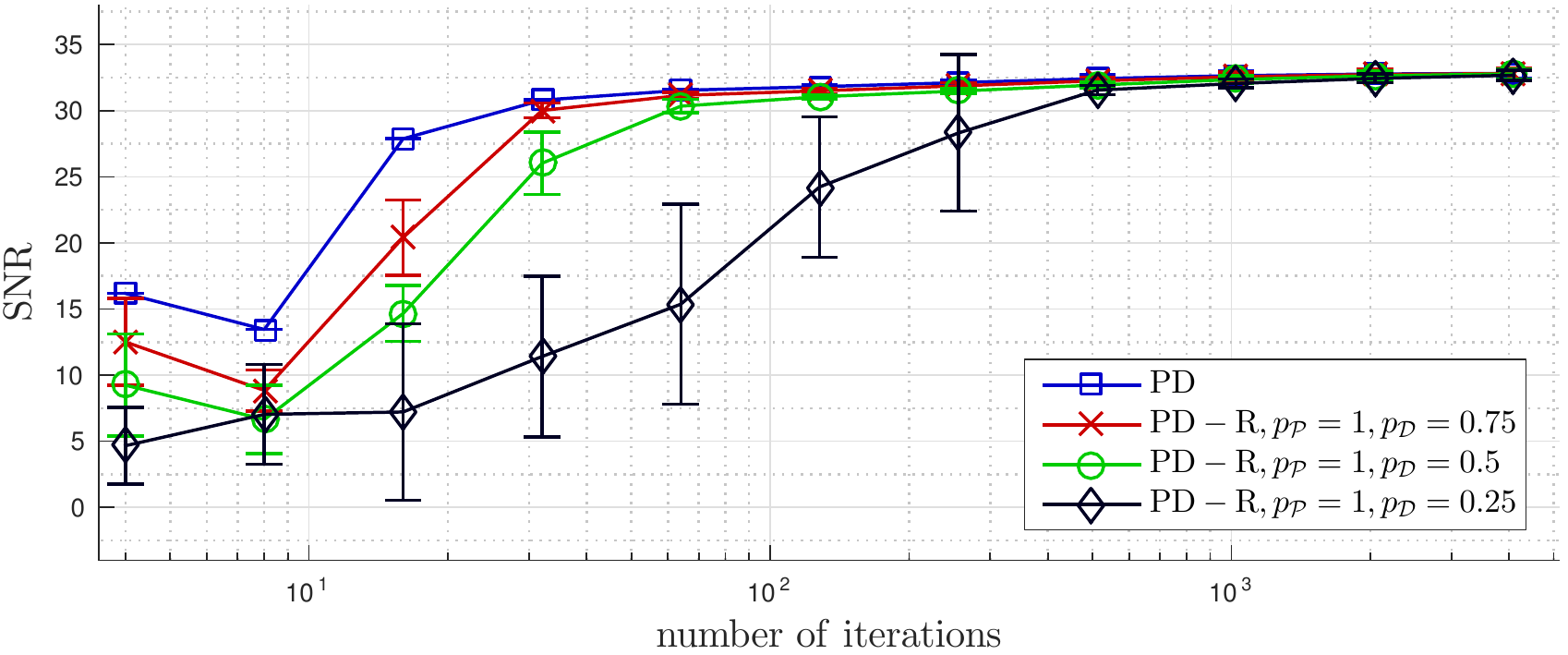}

	\caption{The SNR for the reconstruction of the Cygnus A image from $M=N$ visibilities for the \ac{pd}  and \ac{pd-r} algorithms with parameter $\kappa=10^{-3}$. The algorithms split the input data into: (top) $4$ blocks, (middle) $16$ blocks, (bottom) $64$ blocks.}
	\label{fig-cyn-1M-results-f-prob-L2}
\end{figure}

\begin{figure}
	\centering

  	\includegraphics[trim={0px 0px 0px 0}, clip, width=0.99\linewidth]{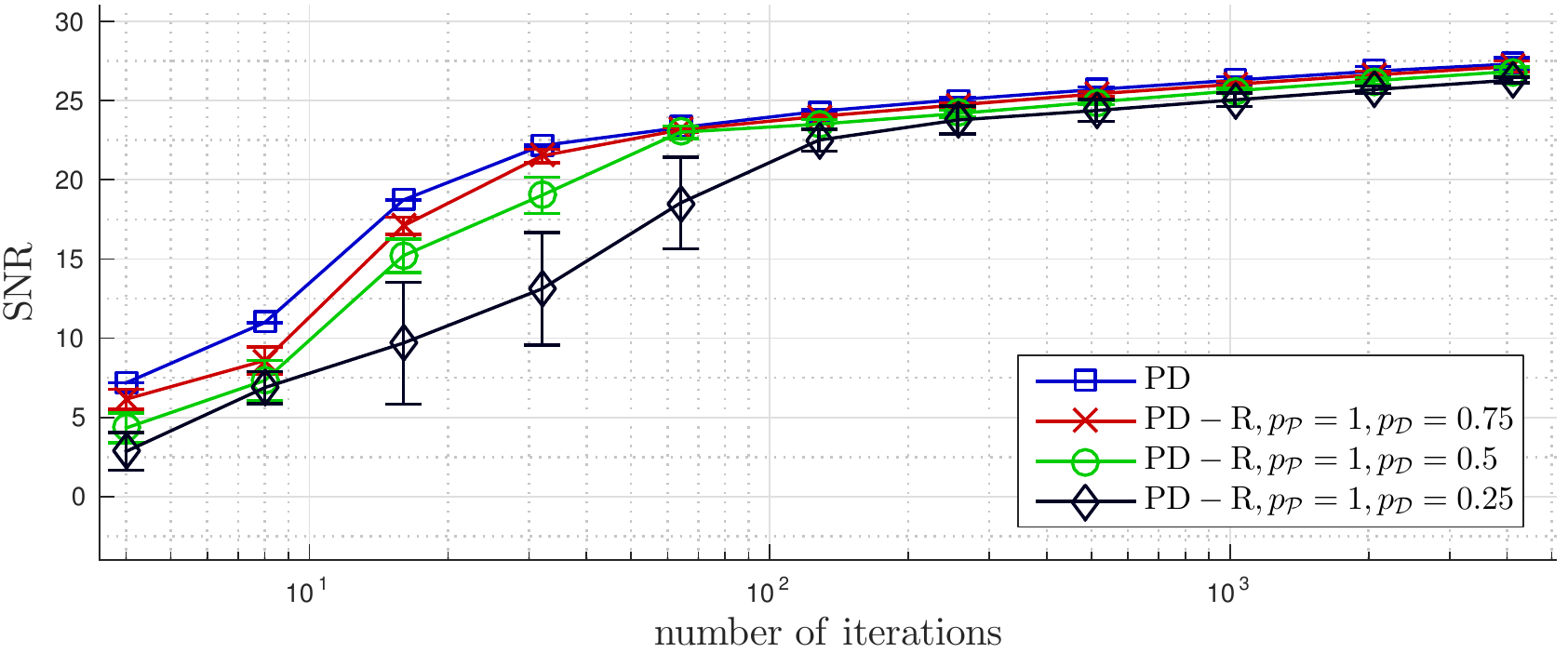}
  	\includegraphics[trim={0px 0px 0px 0}, clip, width=0.99\linewidth]{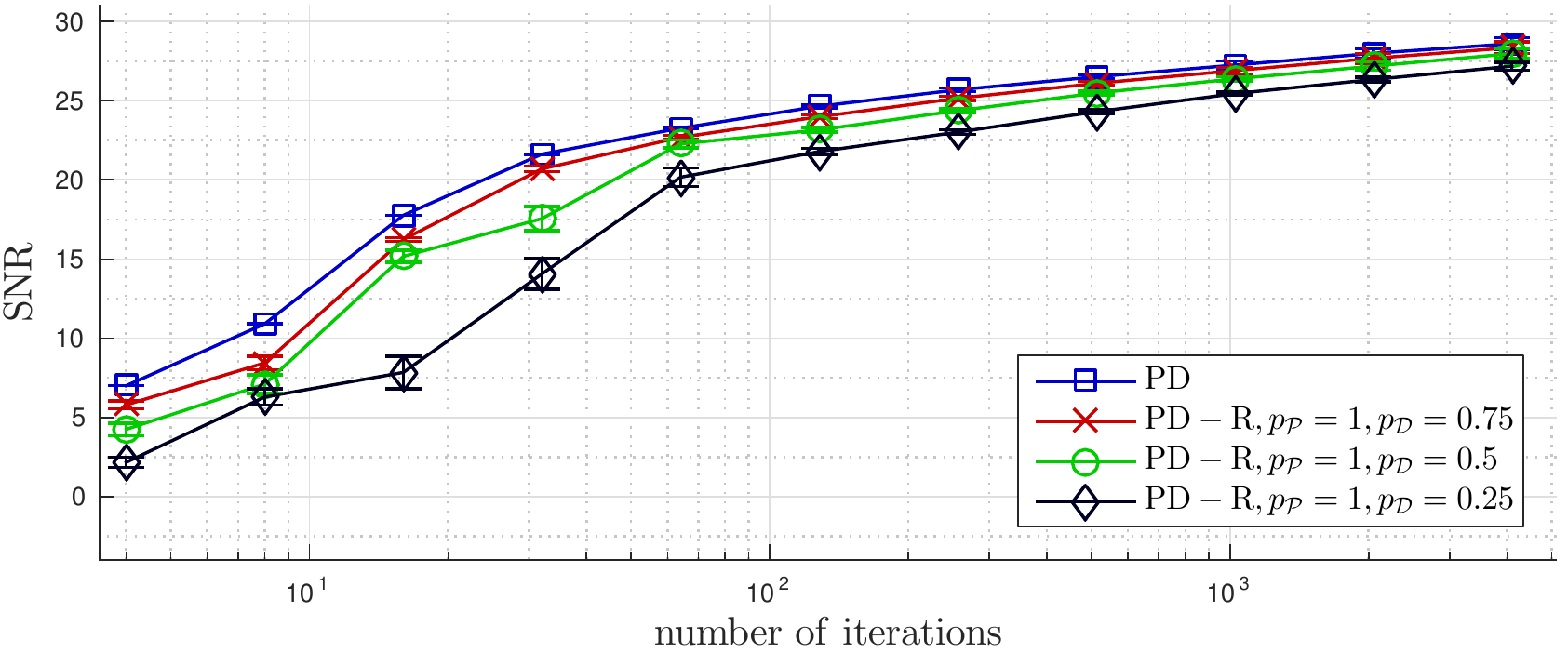}
  	\includegraphics[trim={0px 0px 0px 0}, clip, width=0.99\linewidth]{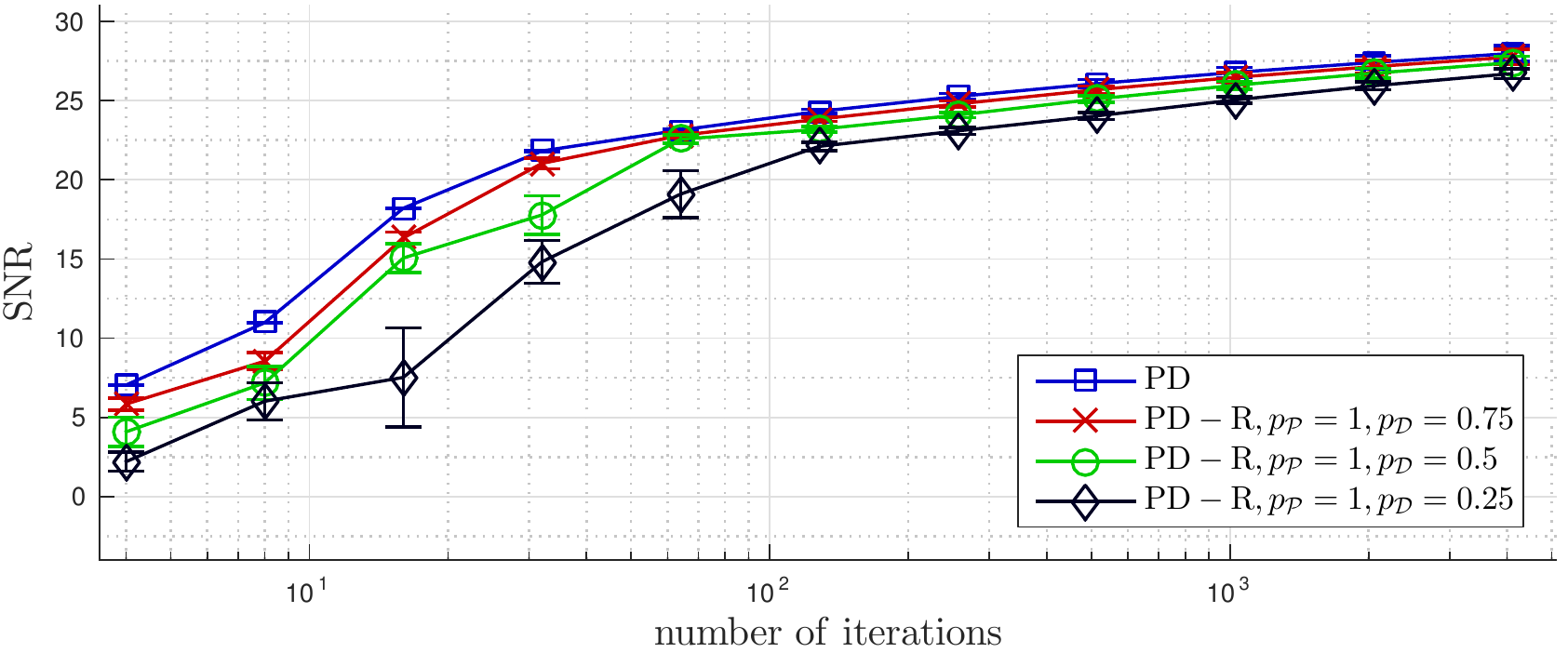}

	\caption{\bc The SNR for the reconstruction of the galaxy cluster image from $M=2N$ visibilities for the \ac{pd}  and \ac{pd-r} algorithms with parameter $\kappa=10^{-3}$. The algorithms split the input data into: (top) $4$ blocks, (middle) $16$ blocks, (bottom) $64$ blocks. \ec}
	\label{fig-gc-2M-results-f-prob-L2}
\end{figure}

The convergence speed of the randomised algorithm, \ac{pd-r}, is studied in Figures \ref{fig-m31-10M-results-f-prob-L2}, \ref{fig-cyn-1M-results-f-prob-L2} and \ref{fig-gc-2M-results-f-prob-L2} for the M31, Cygnus A  and galaxy cluster test images, with three choices for the data splitting.
As expected, the convergence speed decrease when the probability of update $p_{\mc{D}}$ is lowered.
The number of iterations required for convergence increases greatly for probabilities below $0.25$.
Similar behaviour is achieved for the reconstruction of the test images from a smaller number of measurements.
Again, the convergence speed for the galaxy cluster test image is slower.
There is also a very small decrease in the convergence speed for all tests when the data are split into a larger number of blocks.
This is due to the fact that, in order to reach the same global $\epsilon$, the resulting bounds imposed per block are more constraining and due to the fact that achieving a consensus between a larger number of blocks is more difficult.

\begin{figure}
	\centering
  	\includegraphics[trim={0px 0 0px 0}, clip, width=0.99\linewidth]{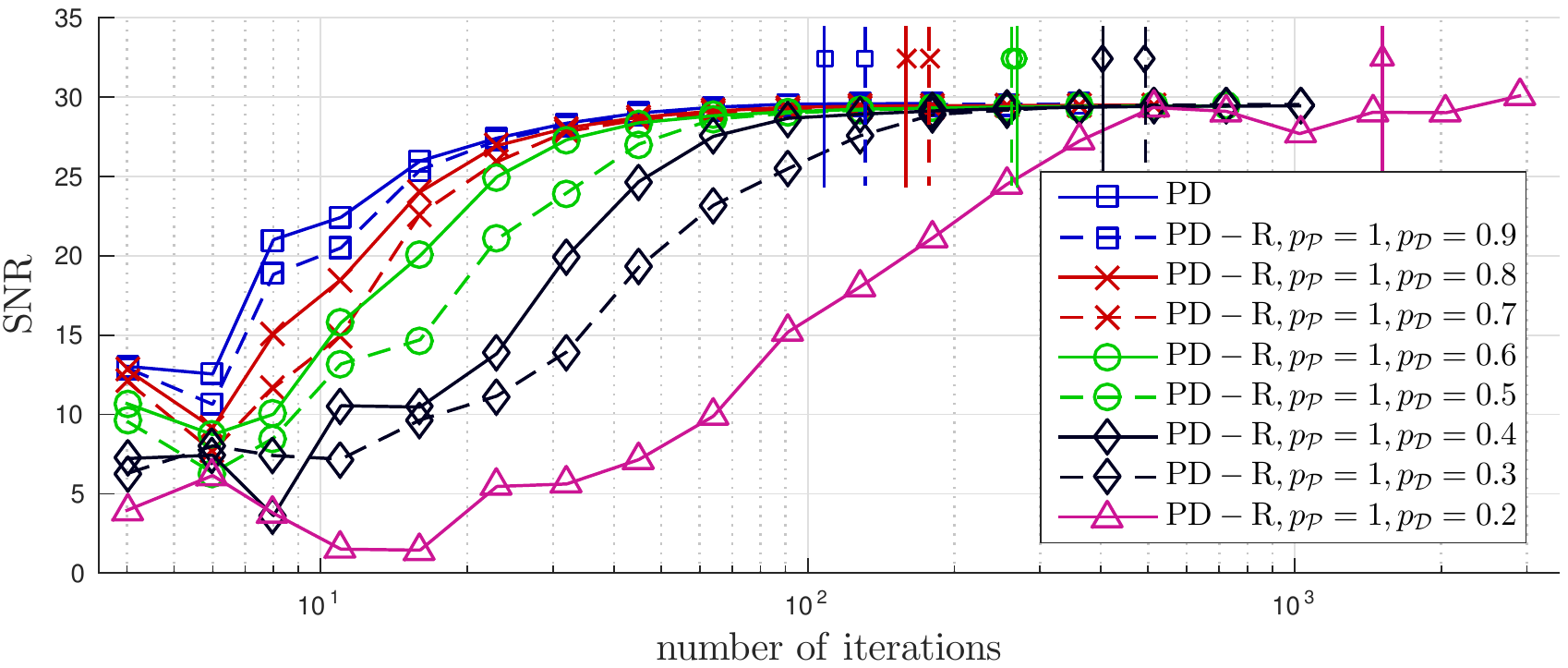}				
	\includegraphics[trim={-4px 0 0px 0}, clip, width=0.99\linewidth]{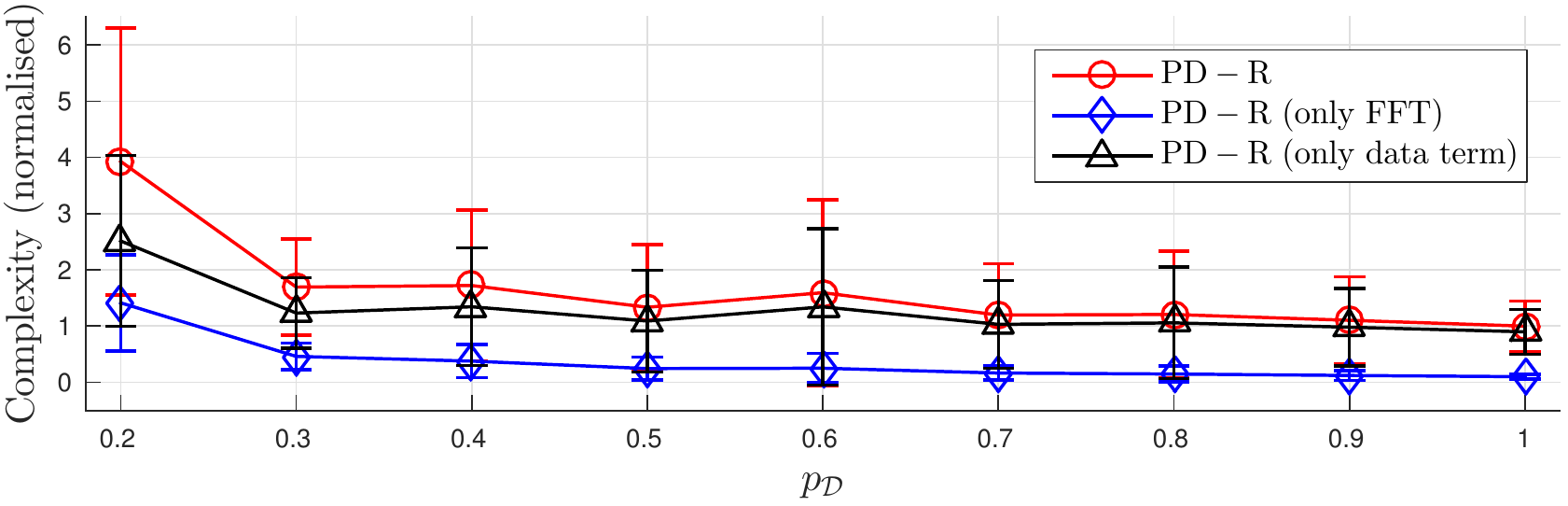}
	\caption{(top) The evolution of the $\rm SNR$ for \ac{pd-r} for different probabilities for the reconstruction of the M31 test image from $M=10N$ measurements. The average number of iterations performed \bc for $\kappa=10^{-3}$, $\bar{\delta}=10^{-4}$ and $\bar{\epsilon}^2 = \left(2M + 3\sqrt{4M}\right)\sigma_{\chi}^2$ is marked by a vertical line\ec. (bottom) The total complexity of \ac{pd-r} and the parts of its total complexity due to the \ac{fft} and the data term computations, all normalised with respect to the average total complexity of \ac{pd}. The visibilities are split into $64$ equal size blocks.}
	\label{fig-m31-10M-results-f-prob-delta}
\end{figure}

Generally, the convergence speed decreases gradually as the probability $p_{\mc{D}}$ gets lower, \ac{pd-r} remaining competitive and able to achieve good complexity as can be seen in Figure~\ref{fig-m31-10M-results-f-prob-delta}.
Here, we exemplify the performance in more detail when using the $64$ blocks with parameter $\kappa=10^{-3}$, the stopping threshold $\bar{\delta} = 10^{-4}$ and the $\ell_2$ ball stopping threshold $\bar{\epsilon}^2 = \left(2M + 3\sqrt{4M}\right)\sigma_{\chi}^2$.
Our tests show that the total number of iterations performed is roughly inversely proportional to the probability $p_{\mc{D}}$.
Additionally, we provide a basic estimate of the overall global complexity given the data from Table~\ref{complexity-table} and the number of iterations required. 
We only take into account the computationally heaviest operations, the \ac{fft} and the operations involving the data fidelity terms. The computations involving the sparsity priors are performed in parallel with the data fidelity computations and are much lighter.
Since the analysis is made up to a scaling factor, for better consistency, we normalised the complexity of \ac{pd-r} with respect to that of the \ac{pd}.

The total complexity of \ac{pd-r} remains similar to that of the non-randomised \ac{pd} which makes \ac{pd-r} extremely attractive.
Generally, if the main computational bottleneck is due to the data term and not to the \ac{fft} computations it is expected that the total complexity of \ac{pd-r} will remain comparable to that of the non-randomised \ac{pd}.
This is of great importance since, for a very large number of visibilities when the data does not fit in memory on the processing nodes, \ac{pd-r} may be the only feasible alternative.
When a more accurate stopping criterion is used, either with a smaller $\bar{\epsilon}_j$ or relative variation of the solution $\bar{\delta}$, the randomised algorithms start to require increasingly more iterations to converge and their relative complexity grows.
Randomisation over the sparsity bases is also possible but, due to the low computational burden of the priors we use, it is not of interest herein.
However, randomisation over the prior functions can become an important feature when computationally heavier priors are used or when the images to be reconstructed are very large.

\subsection{Results with the VLA and SKA coverages}

\begin{figure}
	\centering
	\includegraphics[trim={0px 0px 0px 0px}, clip, width=0.99\linewidth]{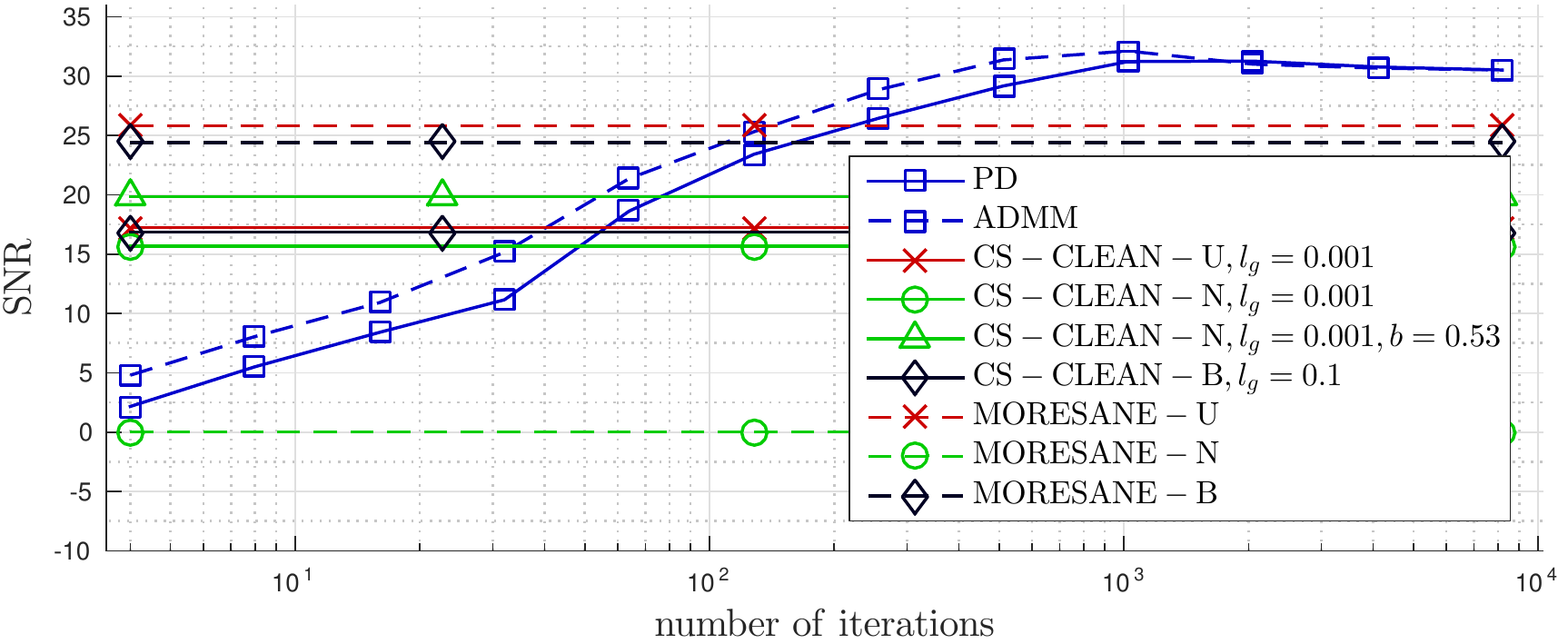}
	
	\includegraphics[trim={0px 0px 0px 0px}, clip, width=0.99\linewidth]{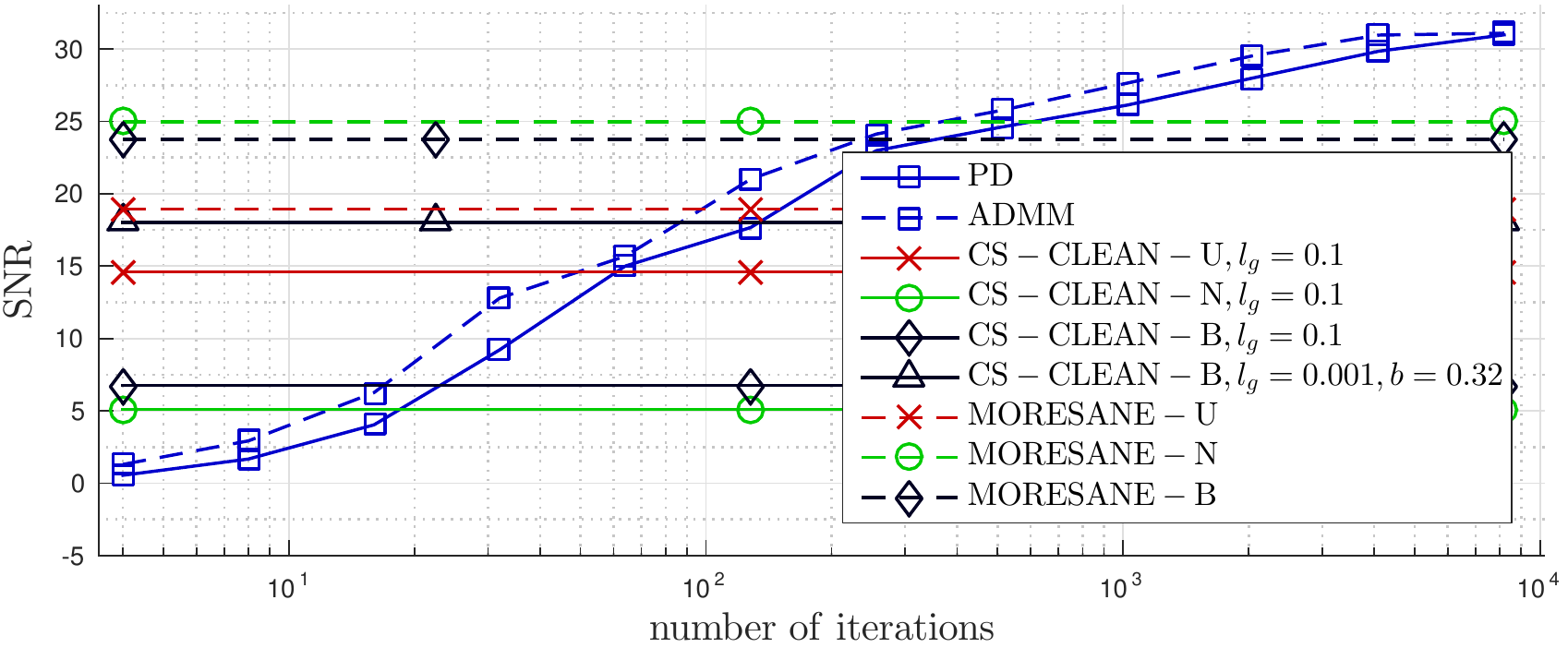}

	\includegraphics[trim={0px 0px 0px 0px}, clip, width=0.99\linewidth]{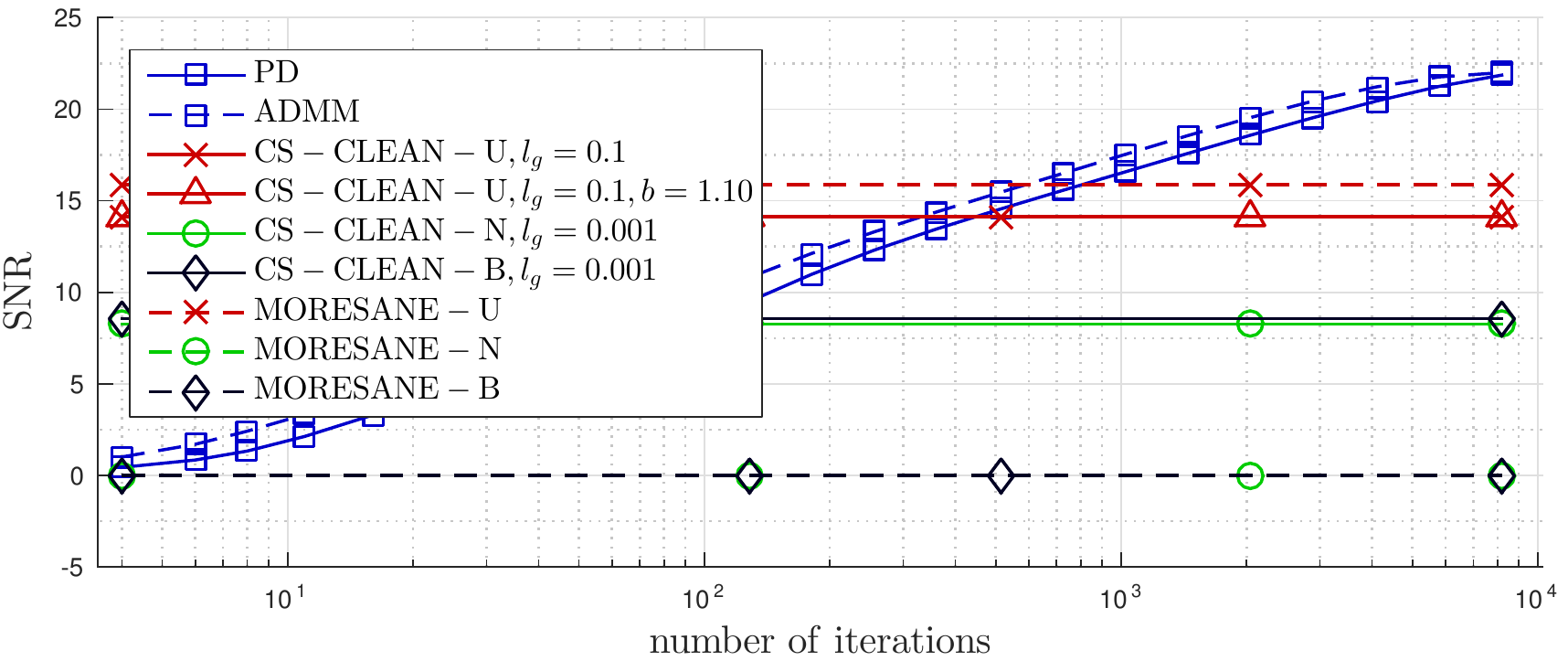}

	\caption{\bc The $\rm SNR$ achieved by the algorithms studied for the reconstruction of (from top to bottom) the Cygnus A and the galaxy cluster images using the VLA coverage, and of the W28 supernova remnant image using SKA the coverage. For the \ac{pd} and \ac{admm} algorithms we report the evolution of the SNR as a function of the iteration number. They use $\kappa = 10^{-5}$ and the data split into $64$ equal size blocks. The horizontal lines represent the final SNR achieved using CS-CLEAN and MORESANE. \ec}
	\label{vla-ska-results}
\end{figure}

In Figure~\ref{vla-ska-results} we present the $\rm SNR$ evolution as a function of the number of iterations for the \ac{pd} and \ac{admm} algorithms for the reconstruction of the Cygnus A and galaxy cluster images using the VLA coverage, and of the W28 supernova remnant test image using the SKA coverage.
The visibilities are split into $64$ equal size blocks and the parameter $\kappa=10^{-5}$.
We also overlay on the figures the $\rm SNR$ achieved using \ac{cs-clean} and \ac{moresane} with the different types of weighting.

\begin{figure}

	\centering
	\hspace{-6pt}
	\begin{minipage}{.98\linewidth}
  		\centering
  		\includegraphics[trim={0px 0px 0px 0px}, clip, height=3.825cm]{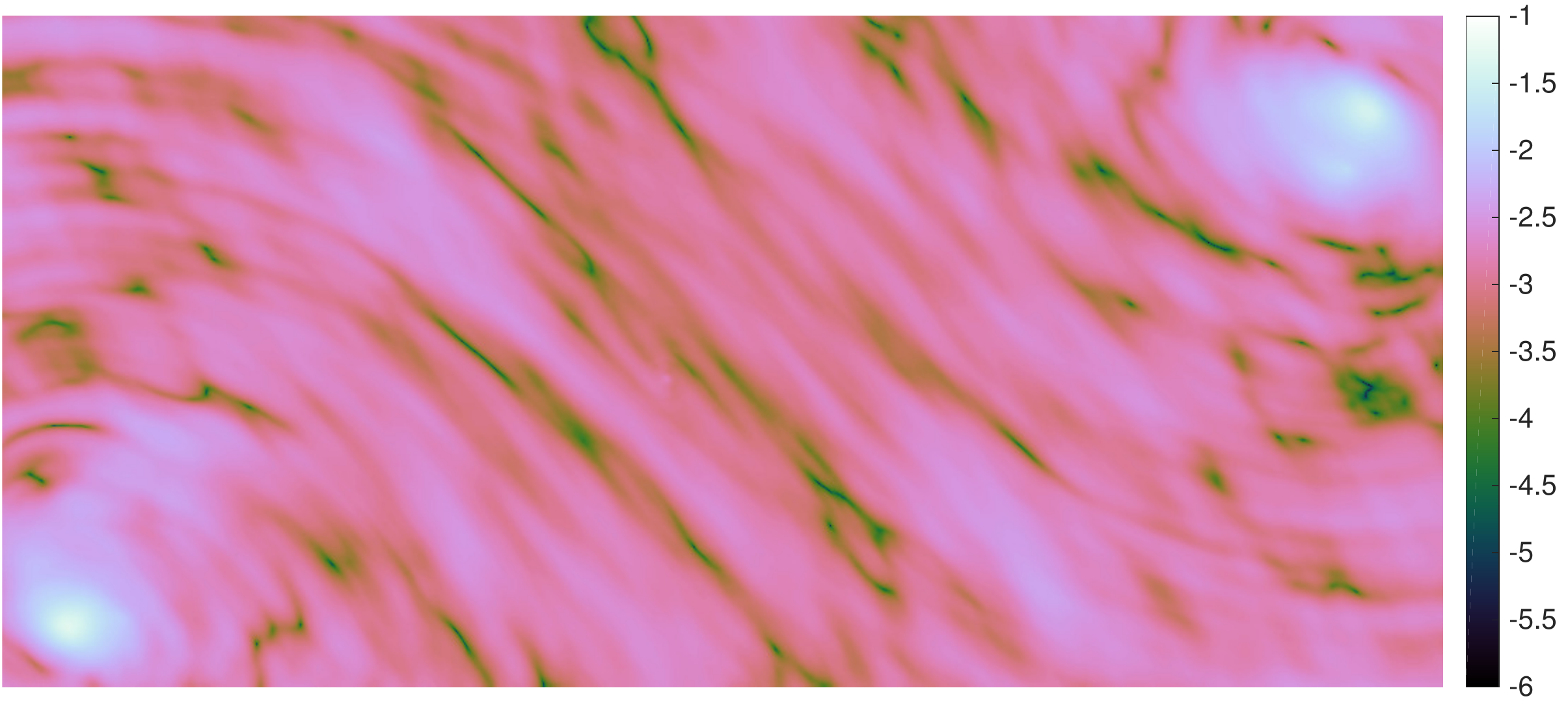}
	\end{minipage}

	\begin{minipage}{.49\linewidth}
	        \centering
  		\includegraphics[trim={0px 0px 0px 0px}, clip, height=3.80cm]{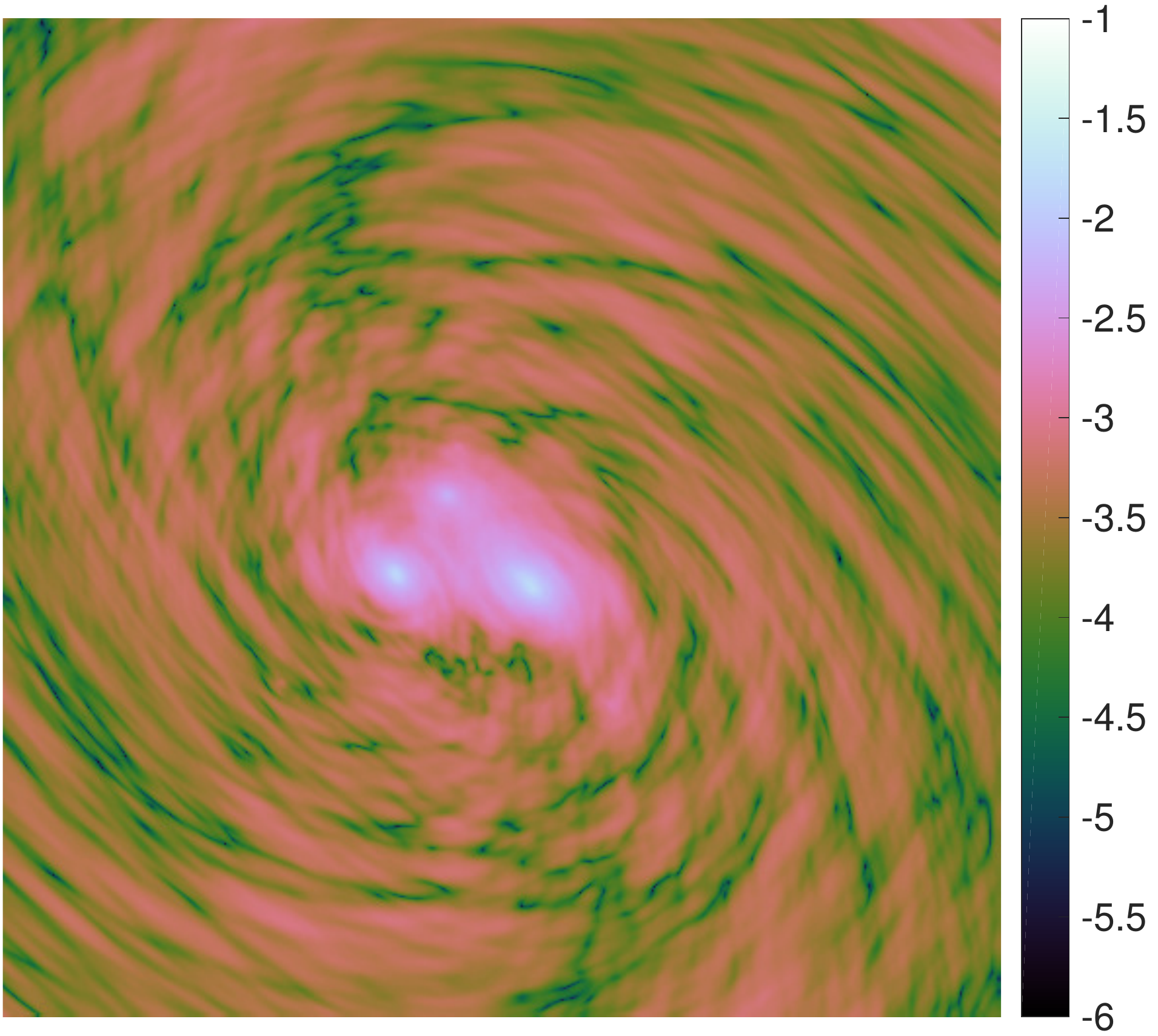}
	\end{minipage}\hspace{0.5pt}
	\begin{minipage}{.49\linewidth}
	        \centering
  		\includegraphics[trim={0px 0px 0px 0px}, clip, height=3.80cm]{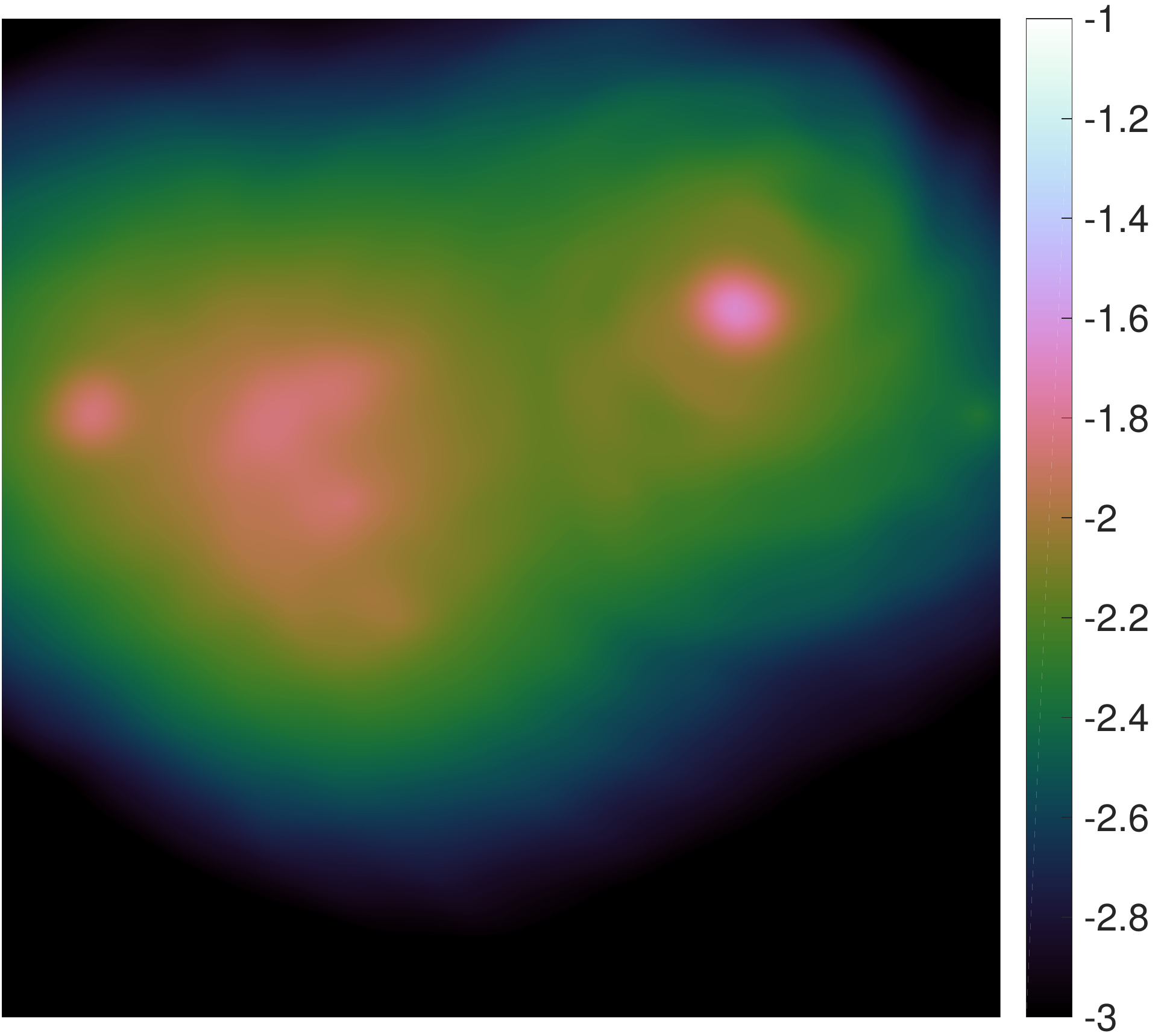}
	\end{minipage}

		\caption{\bc The log scale absolute value of the dirty images using natural weighting corresponding to (top) the Cygnus A and (bottom, left) the galaxy cluster test images, using the VLA coverage, and to (bottom, right) the W28 supernova remnant test image using the SKA coverage.\ec}

	\label{ska-vla-dirty}
\end{figure}

\begin{figure*}
	\centering
	\includegraphics[trim={0px 0px 0px 0px}, clip, width=0.46\linewidth]{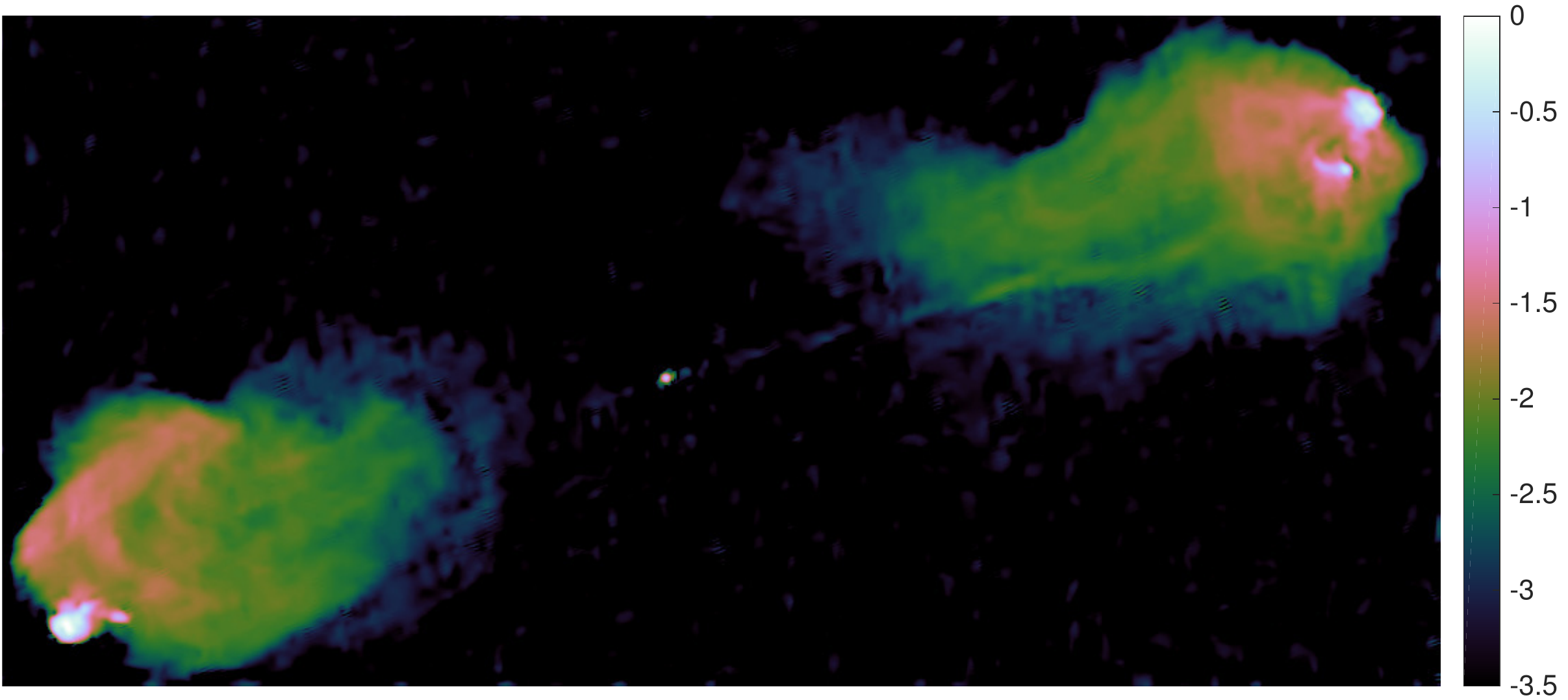}\hspace{25pt}
	\includegraphics[trim={0px 0px 0px 0px}, clip, width=0.46\linewidth]{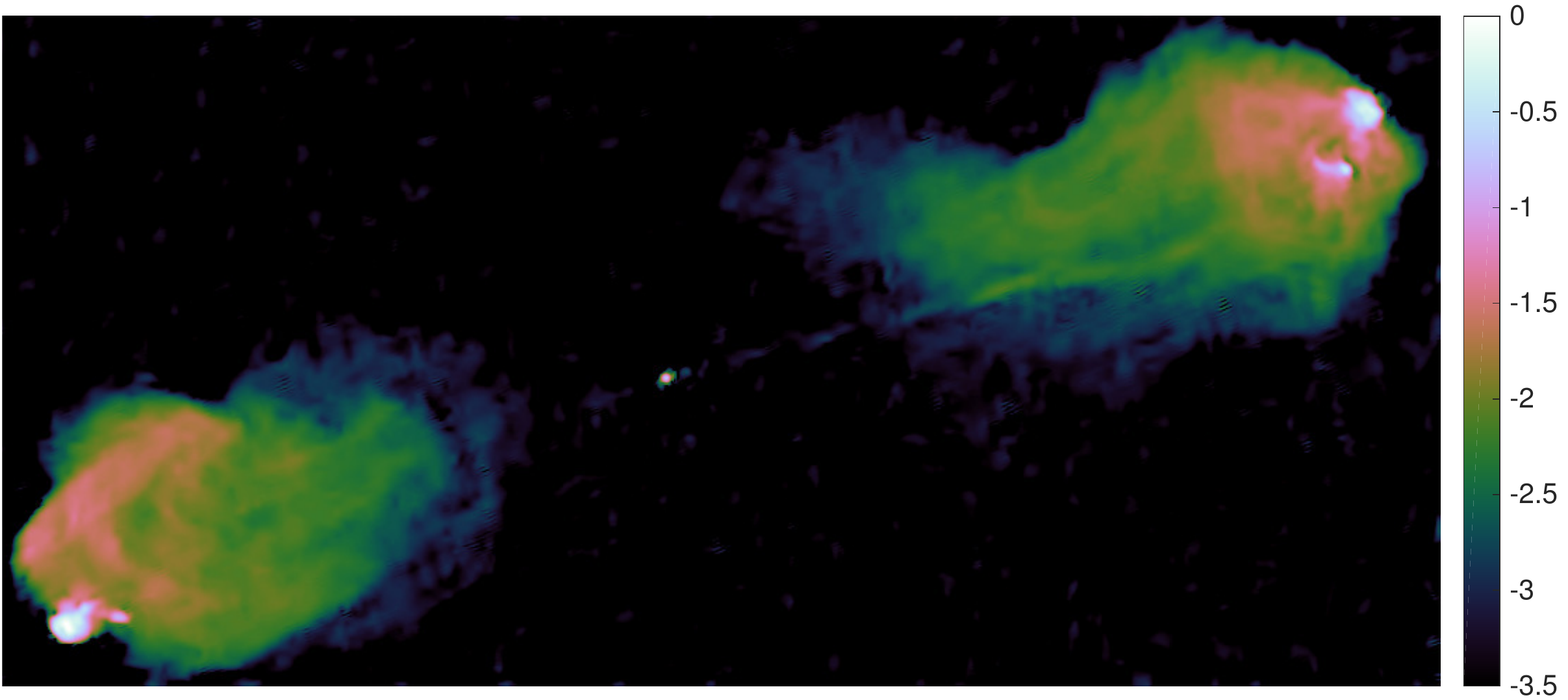}\vspace{-2pt}
	\includegraphics[trim={0px 0px 0px 0px}, clip, width=0.46\linewidth]{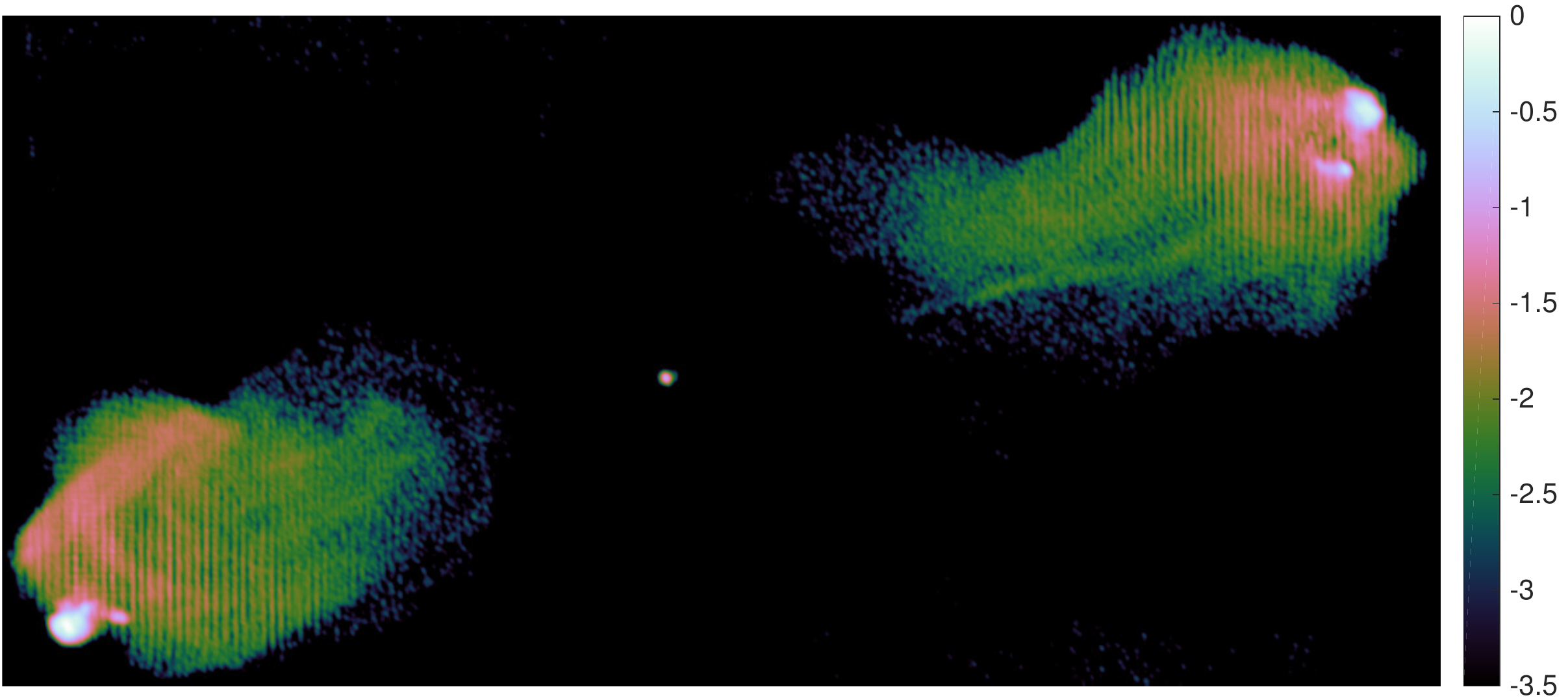}\hspace{25pt}
	\includegraphics[trim={0px 0px 0px 0px}, clip, width=0.46\linewidth]{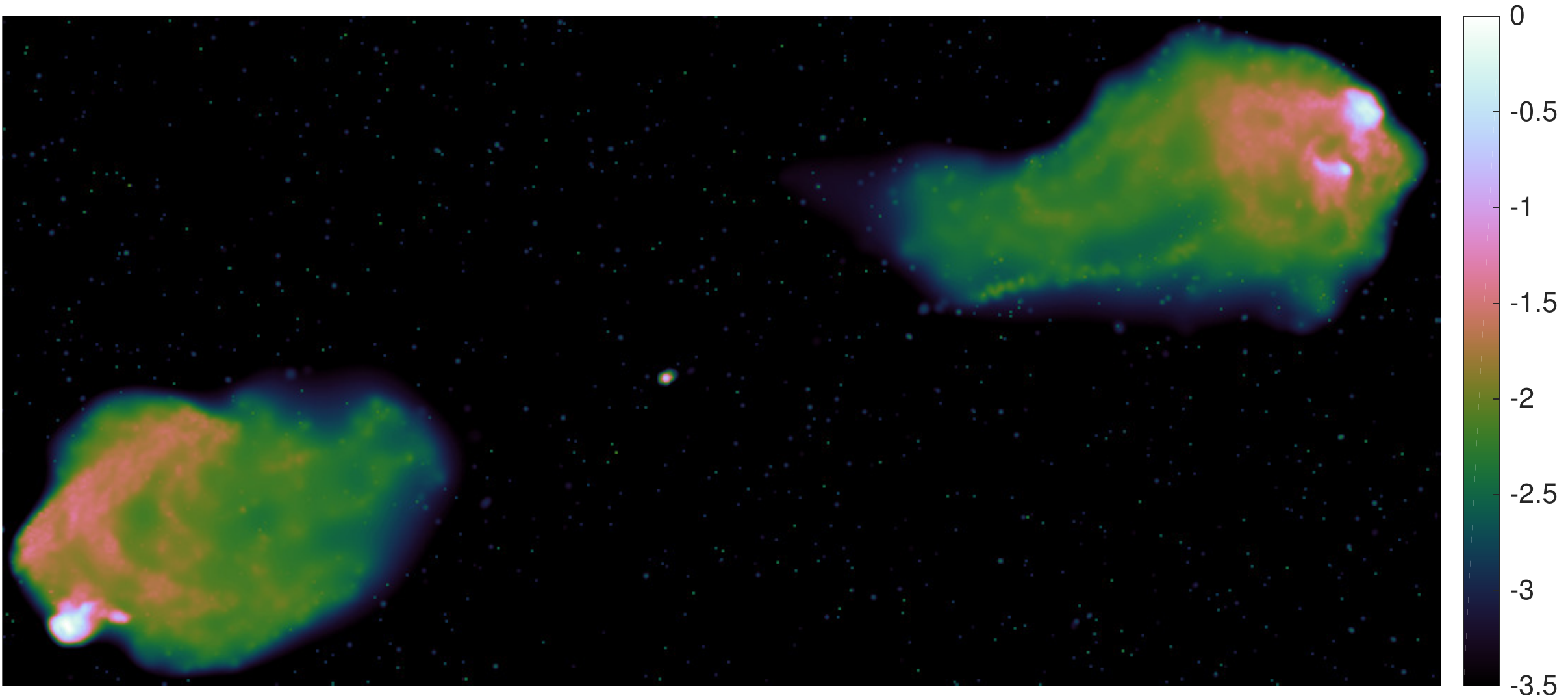}
	
	\includegraphics[trim={0px 0px 0px 0px}, clip, width=0.46\linewidth]{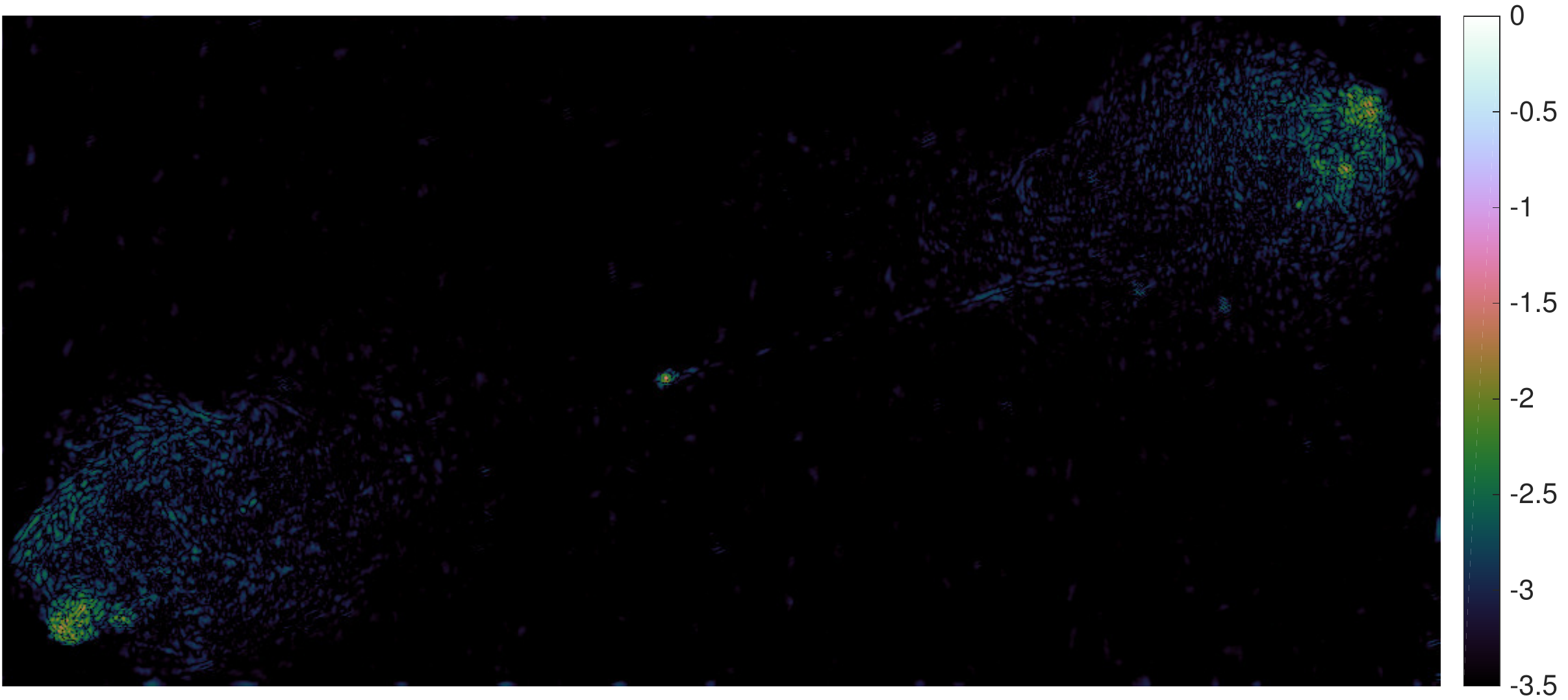}\hspace{25pt}
	\includegraphics[trim={0px 0px 0px 0px}, clip, width=0.46\linewidth]{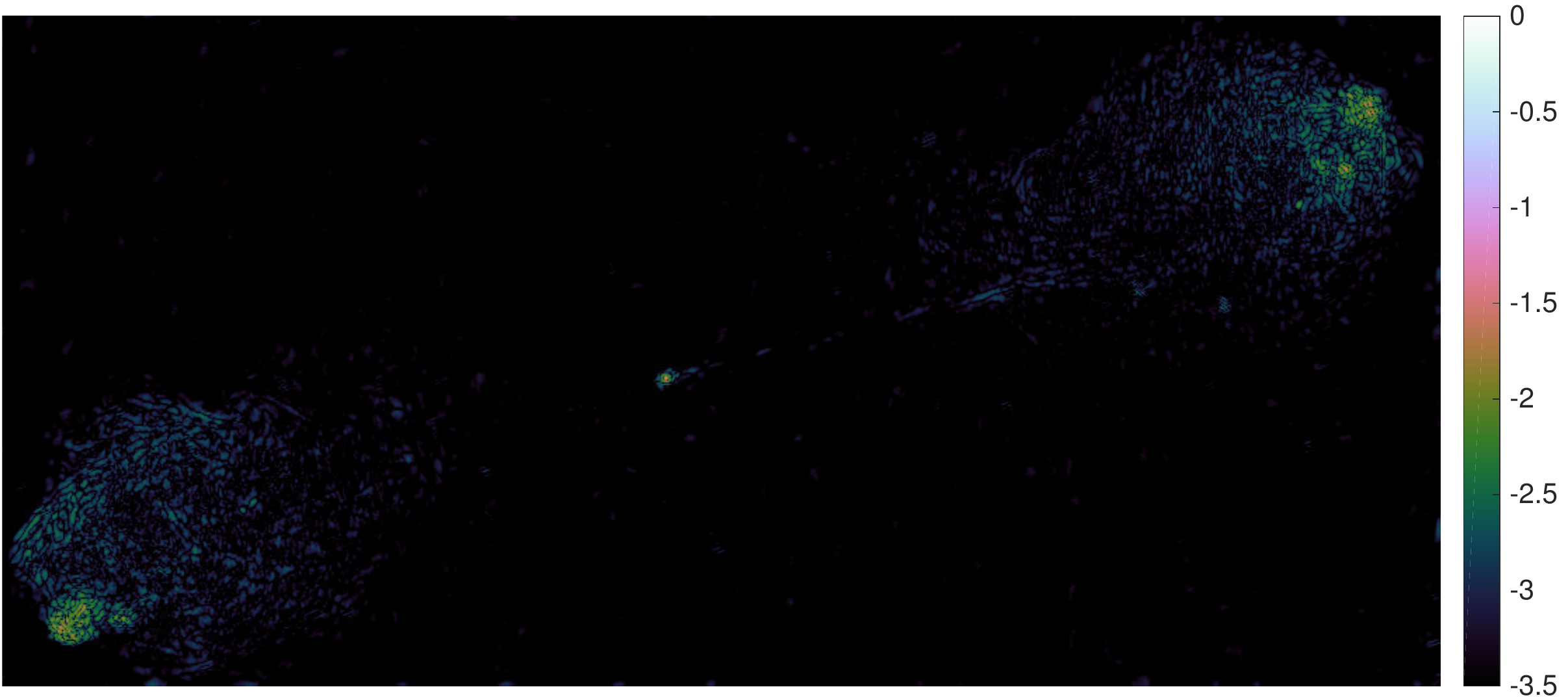}\vspace{-2pt}
	\includegraphics[trim={0px 0px 0px 0px}, clip, width=0.46\linewidth]{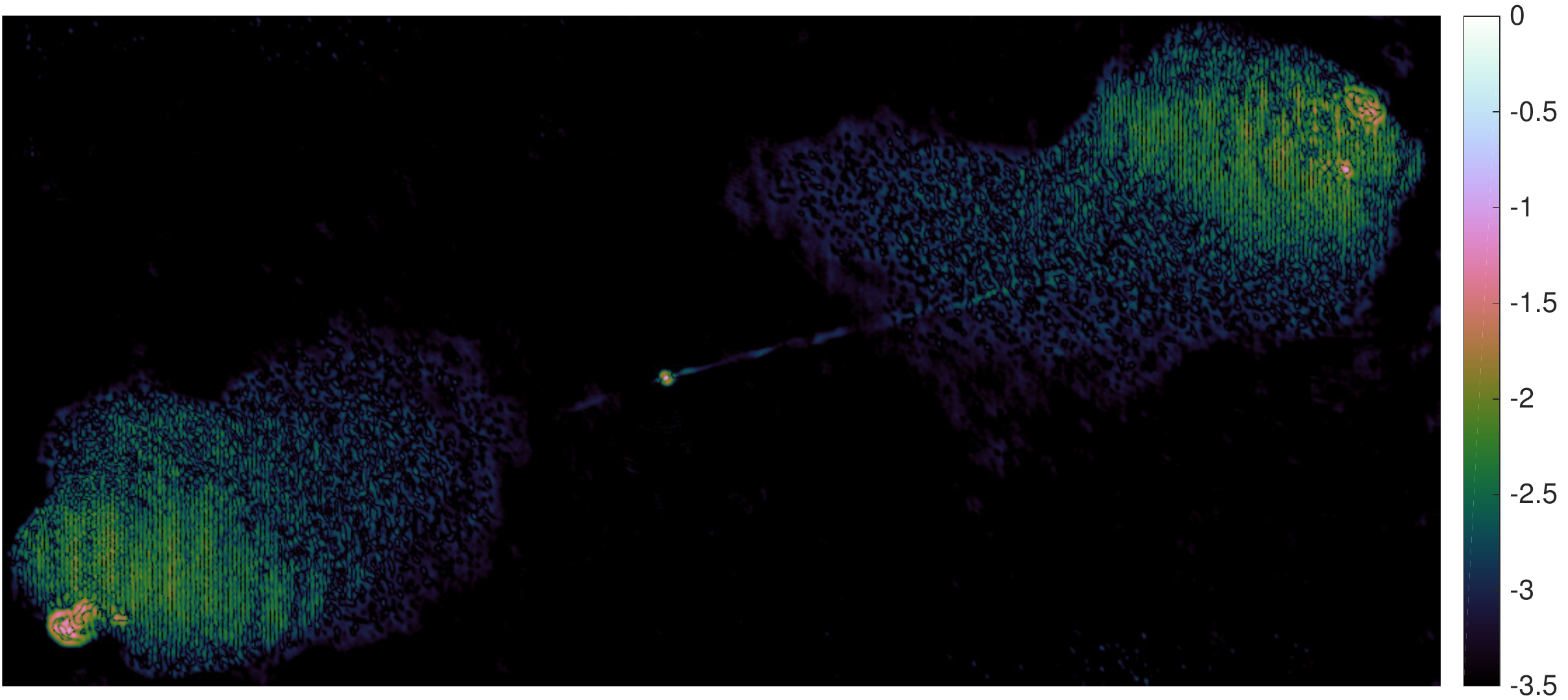}\hspace{25pt}
	\includegraphics[trim={0px 0px 0px 0px}, clip, width=0.46\linewidth]{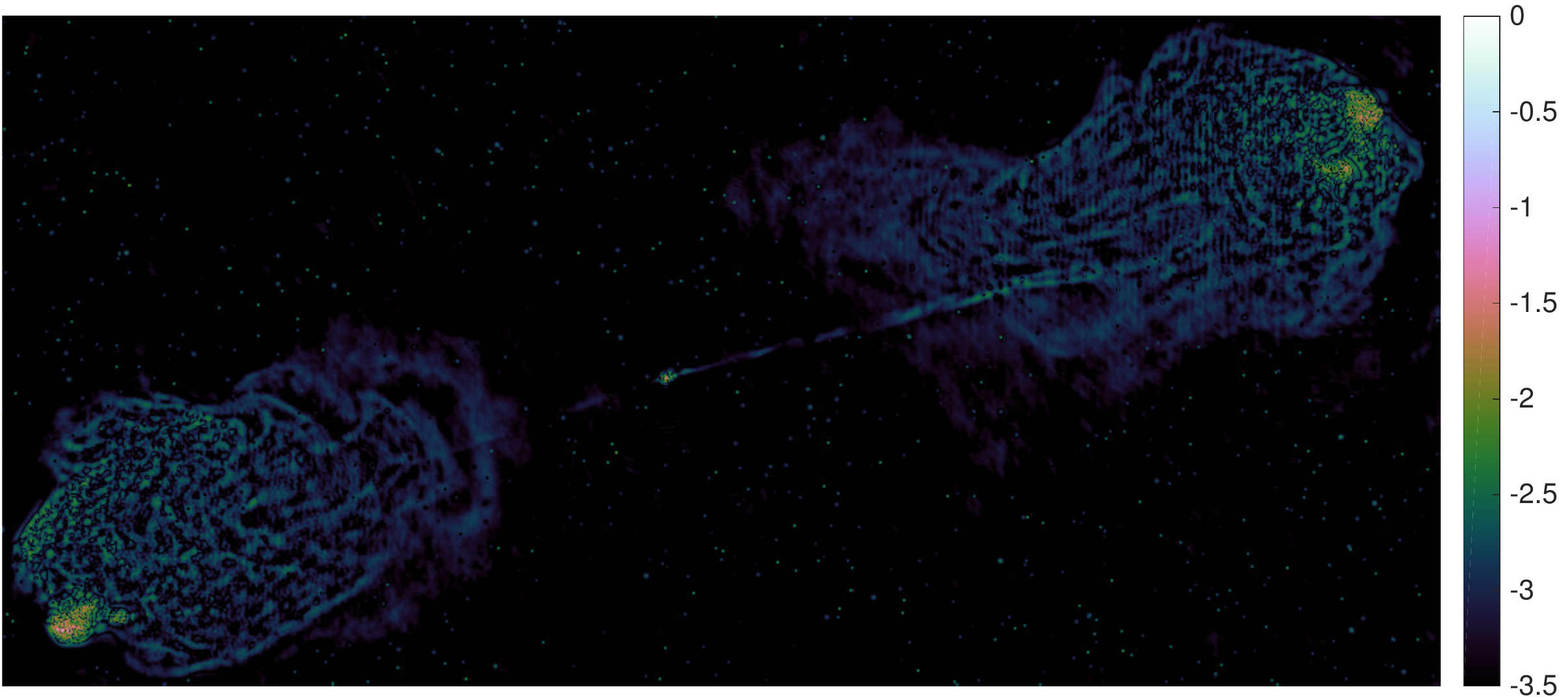}
	
	\includegraphics[trim={0px 0px 0px 0px}, clip, width=0.46\linewidth]{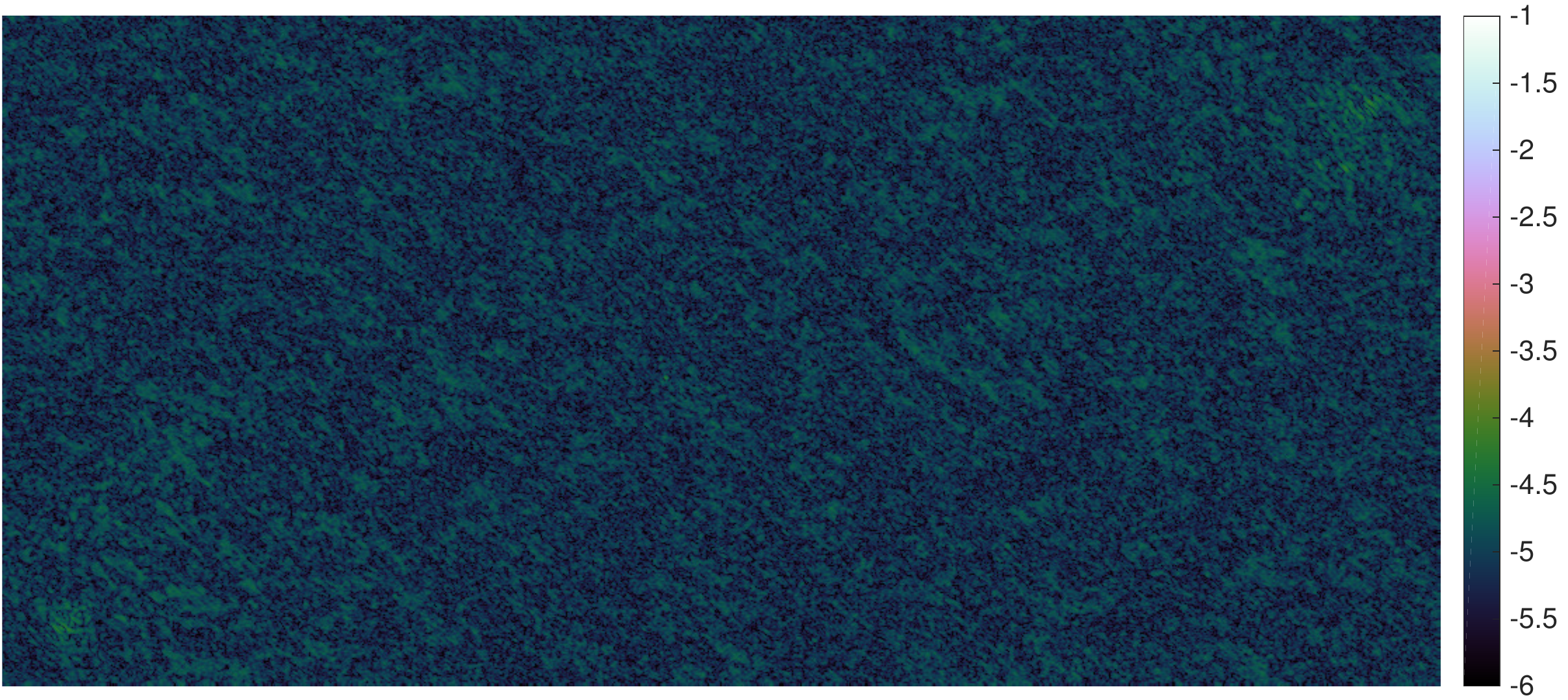}\hspace{25pt}
	\includegraphics[trim={0px 0px 0px 0px}, clip, width=0.46\linewidth]{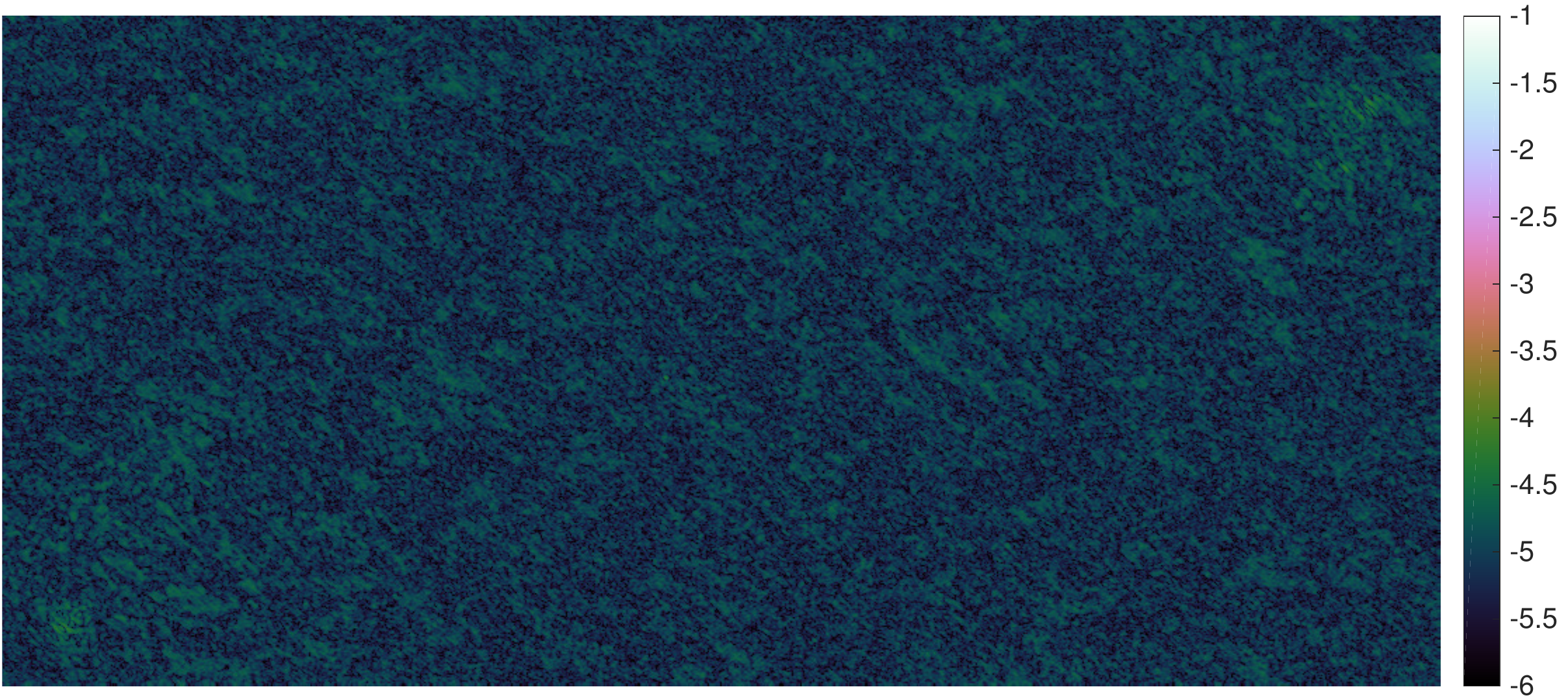}\vspace{-2pt}
	\includegraphics[trim={0px 0px 0px 0px}, clip, width=0.46\linewidth]{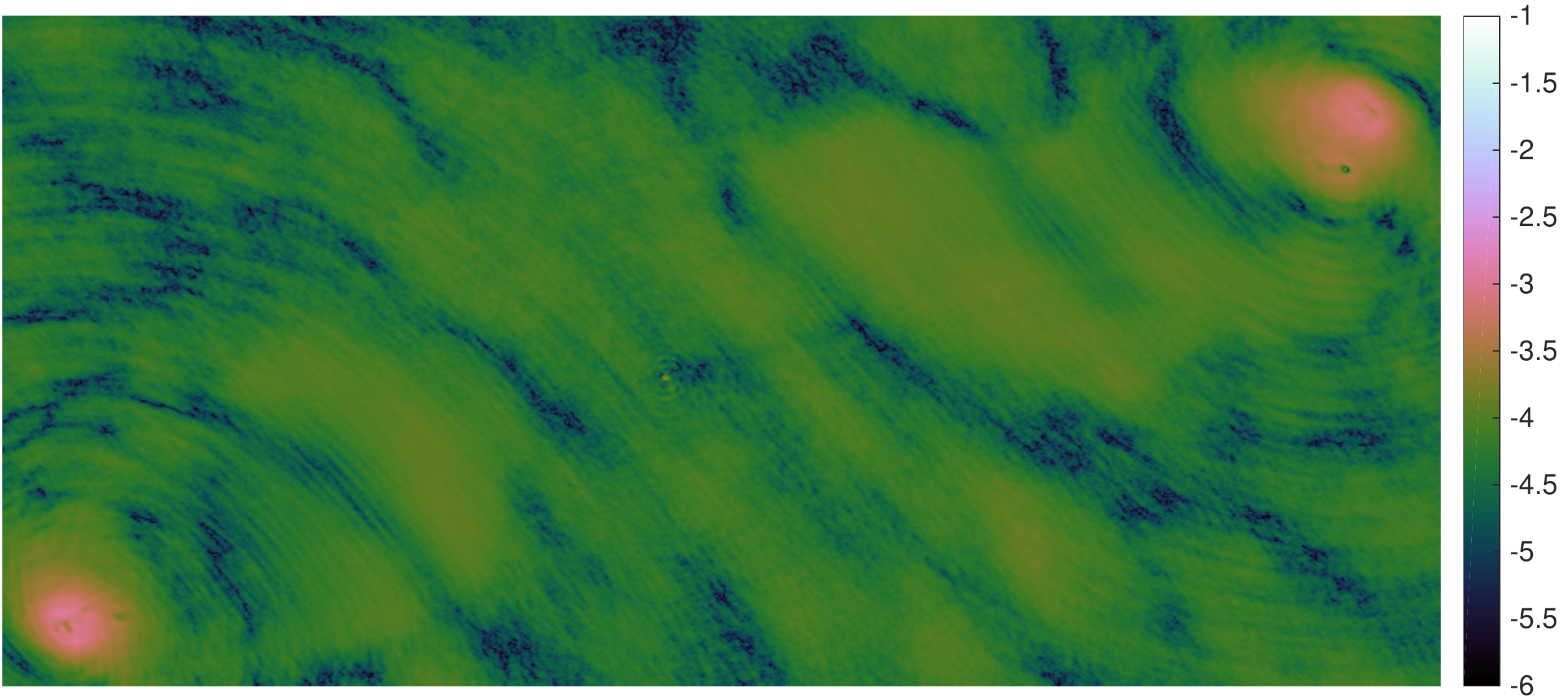}\hspace{25pt}
	\includegraphics[trim={0px 0px 0px 0px}, clip, width=0.46\linewidth]{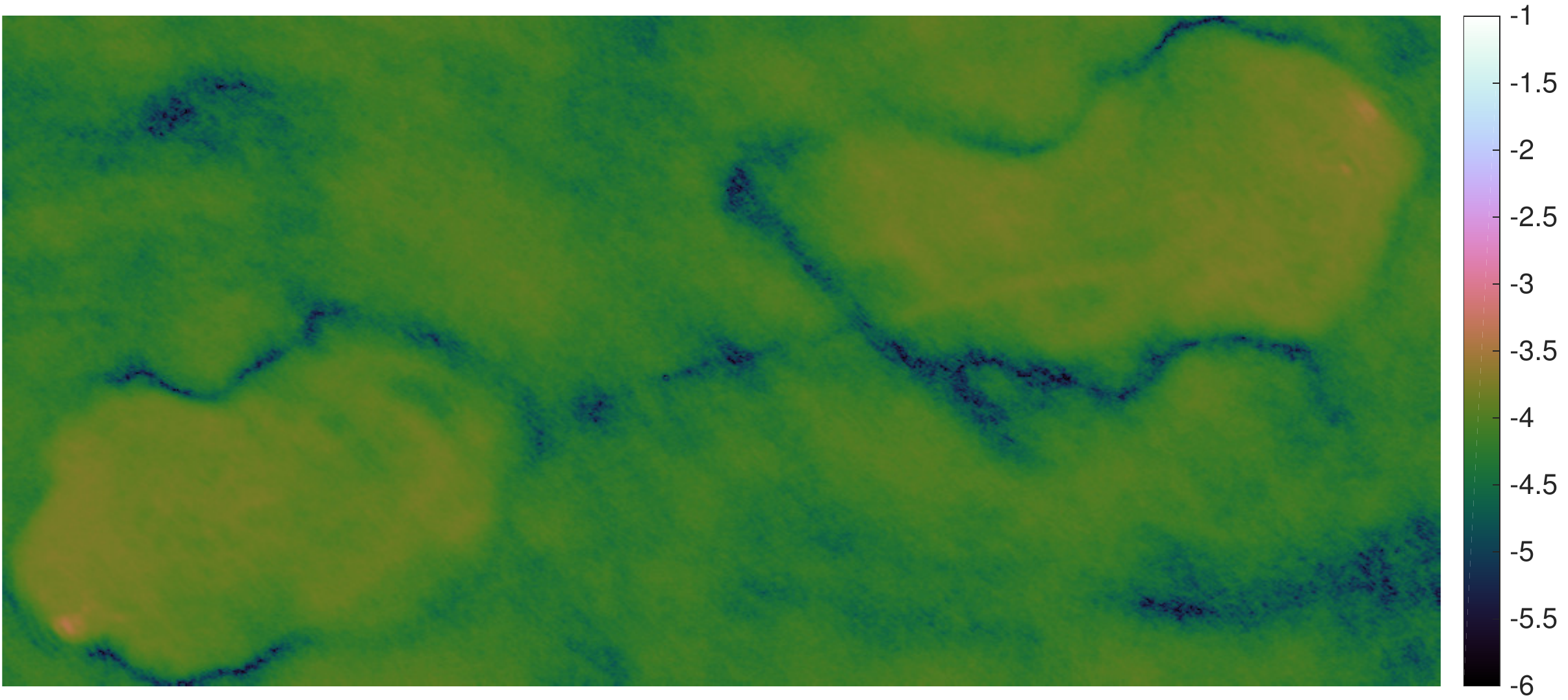}
	
	\caption{\bc (top 4 images) Log scale reconstructed images; (middle 4 images) log scale of the absolute value of the estimation errors; (bottom 4 images) log scale of the absolute value of the naturally weighted residual images, for the $477 \times 1025$ Cygnus A test image using the VLA coverage. For each group, the algorithms are: (top left) \ac{pd} with the reconstruction $\rm{SNR}= 30.51~\rm{dB}$ and the corresponding $\rm{DR}=108620$; (top right) \ac{admm} with the reconstruction $\rm{SNR}= 30.52~\rm{dB}$ and $\rm{DR}=107050$; (bottom left) \ac{cs-clean-n} with $l_g=0.001$ and $b=0.53$ with the reconstruction $\rm{SNR}= 19.95~\rm{dB}$ and $\rm{DR}=10773$; (bottom right) \ac{moresane-u} with the reconstruction $\rm{SNR}=25.82~\rm{dB}$ and $\rm{DR}= 11661$.  The images correspond to the best results obtained by all algorithms as presented in Figure~\ref{vla-ska-results}.
	\ec}
	\label{fig-images-ca}
\end{figure*}

\begin{figure*}
	\centering
	\includegraphics[trim={0px 0px 0px 0px}, clip, height=0.29\linewidth]{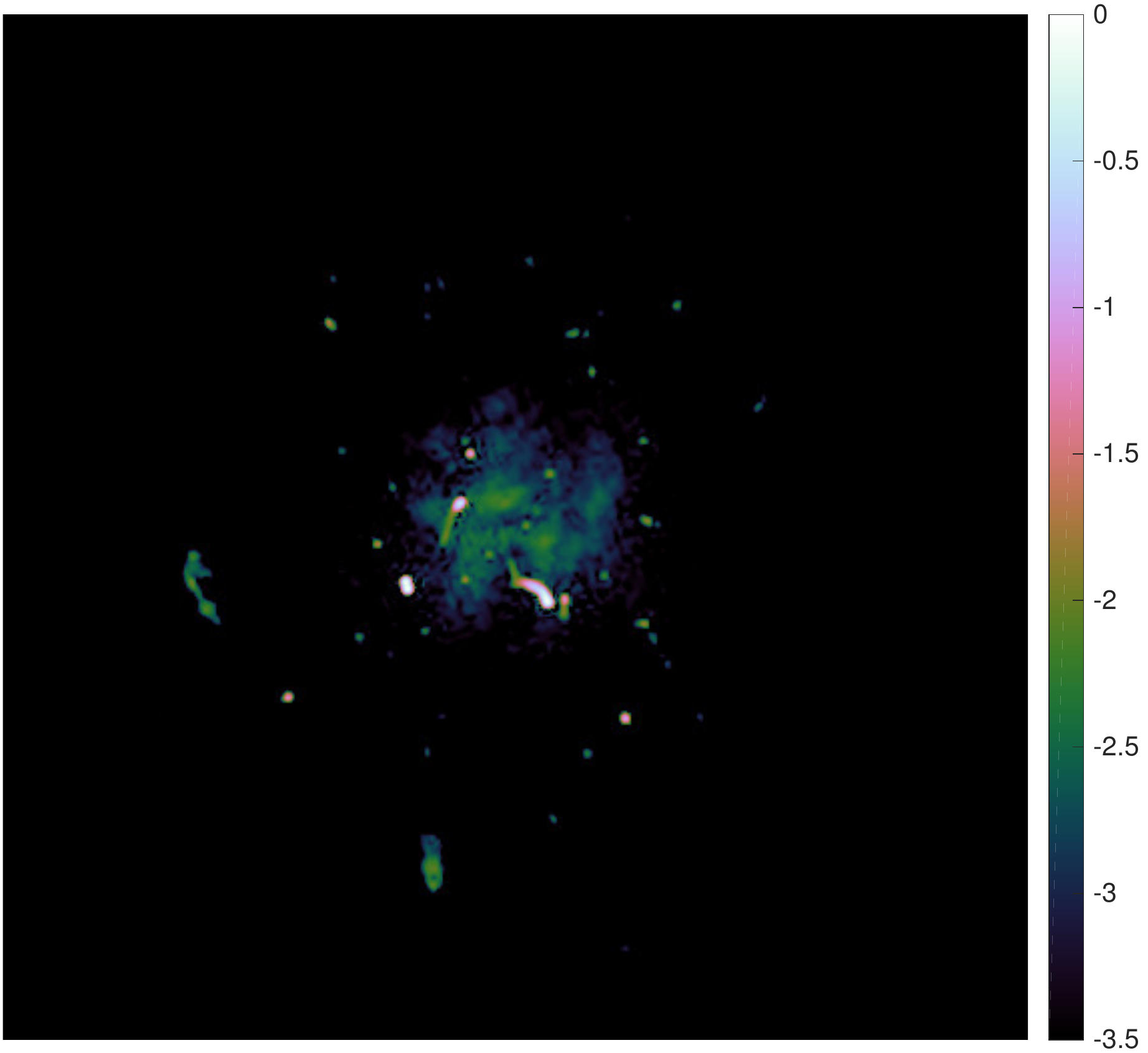}\hspace{2pt}
	\includegraphics[trim={0px 0px 0px 0px}, clip, height=0.29\linewidth]{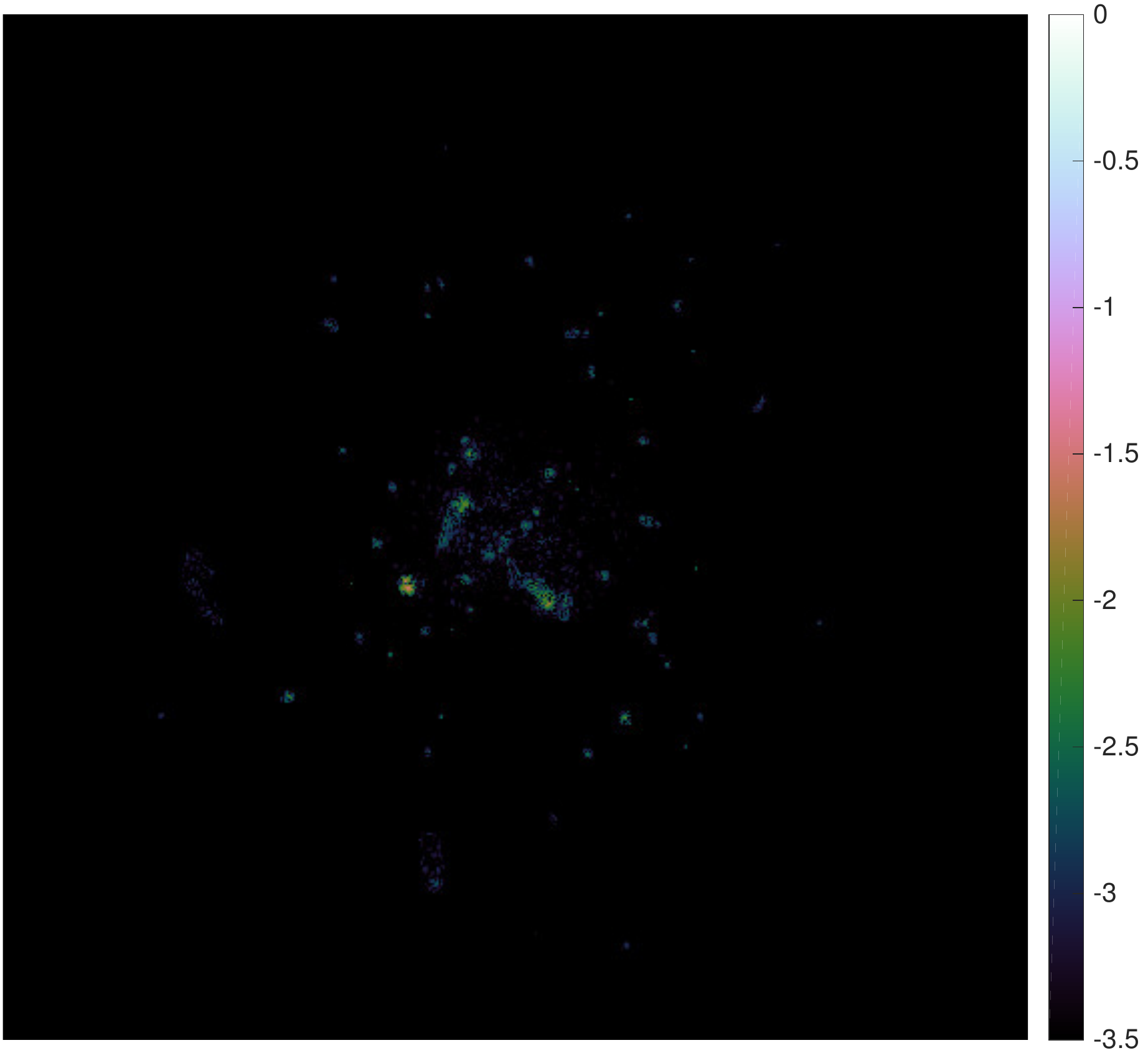}\hspace{2pt}
	\includegraphics[trim={0px 0px 0px 0px}, clip, height=0.29\linewidth]{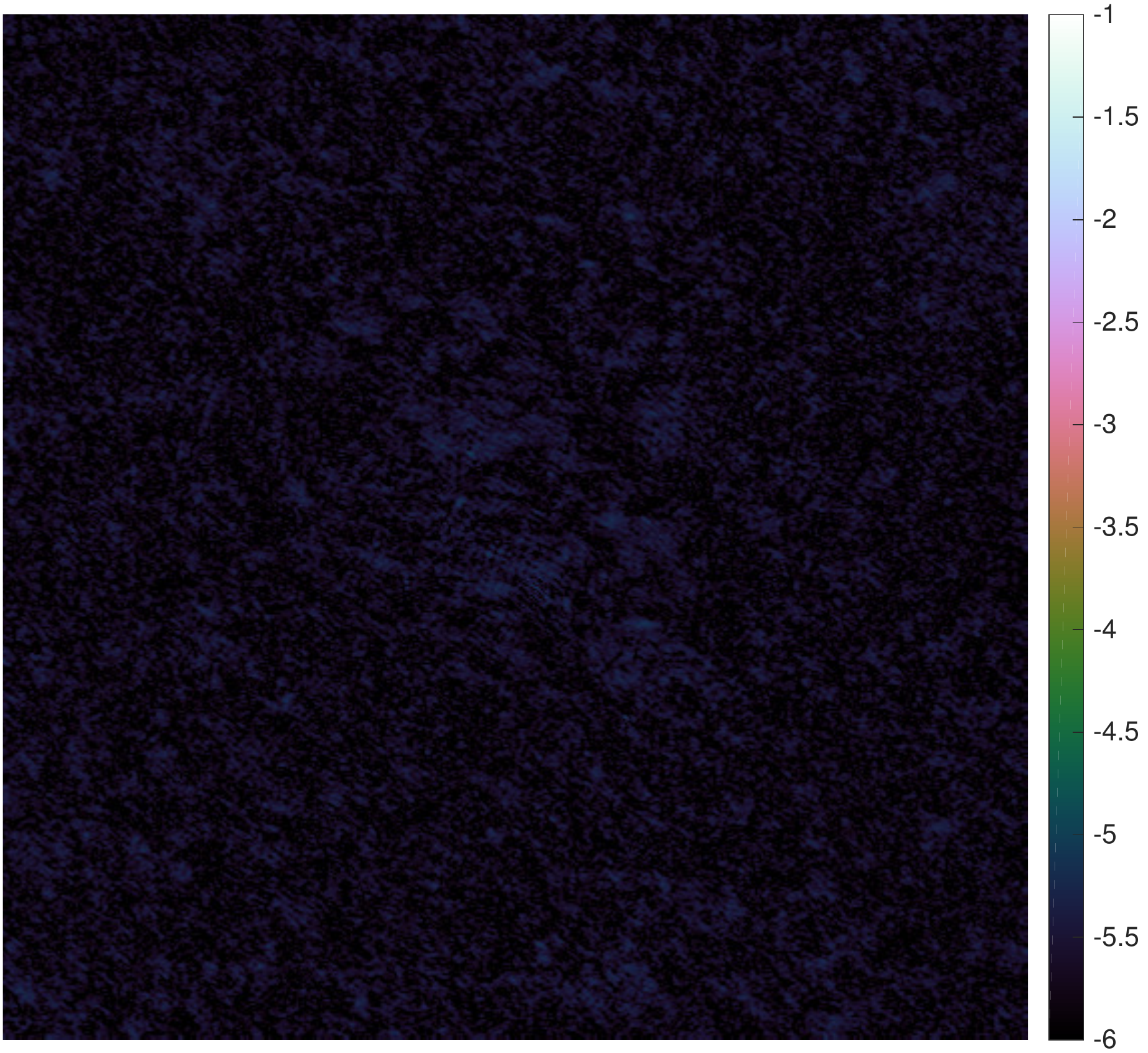}

	\vspace{5pt}
		
	\includegraphics[trim={0px 0px 0px 0px}, clip, height=0.29\linewidth]{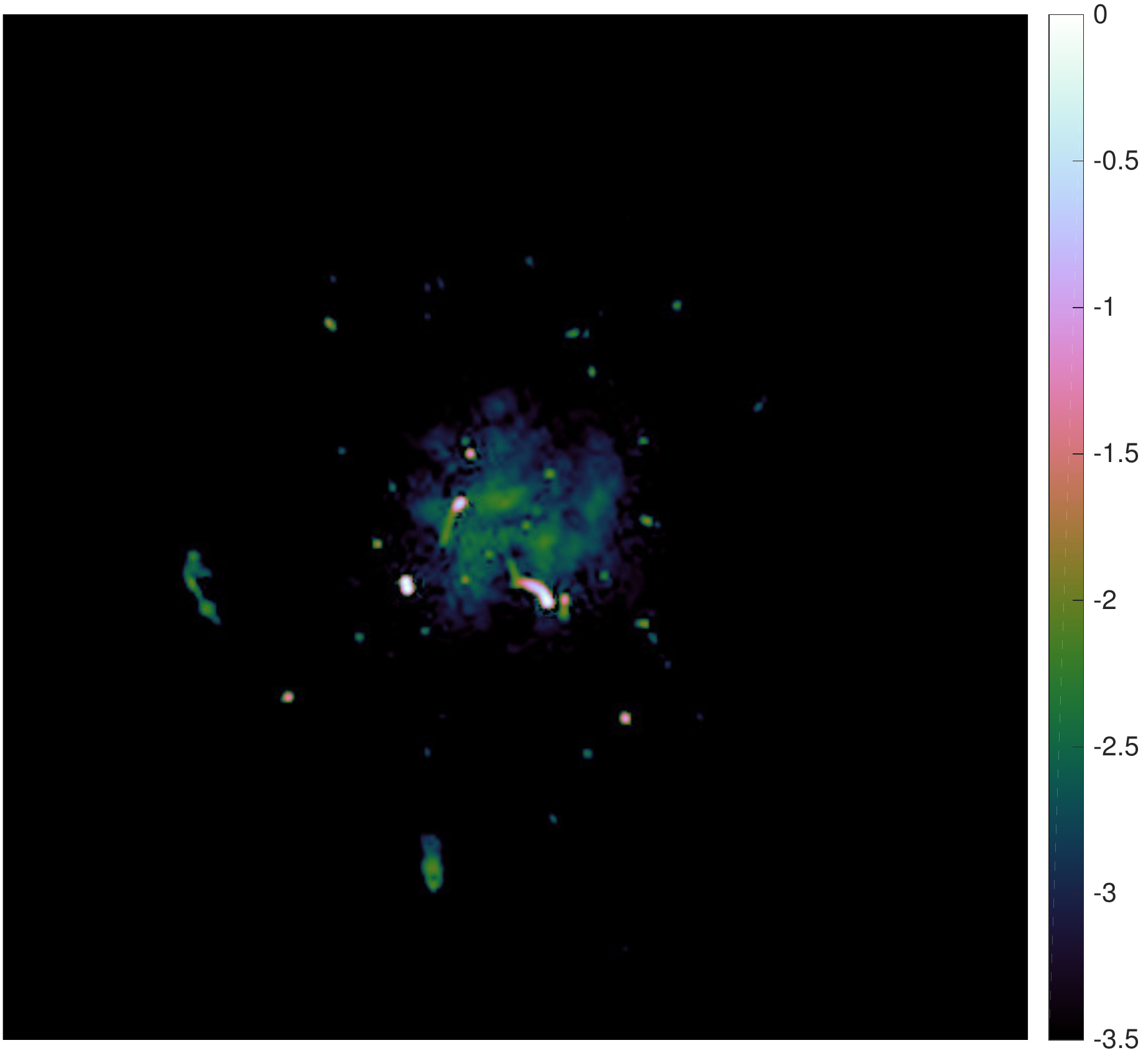}\hspace{2pt}
	\includegraphics[trim={0px 0px 0px 0px}, clip, height=0.29\linewidth]{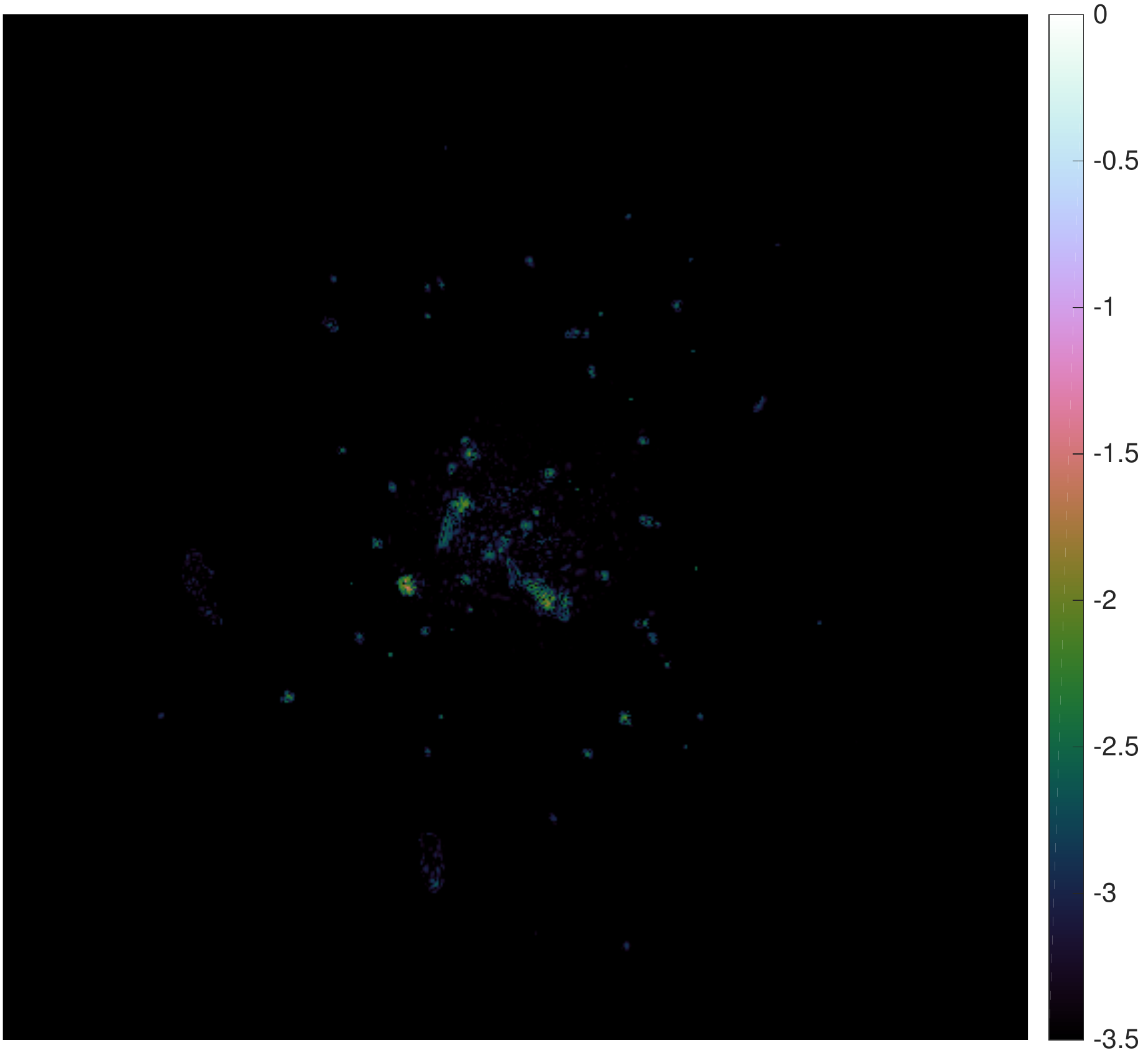}\hspace{2pt}
	\includegraphics[trim={0px 0px 0px 0px}, clip, height=0.29\linewidth]{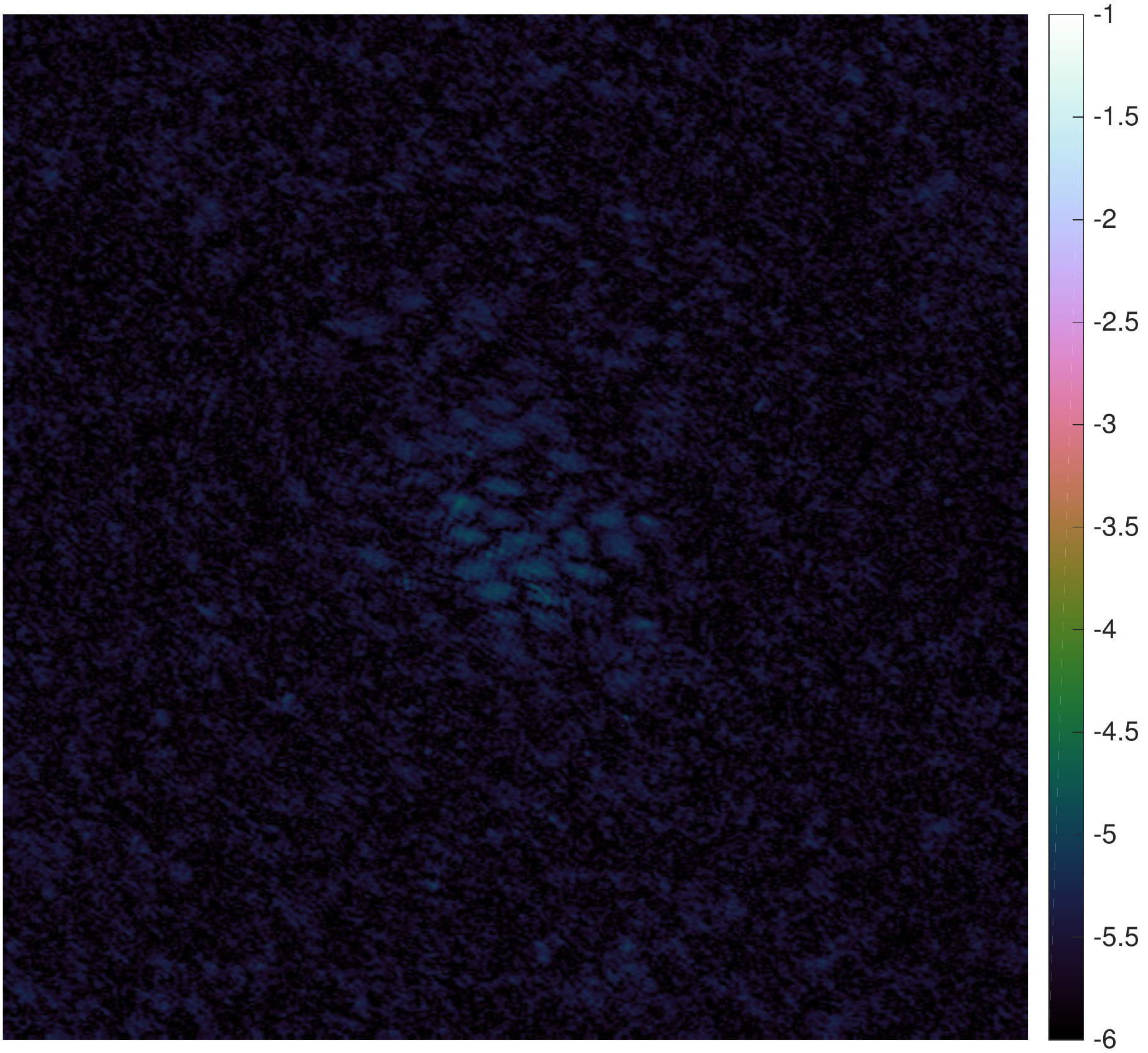}

	\vspace{5pt}
		
	\includegraphics[trim={0px 0px 0px 0px}, clip, height=0.29\linewidth]{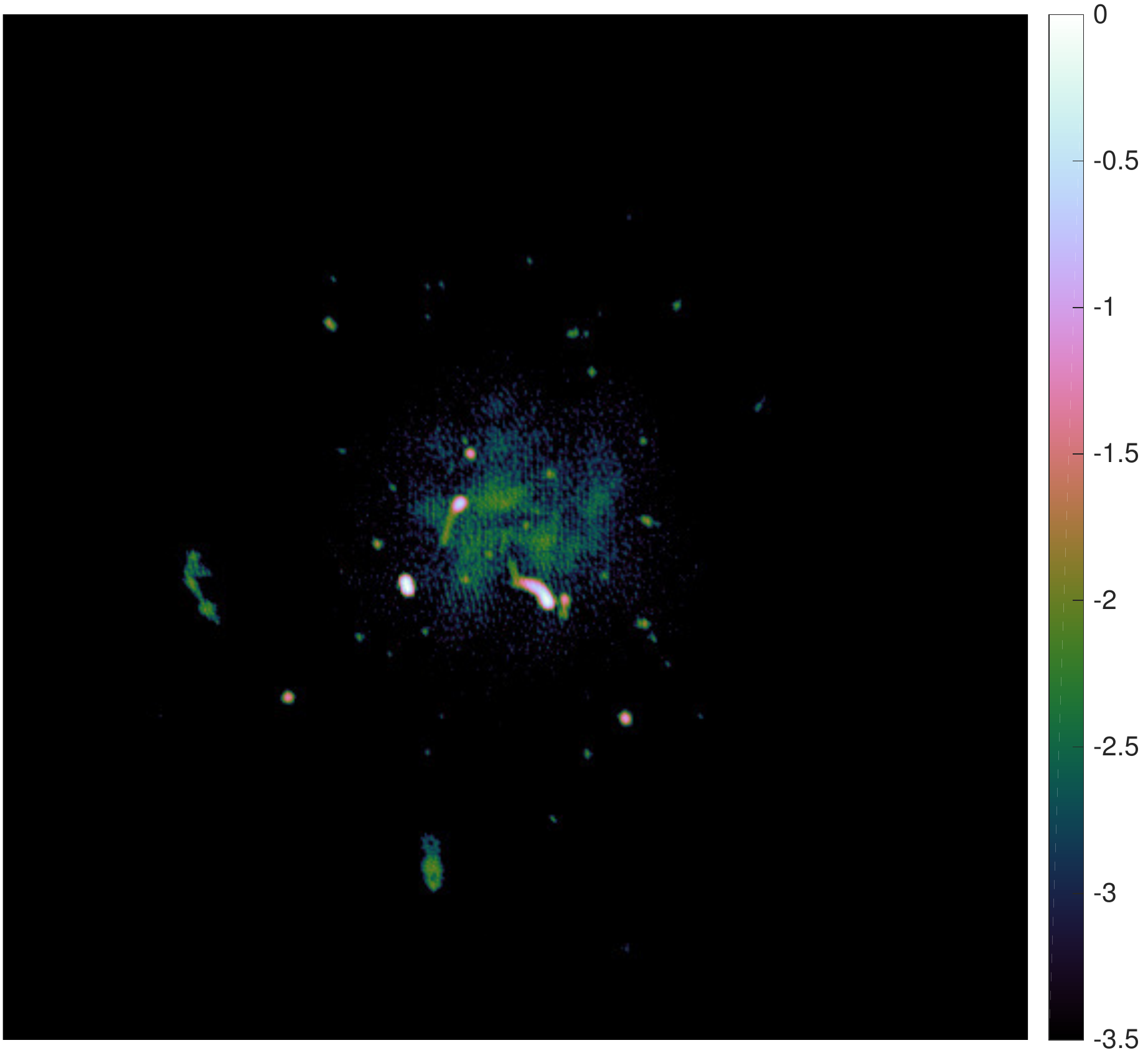}\hspace{2pt}
	\includegraphics[trim={0px 0px 0px 0px}, clip, height=0.29\linewidth]{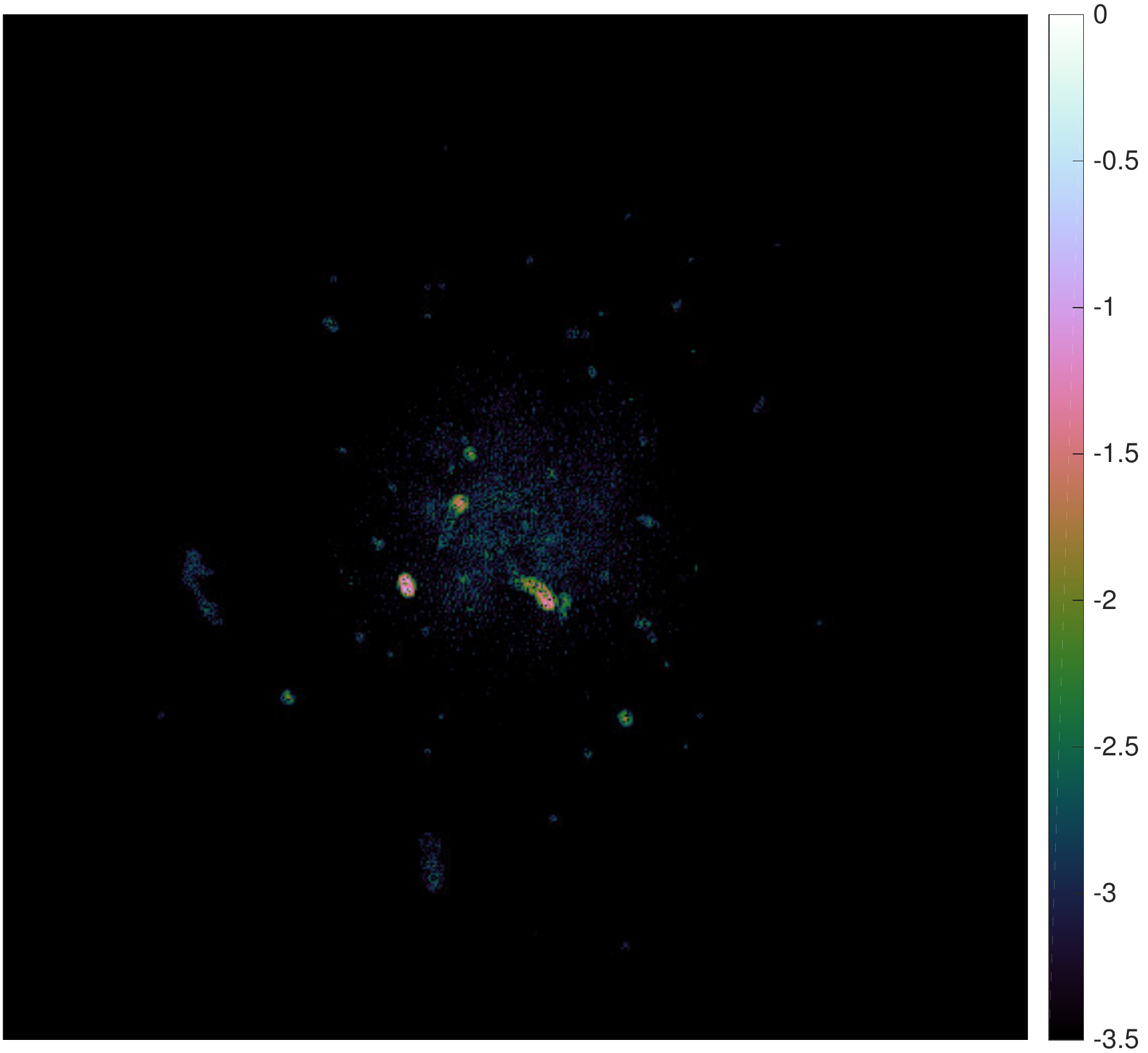}\hspace{2pt}
	\includegraphics[trim={0px 0px 0px 0px}, clip, height=0.29\linewidth]{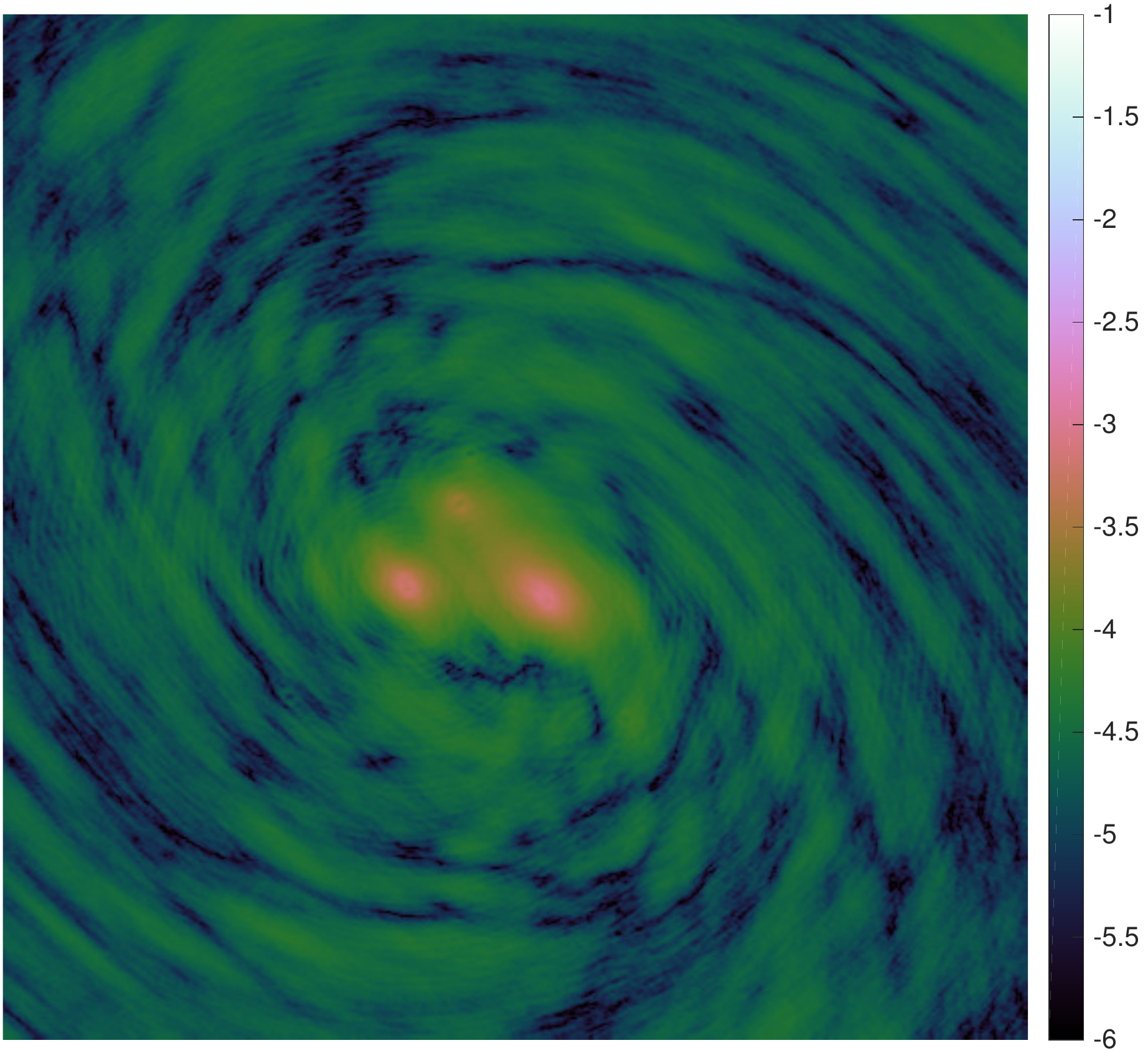}

	\vspace{5pt}
		
	\includegraphics[trim={0px 0px 0px 0px}, clip, height=0.29\linewidth]{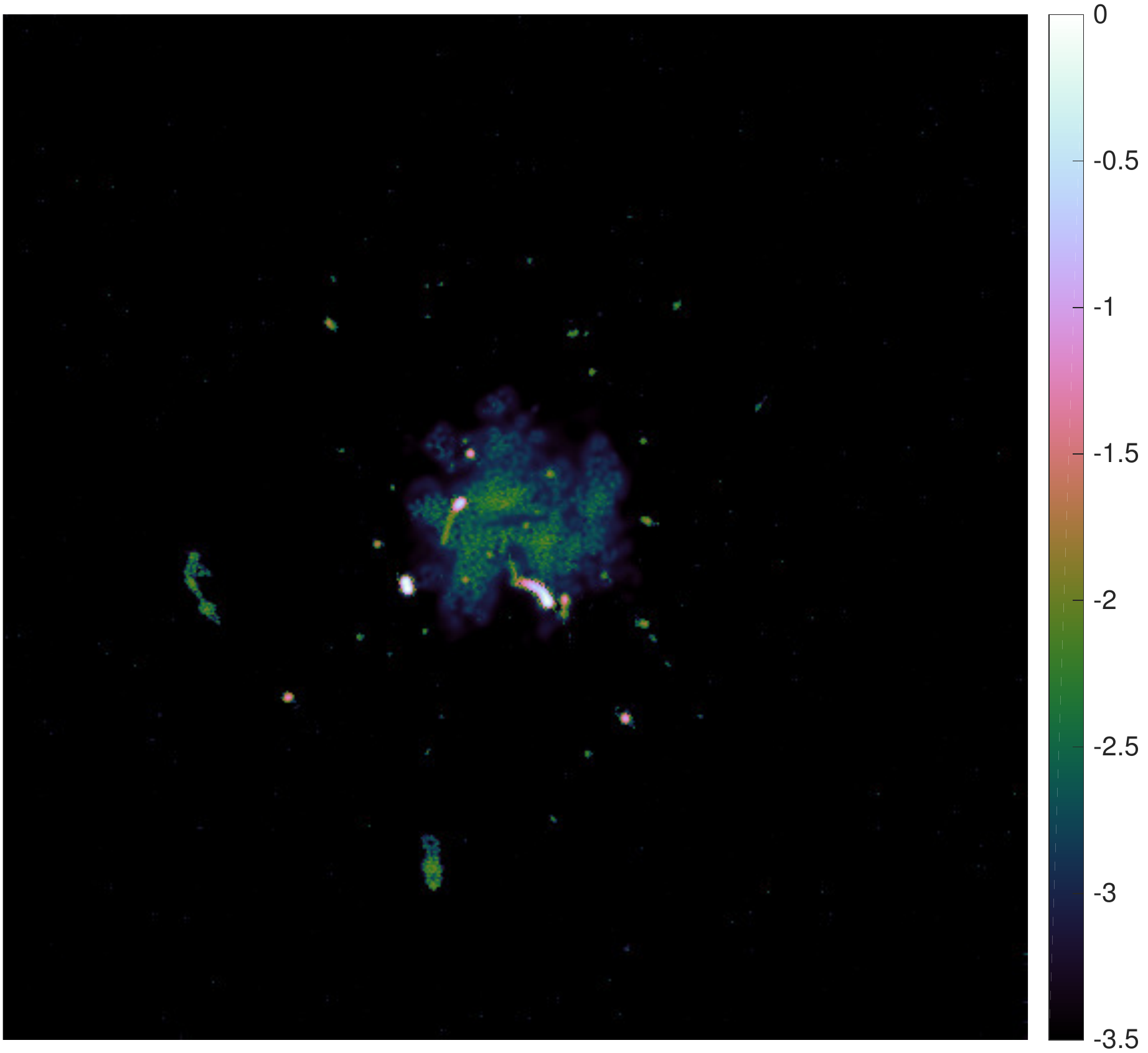}\hspace{2pt}
	\includegraphics[trim={0px 0px 0px 0px}, clip, height=0.29\linewidth]{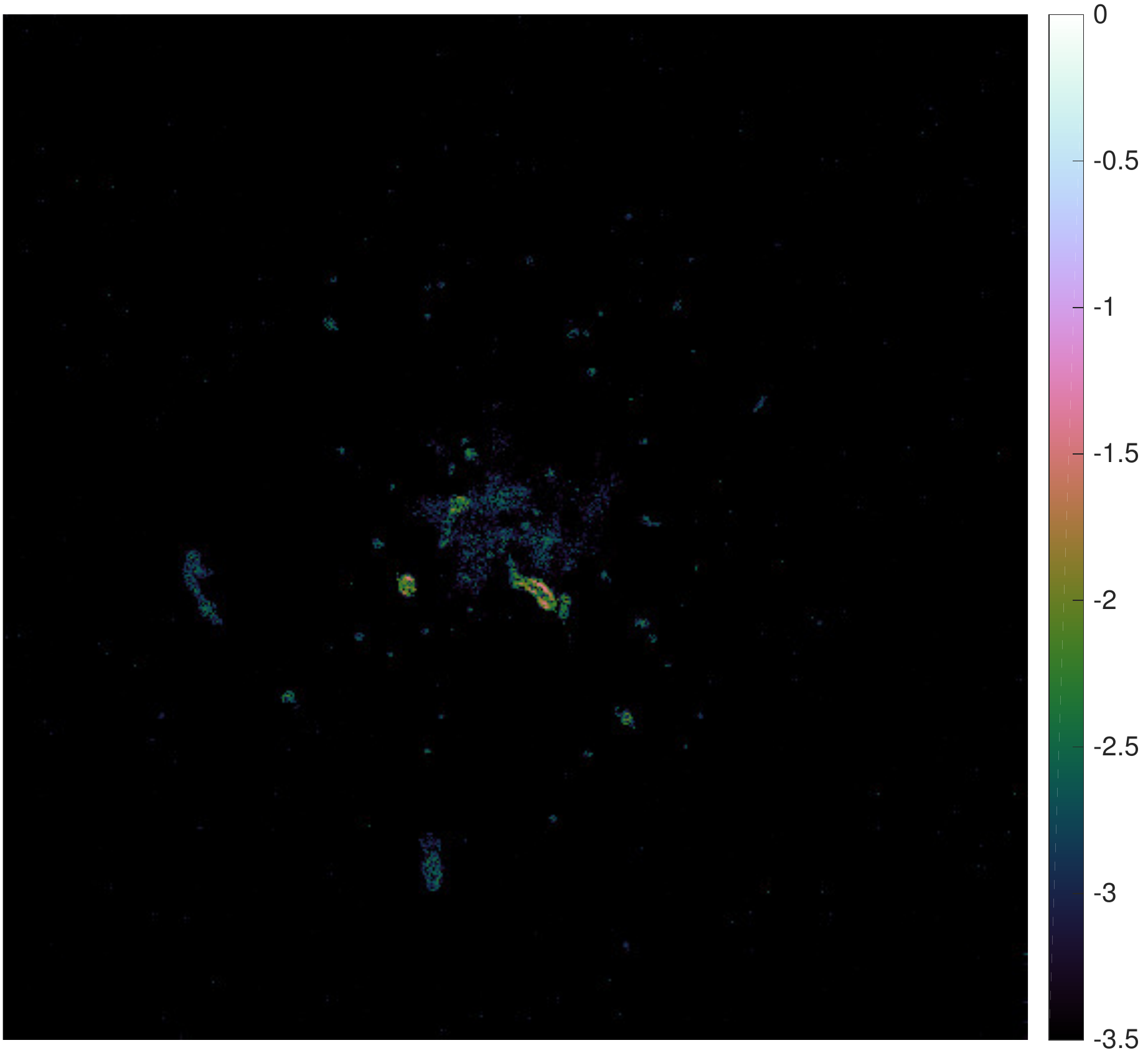}\hspace{2pt}
	\includegraphics[trim={0px 0px 0px 0px}, clip, height=0.29\linewidth]{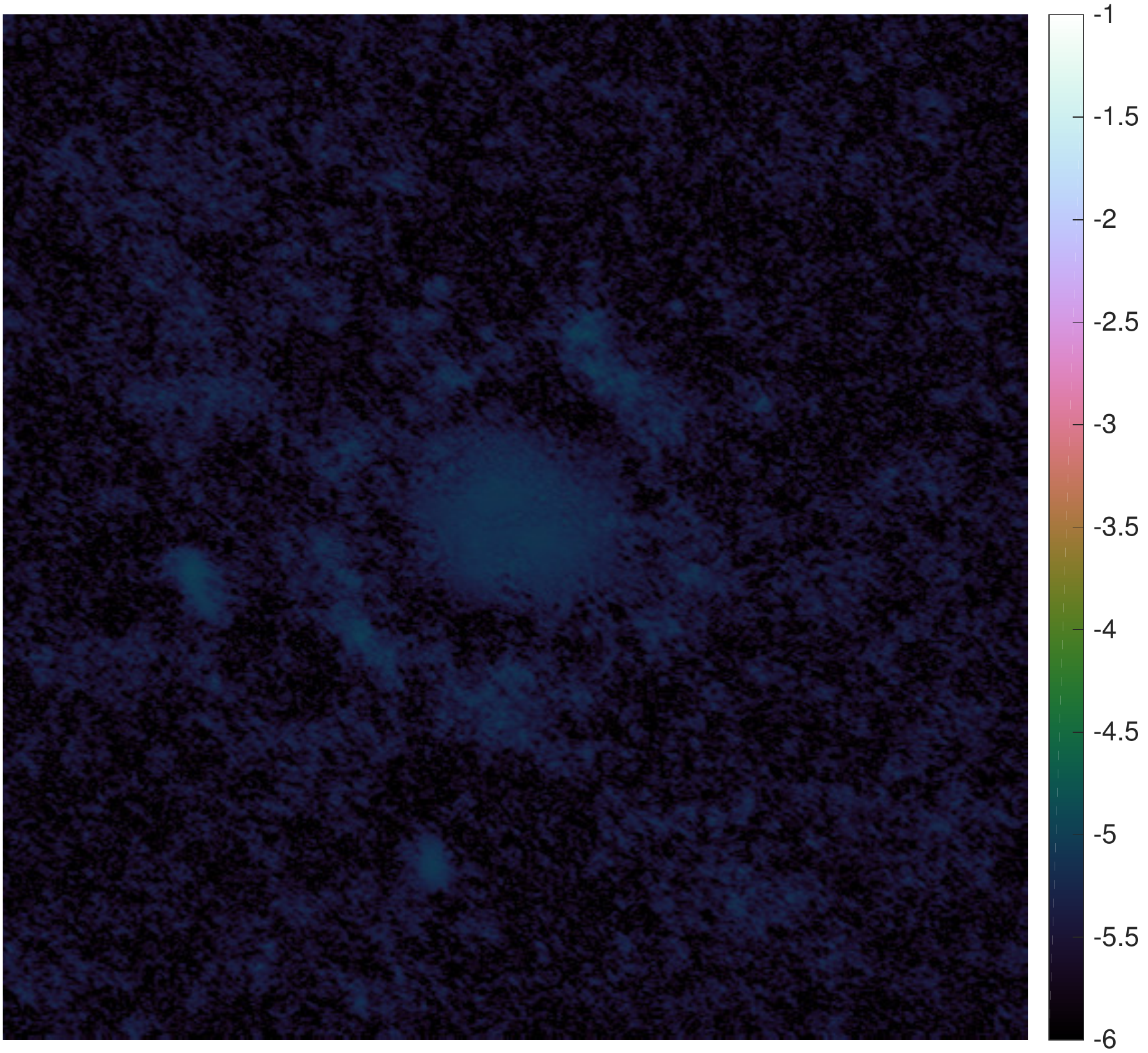}
	
	\caption{\bc (left to right) The reconstructed images, absolute value of the estimation errors, and absolute value of the naturally weighted  residual images, all in log scale, for the $512 \times 512$ galaxy cluster test image using the VLA coverage. The algorithms are: (from top to bottom) \ac{pd} having the reconstruction $\rm{SNR}= 30.98~\rm{dB}$ and the corresponding $\rm{DR}=475300$; \ac{admm} having the reconstruction $\rm{SNR}= 31.08~\rm{dB}$ and $\rm{DR}=432070$; \ac{cs-clean-n} with $l_g=0.001$ and $b=0.32$ having the reconstruction $\rm{SNR}=18.03~\rm{dB}$ and $\rm{DR}=21884$; \ac{moresane-n} having the reconstruction $\rm{SNR}= 24.96~\rm{dB}$ and $\rm{DR}=351850$.  The images correspond to the best results obtained by all algorithms as presented in Figure~\ref{vla-ska-results}. \ec}
	\label{fig-images-gc}
\end{figure*}

\begin{figure*}
	\centering
	\includegraphics[trim={0px 0px 0px 0px}, clip, height=0.29\linewidth]{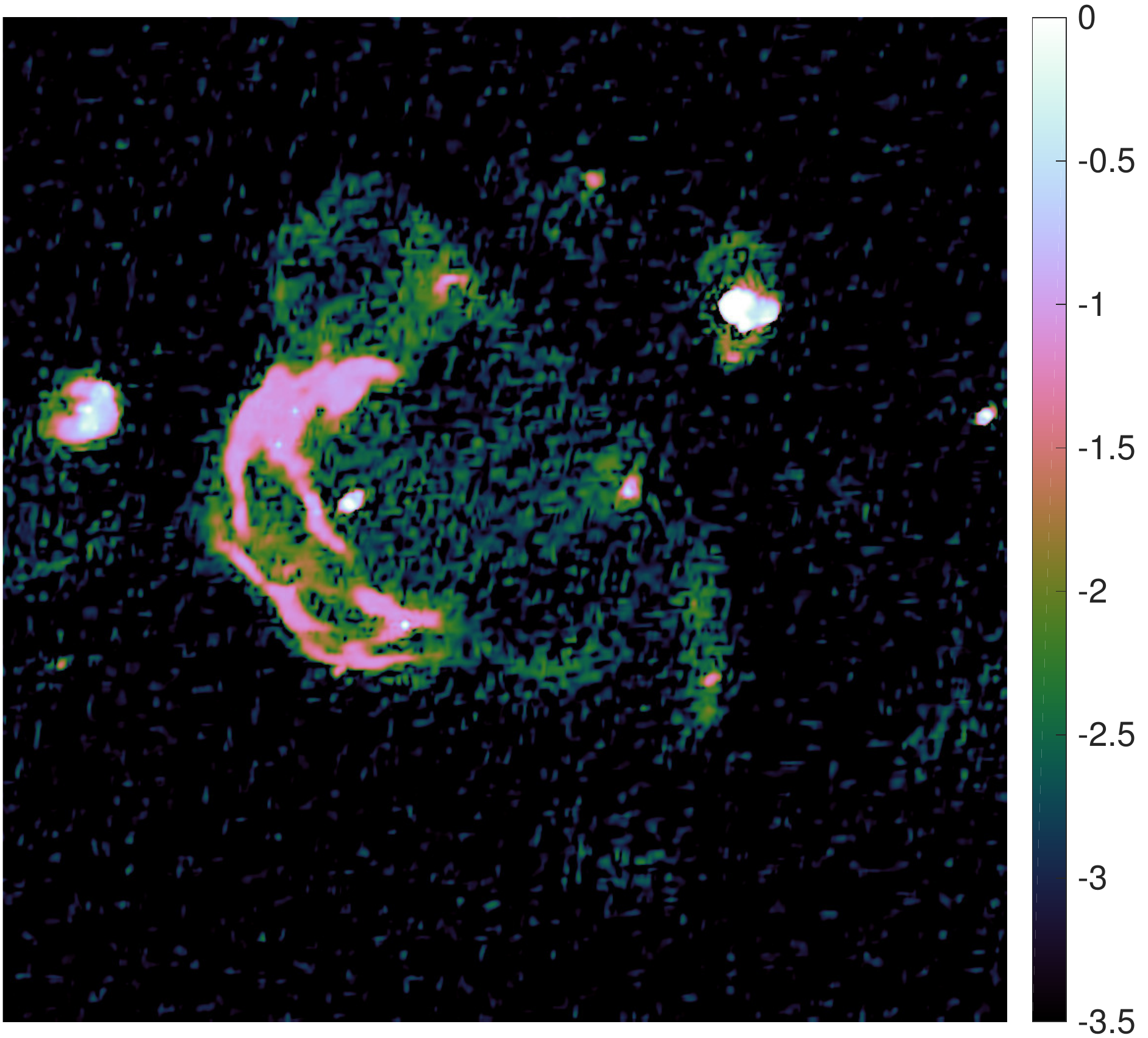}\hspace{2pt}
	\includegraphics[trim={0px 0px 0px 0px}, clip, height=0.29\linewidth]{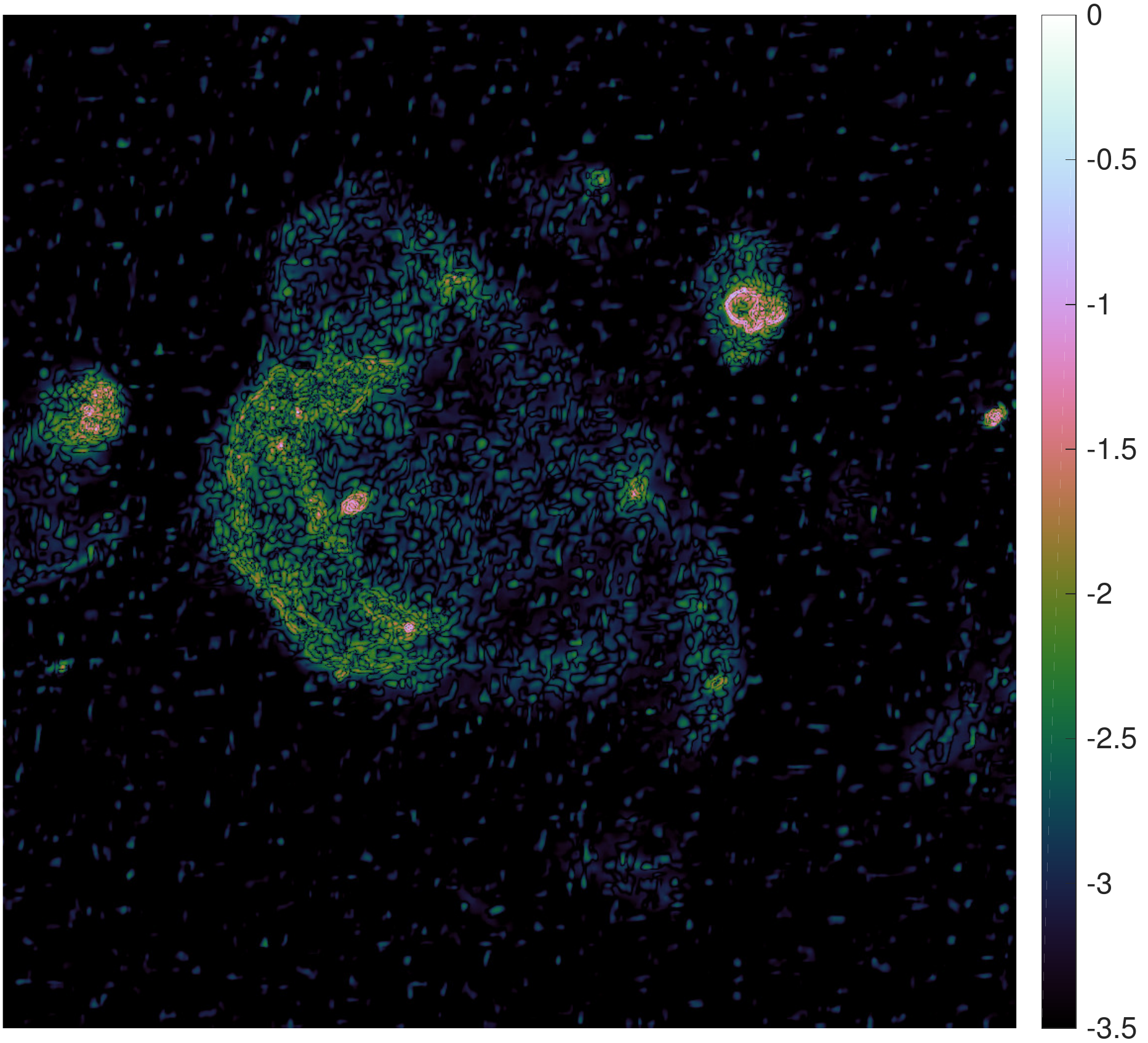}\hspace{2pt}
	\includegraphics[trim={0px 0px 0px 0px}, clip, height=0.29\linewidth]{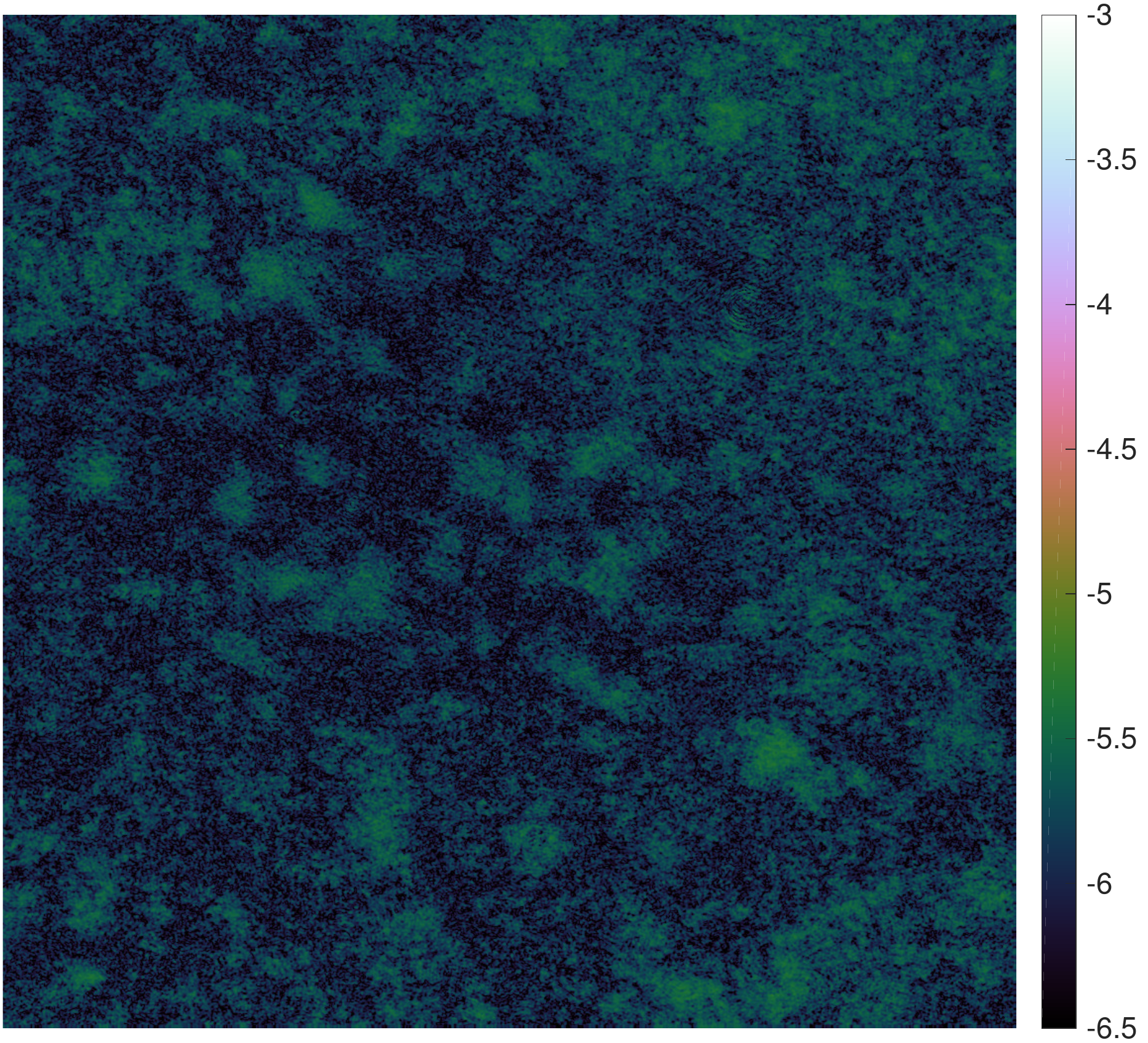}

	\vspace{5pt}
		
	\includegraphics[trim={0px 0px 0px 0px}, clip, height=0.29\linewidth]{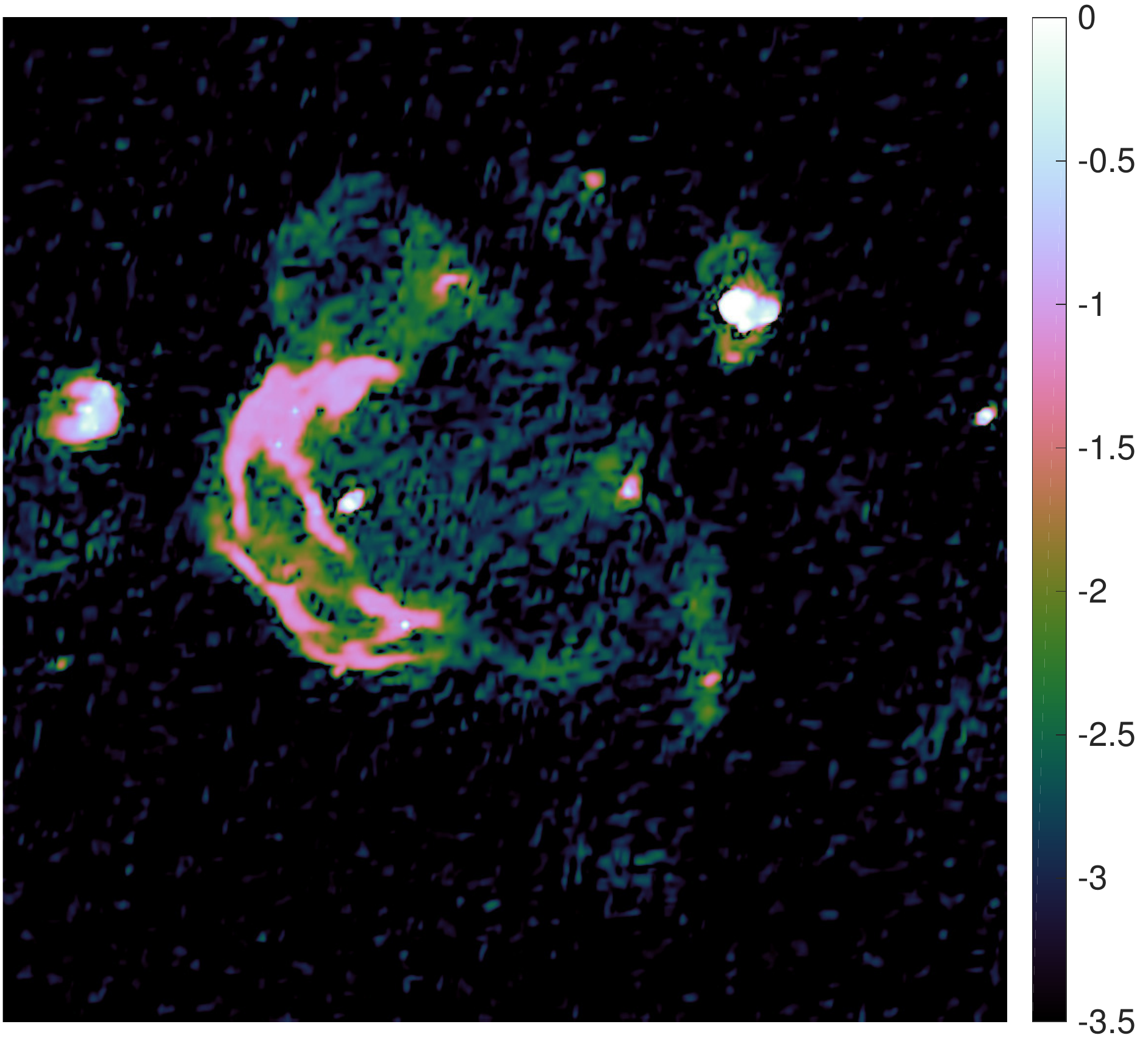}\hspace{2pt}
	\includegraphics[trim={0px 0px 0px 0px}, clip, height=0.29\linewidth]{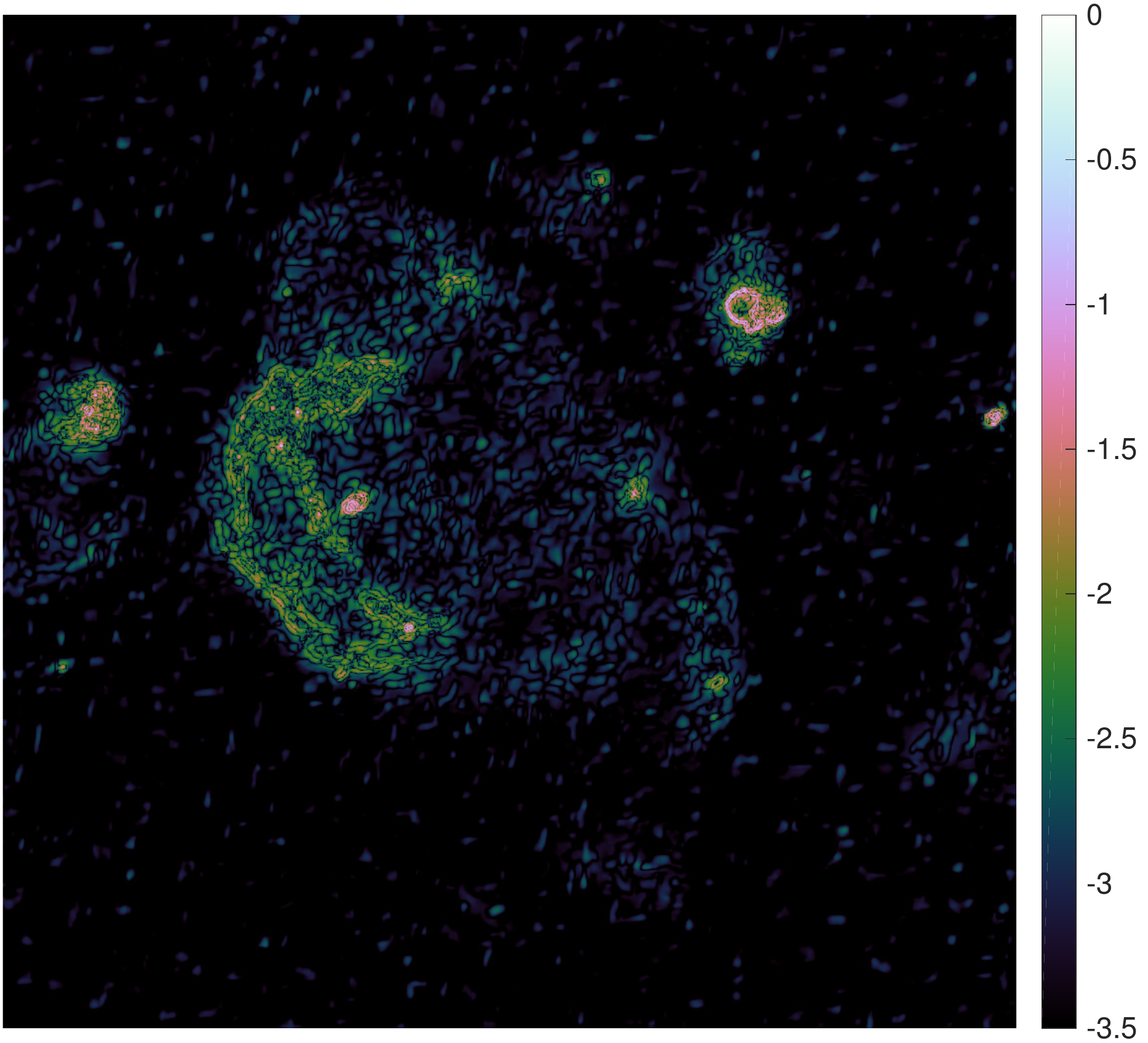}\hspace{2pt}
	\includegraphics[trim={0px 0px 0px 0px}, clip, height=0.29\linewidth]{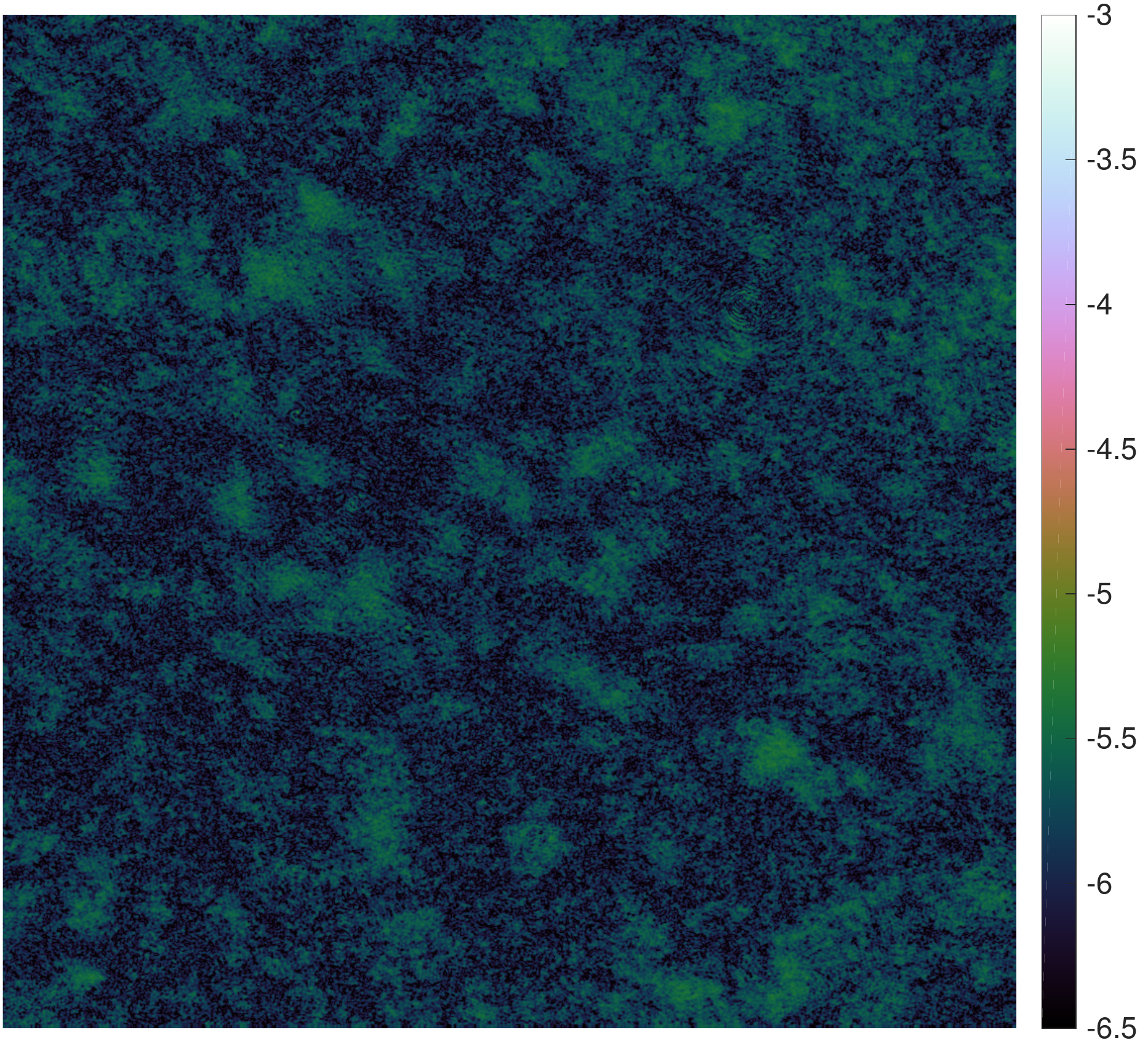}
	
	\vspace{5pt}
	
	\includegraphics[trim={0px 0px 0px 0px}, clip, height=0.29\linewidth]{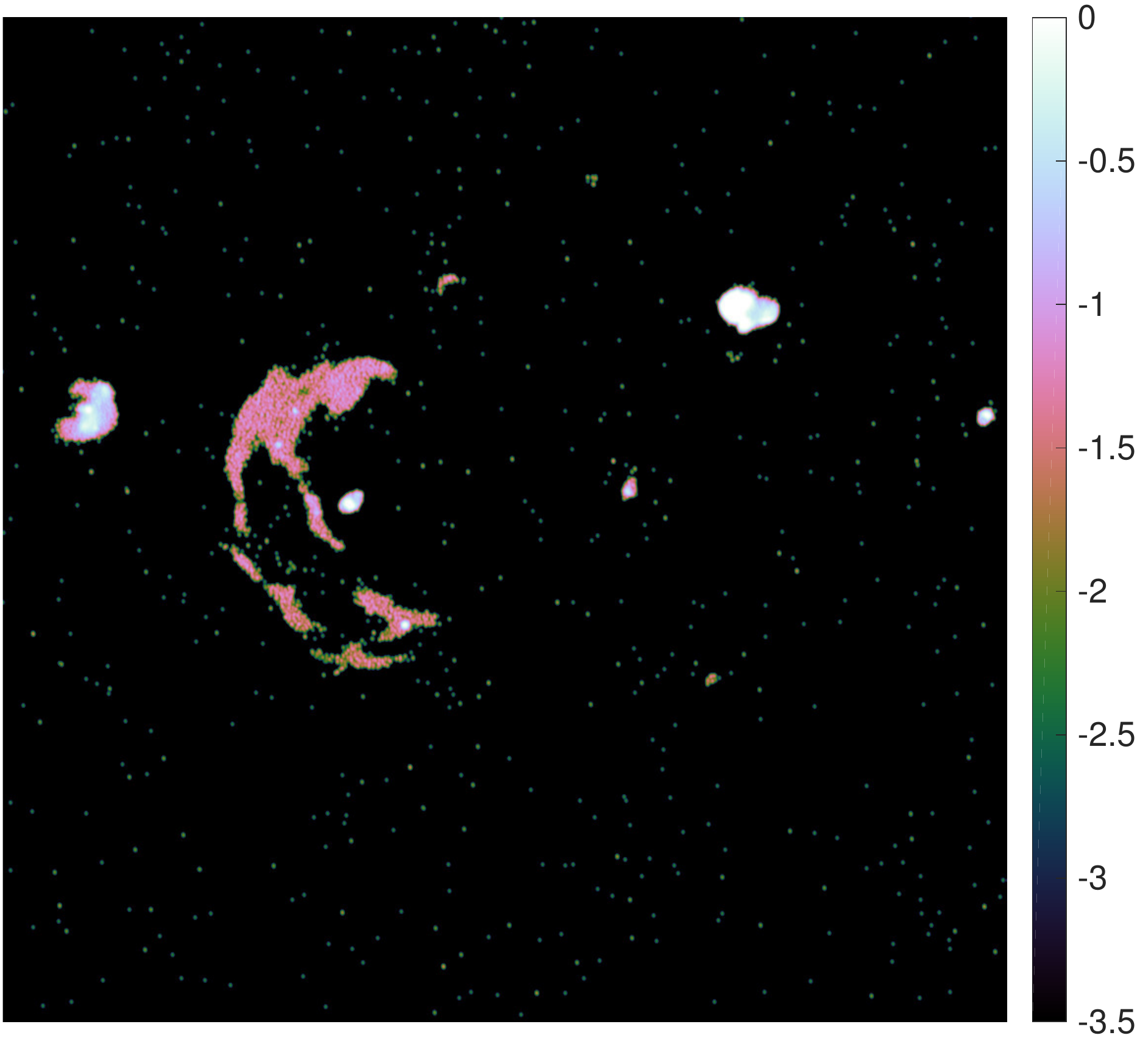}\hspace{2pt}
	\includegraphics[trim={0px 0px 0px 0px}, clip, height=0.29\linewidth]{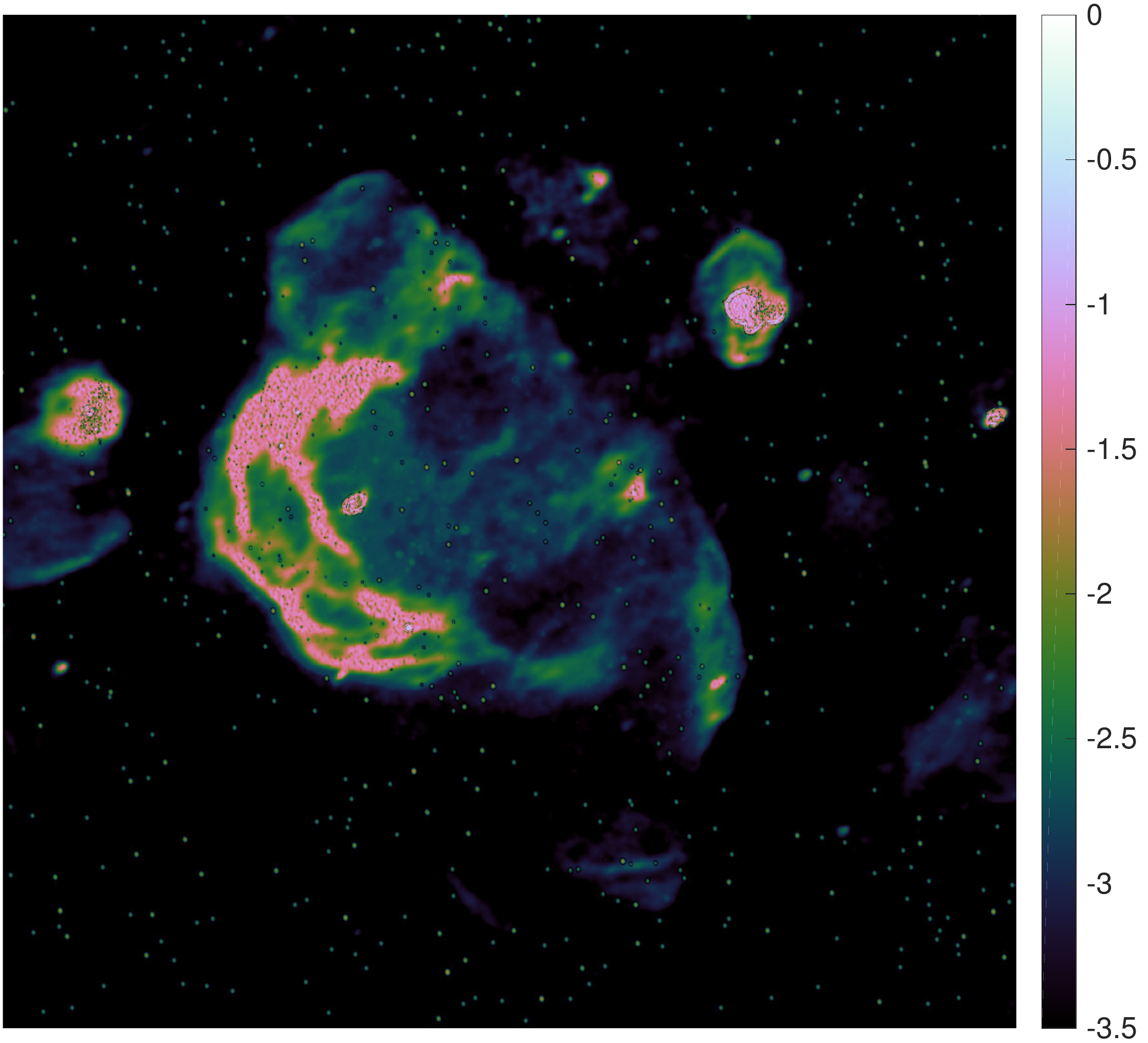}\hspace{2pt}
	\includegraphics[trim={0px 0px 0px 0px}, clip, height=0.29\linewidth]{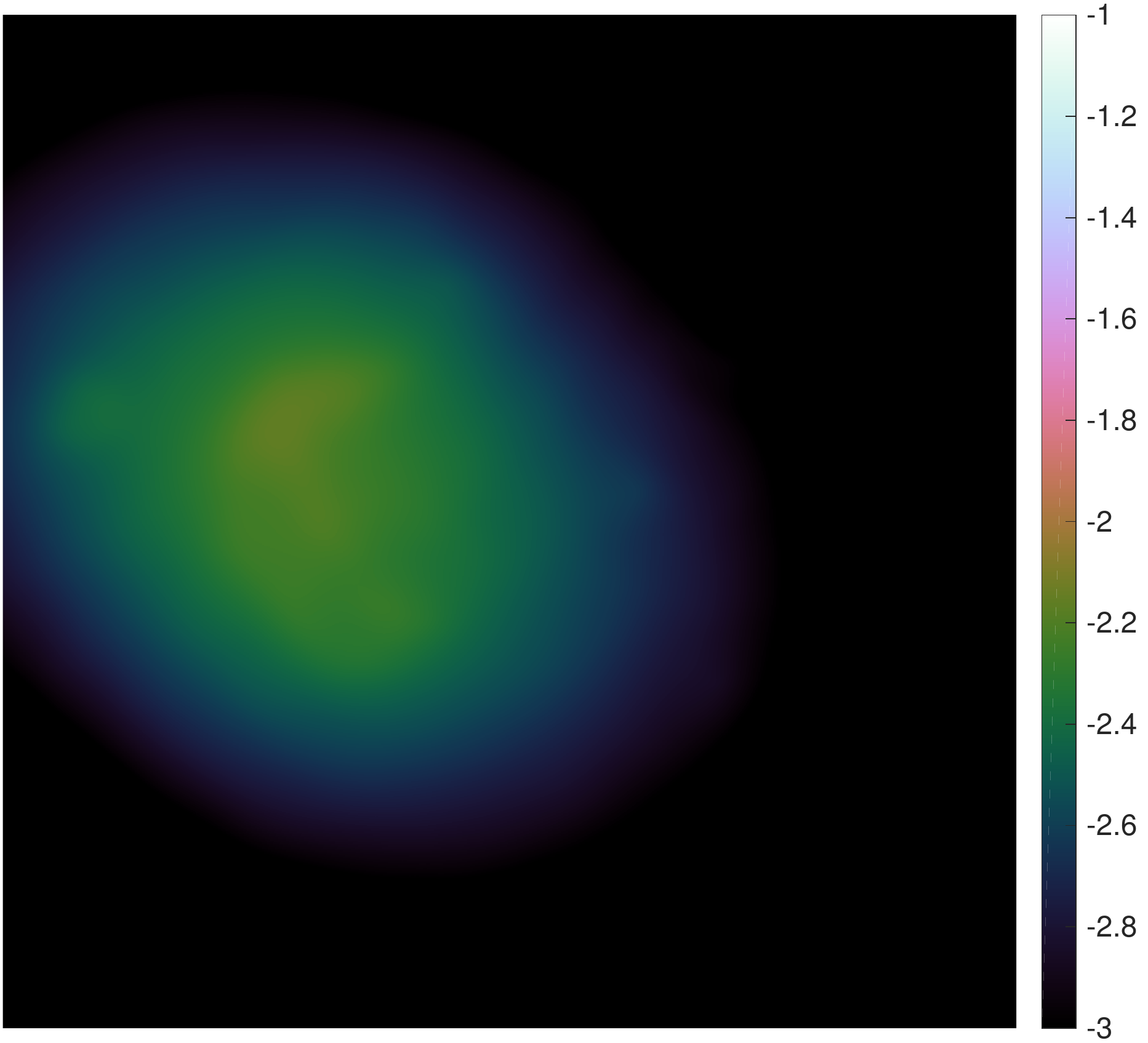}
	
	\vspace{5pt}
	
	\includegraphics[trim={0px 0px 0px 0px}, clip, height=0.29\linewidth]{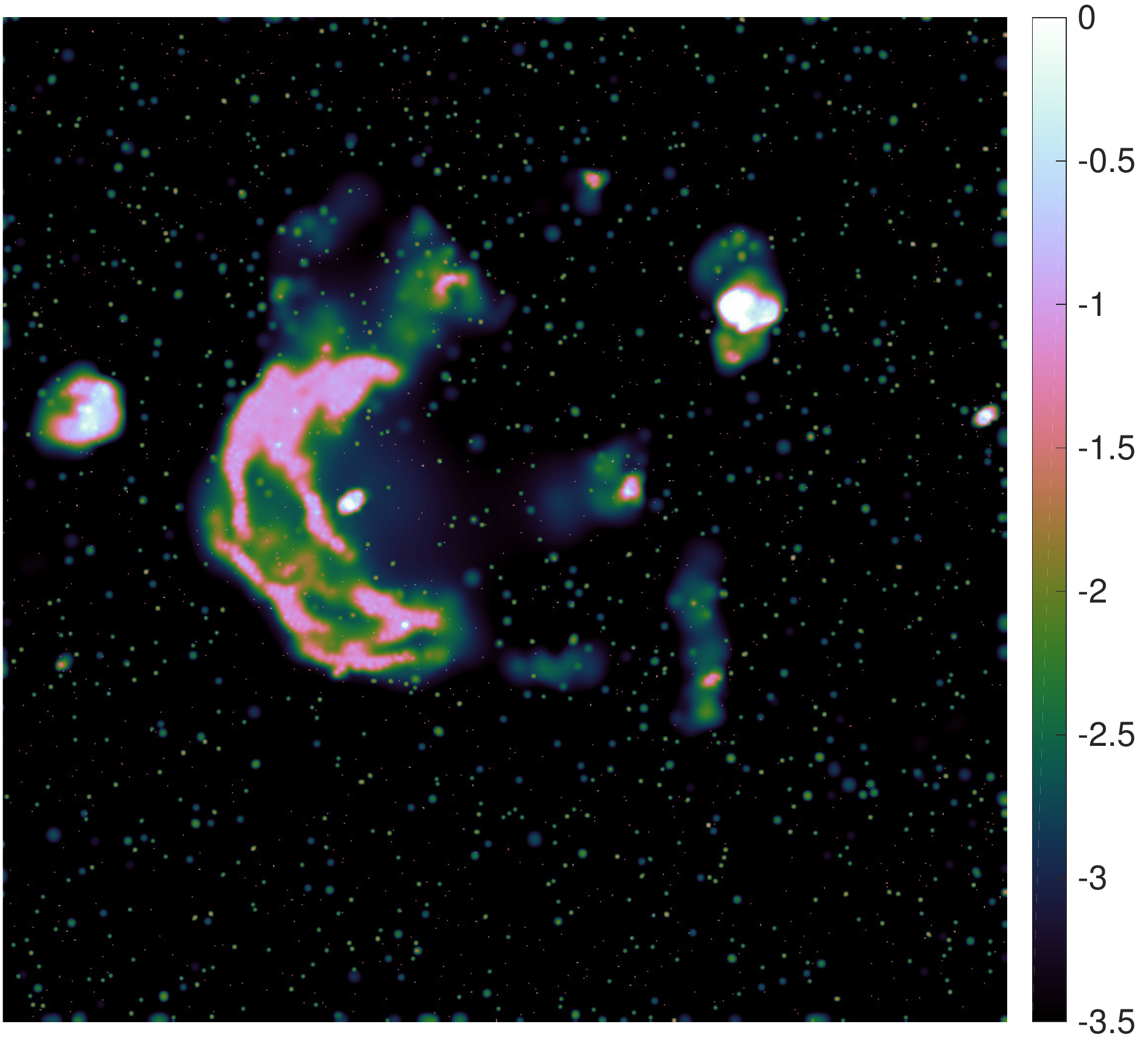}\hspace{2pt}
	\includegraphics[trim={0px 0px 0px 0px}, clip, height=0.29\linewidth]{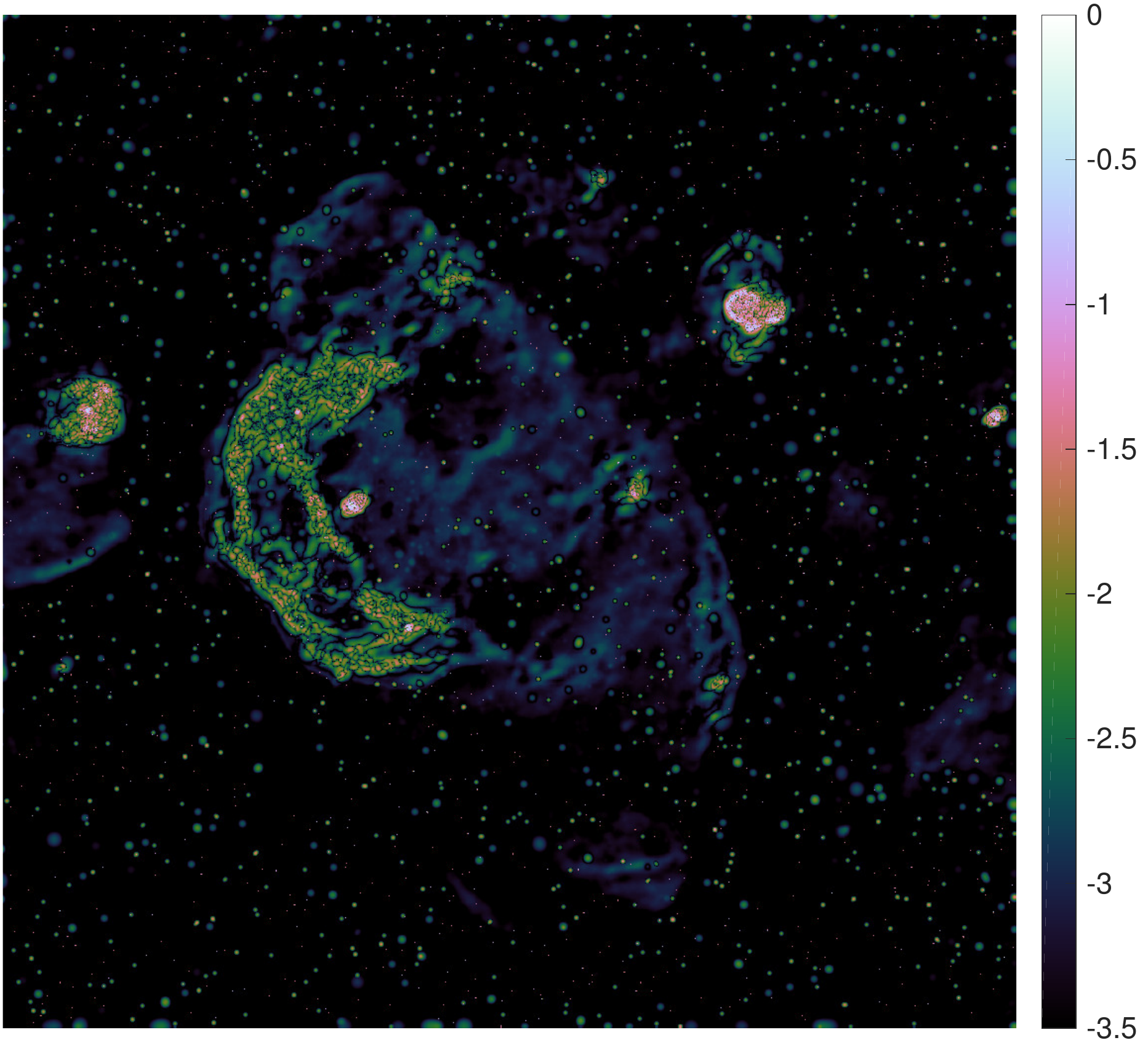}\hspace{2pt}
	\includegraphics[trim={0px 0px 0px 0px}, clip, height=0.29\linewidth]{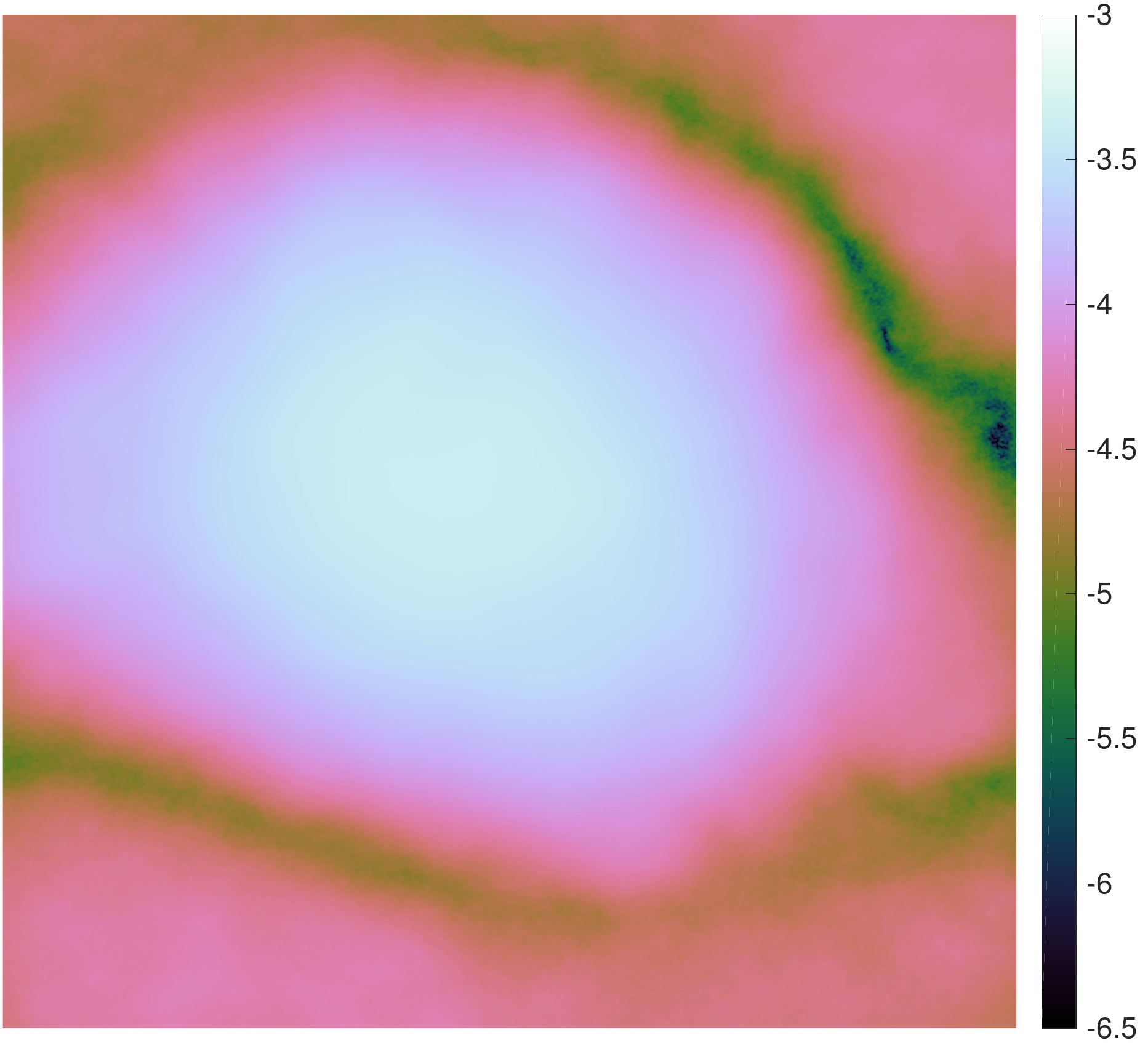}

	\caption{\bc (left to right) The reconstructed images, absolute value of the estimation errors, and absolute value of the naturally weighted  residual images, all in log scale, for the $1024 \times 1024$ W28 test image using the SKA coverage. The algorithms are: (from top to bottom) \ac{pd} having the reconstruction $\rm{SNR}= 21.86~\rm{dB}$ and the corresponding $\rm{DR}=737720$; \ac{admm} having the reconstruction $\rm{SNR}=  21.99~\rm{dB}$ and $\rm{DR}=735620$; \ac{cs-clean-u} with $l_g=0.1$ and $b=1.1$ having the reconstruction $\rm{SNR}=14.14~\rm{dB}$ and $\rm{DR}=515$; \ac{moresane-b} having the reconstruction $\rm{SNR}= 15.89~\rm{dB}$ and $\rm{DR}=10990$.  The images correspond to the best results obtained by all algorithms as presented in Figure~\ref{vla-ska-results}. Note that the scale for the residual image of \ac{cs-clean-u} is in the same range as the dirty image presented in Figure~\ref{ska-vla-dirty} while for \ac{pd}, \ac{admm} and \ac{moresane} the scale of the residual image is below that. \ec}
	\label{fig-images-w28}
\end{figure*}

The dirty images produced using natural weighting for the same tests are presented in Figure~\ref{ska-vla-dirty}.
For all three test cases, we showcase the reconstructed images, the reconstruction error images and the dirty residual images in Figures \ref{fig-images-ca}, \ref{fig-images-gc}, and \ref{fig-images-w28}.
We present the naturally weighted residual images for all methods even when they perform the deconvolution using a different weighting.
Since any other type of weighting essentially biases the data and decreases the sensitivity of the reconstruction, this is the more natural choice of visualising the remaining information in the residual image.
Although both \ac{cs-clean} and \ac{moresane} generally achieve better reconstruction for other weighting types, we present the naturally weighted dirty residual since it represents an unbiased estimation of the remaining structures.

For the reconstruction of the Cygnus A and galaxy cluster images, the methods developed herein outperform \ac{moresane}, using the best performing type of weighting, by  approximately $5~\rm{dB}$. Comparing against \ac{cs-clean} with the best weighting and beam size $b$, the $\rm{SNR}$ is around $10~\rm{dB}$ in favour of the reconstruction performed by the \ac{pd} and \ac{admm} methods.
Visually, both \ac{cs-clean} and \ac{moresane} fail to recover properly the jet present in the Cygnus A image while for \ac{pd} and \ac{admm} it is clearly visible.
It should be noted that the residual images show also very little structure for \ac{pd} and \ac{admm} while \ac{cs-clean} and \ac{moresane} still allow for a more structured residual image.
This is partially due to the biasing of the data when the uniform and Briggs weighting is performed.
\ac{pd} and \ac{admm} also achieve a better reconstruction of the galaxy cluster image. They are able to better estimate the three bright sources in the centre of the image. They are however slower to converge if compared to the recovery of the Cygnus A image.
\ac{moresane-n} also performs well for this test image and is able to produce a relatively smoother residual image in comparison to the Cygnus A case.
Note also that the performance of both \ac{cs-clean} and \ac{moresane} is inconsistent and varies greatly with the weighting type.

The last test is performed for the reconstruction of the W28 supernova remnant image using the \ac{ska} coverage.
In this case, the coverage is dominated by the low frequency points and lowers the convergence speed of both \ac{pd} and \ac{admm} algorithms.
Both \ac{pd} and \ac{admm} achieve good $\rm{SNR}$, again around $5~\rm{dB}$ over that reached by \ac{moresane}.
\ac{cs-clean} is $2~\rm{dB}$ worse than \ac{moresane} and is only able to recover the brightest sources as can be seen in Figure \ref{fig-images-w28}.
Again, both of our methods are able to recover more of the faint regions surrounding the bright sources.
The dirty residual images show less structure for the methods developed herein since they work directly with the naturally weighted visibilities.
Note that in Figure \ref{fig-images-w28}, in order to achieve a better visualisation, the scale of the dirty residual images for \ac{cs-clean} is different than that of the other methods.
Also, the performance of both \ac{cs-clean} and \ac{moresane} is again very inconsistent and varies greatly with the weighting type.

Both \ac{pd} and \ac{admm} methods show decreased convergence speed for the recovery of the galaxy cluster and W28 supernova remnant images.
A future study should, possibly by using generalised proximity operators \citep{Pesquet2014}, address the acceleration of the convergence which is influenced by the relative distribution of the visibilities in frequency. 
Coverages dominated by low frequency points, like the \ac{ska} one, generally produce slower convergence speed.
Furthermore, if a faster convergence is achieved, a reweighing $\ell_1$ approach becomes more attractive and should increase the reconstruction quality significantly.

\ec 

\section{Conclusions}
\label{sec-conc}
We proposed two algorithmic frameworks based on \ac{admm} and \ac{pd} approaches for solving the \ac{ri} imaging problem.
Both methods are highly parallelisable and allow for an efficient distributed implementation which is fundamental in the context of the \bc high dimensionality \ec problems associated with the future \ac{ska} radio telescope.
The structure of \ac{admm} is sub-iterative, which for much heavier priors than the ones used herein may become a bottleneck.
The \ac{pd} algorithm achieves greater flexibility, in terms of memory requirements and computational burden per iteration, by using full splitting and randomised updates.
\bc Through the analogy between the \alg{clean} major-minor loop and a \ac{FB} iteration, both methods can be understood as being composed of sophisticated \alg{clean}-like iterations running in parallel in multiple data, prior, and image spaces.\ec

The reconstruction quality for both \ac{admm} and \ac{pd} methods is similar to that of \ac{sdmm}.
The computational burden is much lower.
Experimental results with realistic coverages show impressive performance in terms of parallelisation and distribution, suggesting scalability to extremely large data sets.
We give insight into the performance as a function of the configuration parameters and provide a parameter setup, with the normalised soft-thresholding values \bc between $10^{-3}$ and $10^{-5}$\ec, that produce consistently stable results for a broad range of tests.
The solution to the optimisation problem solved herein was shown to greatly outperform the standard methods in \ac{ri} which further motivates the use of our methods.
Our tests also confirm the reconstruction quality in the high dynamic range regime.

\bc Our Matlab code is available online on GitHub, \url{http://basp-group.github.io/pd-and-admm-for-ri/}. \ec
In the near future, we intend to provide an efficient implementation, using the \alg{mpi} communication library, for a distributed computing infrastructure.
This will be included in the \alg{purify} C++ package, which currently only implements a sequential version of \ac{sdmm}.
\bc The acceleration of the algorithms for coverages dominated by low frequency points will also be investigated, by leveraging a generalised proximal operator\ec.
Additionally, recent results suggest that the conditions for convergence for the randomised \ac{pd} can be relaxed, which would accelerate the convergence speed making these methods to be even more competitive.
We also envisage to use the same type of framework to image in the presence of DDEs, such as the $w$ component, as well as to jointly solve the calibration and image reconstruction problems.

\section*{Acknowledgements}

This work was supported by the UK Engineering and Physical Sciences Research Council (EPSRC, grants EP/M011089/1 and EP/M008843/1) \bc and UK Science and Technology Facilities Council (STFC, grant ST/M00113X/1) \ec, as well as by the Swiss National Science Foundation (SNSF) under grant 200020-146594.
We would like to thank Federica Govoni and Matteo Murgia for providing the simulated galaxy cluster image.

\bibliographystyle{mnras.bst}

\appendix

\section{Parameter overview}
\label{sec:param-overview}
\bc
An overview of the parameters used to define the minimisation problems is presented in Table \ref{param-table-opt}.
The configuration parameters for the algorithms are presented in Table \ref{param-table-alg}.
\ec

\begin{table*}
	\bc
  	\caption{Overview of the parameters for defining the optimisation problem (\ref{basic-min-problem}).}
  	\label{param-table-opt}
  	\centering
	\small
	\begin{tabular}{cp{14cm}}
	\hline
	\multicolumn{2}{c}{Optimisation problem definition} \\ \hline
	$\bm{\Psi}_i$ & the $n_{\rm{b}}$ wavelet bases in which the signal is considered sparse; other priors can be incorporated as well by redefining the functions $l_i$ and their associated proximity operators\\[3px]
	$n_b$ & the number of data blocks generally linked to the computing infrastructure \\[3px]
	$\mc{B}_j$ & the $\ell_2$ balls imposing data fidelity; they are linked to the modality in which the data are split into blocks $y_j$ \\[3px]
	$\epsilon_j$ & the size of the $\ell_2$ balls defining the data fidelity; they are linked to the statistics of the noise; herein $\epsilon_j$ are set based the $\chi^2$ distribution associated with the noise \\ \hline
  	\end{tabular}%
	\ec
\end{table*}

\begin{table}
	\bc
  	\caption{The configuration parameters for the \ac{admm} (top) and \ac{pd} (bottom) algorithms.}
  	\label{param-table-alg}
  	\centering
	\small
  	\begin{tabular}{p{1.4cm}p{6cm}}
	\hline
	\multicolumn{2}{c}{Algorithm~\ref{alg-admm}~\acp{admm}} \\ \hline
	$\kappa > 0$ & configurable; influences the convergence speed \\[6px]
	$\bar{\delta} \leq 10^{-3}$ & \multirow{2}{*}{\parbox{6cm}{ configurable; stopping criteria; linked to the accuracy of the desired solution}} \\
	$\bar{\epsilon}_j$ &  \\[6px]
	$\bar{\delta}_{\bar{f}}  \leq 10^{-3}$ & \multirow{2}{*}{\parbox{6cm}{ configurable; sub-iteration stopping criteria; linked to the accuracy of the desired solution}} \\
	$\mathrm{n}_{\bar{f}}$ &  \\[12px]
	$\varrho = 0.9$ & \multirow{2}{*}{\parbox{6cm}{fixed; algorithm convergence parameters; need to satisfy (\ref{convergence-req-admm})}} \\
	$\rho = \frac{1}{\|\bm{\Phi}\|_{\rm{S}}^2}$ & \\[6px]
	$\eta = \frac{1}{\|\bm{\Psi}\|_{\rm{S}}^2}$ & fixed; algorithm convergence parameter \\ \hline 
  	\end{tabular}%
	\vspace{10px}
	\begin{tabular}{p{1.4cm}p{6cm}}
	\hline
	\multicolumn{2}{c}{Algorithm~\ref{alg-primal-dual}~\acp{pd}} \\ \hline
	$\kappa > 0$ & configurable; influences the convergence speed \\[6px]
	$\bar{\delta} \leq 10^{-3}$ & \multirow{2}{*}{\parbox{6cm}{ configurable; stopping criteria; linked to the accuracy of the desired solution}} \\
	$\bar{\epsilon}_j$ &  \\[6px]
	$p_{\mc{P}_i} > 0$ & \multirow{2}{*}{\parbox{6cm}{configurable; randomisation probabilities; linked to the computing infrastructure}} \\
	$p_{\mc{D}_j} > 0$ & \\[6px] 
	$\tau = 0.49$ & \multirow{3}{*}{\parbox{6cm}{fixed; algorithm convergence parameters; need to satisfy (\ref{convergence-req-pd})}} \\
	$\varsigma = \frac{1}{\|\bm{\Phi}\|_{\rm{S}}^2}$ &  \\ 
	$\sigma = \frac{1}{\|\bm{\Psi}\|_{\rm{S}}^2}$ & \\ 
	\hline 
  	\end{tabular}%
	\ec
\end{table}

\section{SDMM Algorithm}
\label{sdmm}

\bc
The structure of \ac{SDMM}, solving the specific \ac{RI} problem (\ref{split-min-problem}), is presented for completeness in Algorithm \ref{alg-sdmm}.
\ec
\begin{algorithm}[h]
\caption{SDMM.}
\label{alg-sdmm}

\begin{algorithmic}[1]
\small
\Given{$\bs{x}^{(0)}, \tilde{\bs{r}}_j^{(0)}, \bar{\bs{r}}_j^{(0)}, \check{\bs{r}}_j^{(0)}, \tilde{\bs{s}}_j^{(0)}, \bar{\bs{s}}_j^{(0)}, \hat{\bs{s}}^{(0)}_i, \kappa$}
\RepeatFor{$t=1,\ldots$}

\State $\ds \tilde{\bs{b}}^{(t)} = \bm{F}\bm{Z} \bs{x}^{(t-1)}$
\Set{$\forall j \in \{1, \ldots, n_{\rm{d}}\}$}
	\State $\ds \bs{b}_j^{(t)} = \bm{M}_j \tilde{\bs{b}}^{(t)}$
\EndSet

\Block{\bf run simultaneously}
	\ParForD{$\forall j \in \{1, \ldots, n_{\rm{d}}\}$}{$\bs{b}_j^{(t)}$}
		\State $\ds \tilde{\bs{r}}_j^{(t)} = \proj_{\mc{B}_j} \bigg( \bm{G}_j \bs{b}_j^{(t)} + \tilde{\bs{s}}_j^{(t-1)} \bigg)$
		\State $\ds \tilde{\bs{s}}_j^{(t)} = \tilde{\bs{s}}_j^{(t-1)} + \bm{G}_j \bs{b}_j^{(t)} - \tilde{\bs{r}}_j^{(t)}$
		\State $\ds \tilde{\bs{q}}_j^{(t)} = \bm{G}_j^\dagger \big( \tilde{\bs{r}}_j^{(t)} - \tilde{\bs{s}}_j^{(t)} \big)$
	\EndParForD{$\tilde{\bs{q}}_j^{(t)}$}
	\ParFor{$\forall i \in \{1, \ldots, n_{\rm{b}}\}$}
		\State $\ds \bar{\bs{r}}_i^{(t)} = \soft_{\kappa \|\bm{\Psi}\|_{\rm{S}}} \bigg( \bm{\Psi}_i^\dagger \bs{x}^{(t-1)} + \bar{\bs{s}}_i^{(t-1)} \bigg)$
		\State $\ds \bar{\bs{s}}_i^{(t)} = \bar{\bs{s}}_i^{(t-1)} + \bm{\Psi}_i^\dagger \bs{x}^{(t-1)} - \bar{\bs{r}}_i^{(t)}$
		\State $\ds \bar{\bs{q}}_i^{(t)} = \bm{\Psi}_i \big( \bar{\bs{r}}_i^{(t)} - \bar{\bs{s}}_i^{(t)} \big)$
	\EndParFor
	\Block{\bf do}
		\State $\ds \hat{\bs{r}}^{(t)} = \proj_{\mc{C}} \bigg( \bs{x}^{(t-1)} + \hat{\bs{s}}^{(t-1)} \bigg)$
		\State $\ds \hat{\bs{s}}^{(t)} = \hat{\bs{s}}^{(t-1)} + \bs{x}^{(t-1)} - \hat{\bs{r}}^{(t)}$
		\State $\ds \hat{\bs{q}}^{(t)} = \hat{\bs{r}}^{(t)} - \hat{\bs{s}}^{(t)} $
	\EndBlock{\bf end}
\EndBlock{\bf end}
\vspace{-5px}
\State $\ds \tilde{\bs{x}}^{(t)} = \hat{\bs{q}}^{(t)} + \frac{1}{\|\bm{\Phi}\|^2_{\rm{S}}} \bm{Z}^\dagger\bm{F}^\dagger\sum_{j=1}^{n_{\rm{d}}} \bm{M}_j^\dagger\tilde{\bs{q}}_j^{(t)} + \frac{1}{\|\bm{\Psi}\|^2_{\rm{S}}} \sum_{i=1}^{n_{\rm{b}}} \bar{\bs{q}}_i^{(t)}$
\State $\ds \bs{x}^{(t)} = \bigg(\frac{1}{\|\bm{\Phi}\|^2_{\rm{S}}} \sum_{j=1}^{n_{\rm{d}}}\bm{\Phi}_j^\dagger \bm{\Phi}_j +  \frac{1}{\|\bm{\Psi}\|^2_{\rm{S}}} \sum_{i=1}^{n_{\rm{b}}} \bm{\Psi}_i \bm{\Psi}_i^\dagger + \bm{I} \bigg)^{-1} \tilde{\bs{x}}^{(t)}$
\Until {\bf convergence}
\end{algorithmic}
\end{algorithm}

\section{Convex optimisation tools}
\bc
\begin{define}
The proximity operator \citep{Moreau1965} applied to any lower-semicontinuous and proper convex function $g$ is defined as
\begin{equation}
	\prox_g (\bs{z}) \overset{\Delta}{=} \argmin_{\bar{\bs{z}}} g(\bar{\bs{z}}) + \frac{1}{2} \| \bs{z} - \bar{\bs{z}}\|_2^2.
	\label{proximity-operator}
\end{equation}
\end{define}

\begin{define}
The indicator function $\iota_{\mc{C}}$ of any set $\mc{C}$ is defined as
\begin{equation}
	(\forall \bs{z}) \qquad \iota_{\mc{C}} (\bs{z}) \overset{\Delta}{=} \left\{ \begin{aligned}
					0 & \qquad \bs{z} \in \mc{C} \\
					+\infty & \qquad \bs{z} \notin \mc{C}.
				   \end{aligned}. \right.
	\label{indicator-function}
\end{equation}
\end{define}
In convex optimisation, it allows the use of an equivalent formulation for constrained problems by replacing the explicit convex constraints with the indicator function of the convex set $\mc{C}$ defined by the constraints.
Its use makes the minimisation task easier to tackle by general convex optimisation solvers.

\begin{define}
The Legendre-Fenchel conjugate function $g^*$ of a function $g$ is%
\begin{equation}%
	(\forall \bs{v}) \qquad g^*(\bs{v}) \overset{\Delta}{=}  \sup_{\bs{z}} \bs{z}^\dagger \bs{v} - g(\bs{z}).
	\label{f-conj}
\end{equation}
\end{define}

\begin{property}[Moreau decomposition]
The Moreau decomposition links the proximity operator of a lower-semicontinuous and proper convex function $g$ to that of its Legendre-Fenchel conjugate $g^*$ as%
\begin{equation}%
	(\forall \bs{z}) ~~ \bs{z} = \prox_{\alpha g} (\bs{z}) + \alpha \prox_{\alpha^{-1}g^*}(\alpha^{-1}\bs{z}), ~0 < \alpha < \infty.
	\label{moreau-decomposition}
\end{equation}
\end{property}
\ec

\section{Algorithm convergence}

\subsection{Alternating Direction Method of Multipliers}

The convergence of Algorithm \ref{alg-admm} is achieved through a careful choice of the parameters $\rho$ and $\varrho$.
The algorithm converges for any choice of the Lagrange parameter $\mu$ satisfying $\mu > 0$. This imposes the same constraint on $\kappa$.
For the convergence of the dual \ac{fb} sub-iterations, the update parameter $\eta$ should satisfy $0 < \eta < \sfrac{2}{\|\bm{\Psi}\|_{\rm{S}}^2}$.

Assuming that the measurement operator $\bm{\Phi}$ is full column rank and that convergence has been reached with the dual \ac{fb} sub-iterations, the convergence for the whole algorithm is achieved in terms of both objective function $\bar{f}(\bs{x})+ \bar{h}(\bm{\Phi}\bs{x})$ and iterates $\bs{x}^{(t)}$, $\bs{r}_j^{(t)}$ and, $\bs{s}_j^{(t)}$ \citep{Komodakis2015, Boyd2011}.
It requires that%
\begin{equation}
	\rho \|\bm{\Phi}\|^2_{\rm{S}} + \varrho < 2,
	\label{convergence-req-admm}
\end{equation}
with $\|\bm{\Phi}\|_{\rm{S}}$ being the spectral norm of the measurement operator and the parameters $\rho$ and $\varrho$ being the update step used for the proximal splitting and the gradient ascent step, respectively.

In practice however, the \ac{ri} imaging problem is very ill-conditioned and the operator $\bm{\Phi}$ is typically not full rank.
Under these relaxed conditions, the convergence is guaranteed only with respect to the objective function and the multipliers $\bs{s}_j^{(t)}$, without any guarantees for the iterates $\bs{x}^{(t)}$ and $\bs{r}^{(t)}$ \citep{Boyd2011}.
A possible way to improve this is to replace $\bar{h}$ with an augmented function $\tilde{h}$,
\begin{equation}
	\tilde{h}\left(\left[\begin{array}{c}\bm{\Phi} \\ \bm{\Gamma}\end{array}\right]\bs{x}\right) = \bar{h}(\bm{\Phi} \bs{x}) + 0(\bs{\bm{\Gamma}x}),
\end{equation}
where $0$ represents the null function, zero for any $\bs{x}$. Such a trick \citep{Pesquet2012} replaces the measurement operator $\bm{\Phi}$ with the augmented operator representing the concatenation of  both $\bm{\Phi}$ and $\bm{\Gamma}$. The new resulting operator is full rank for a proper choice of the matrix $\bm{\Gamma}$.
In practice Algorithm \ref{alg-admm} produces reliable performance and we did not employ such a trick herein.

\balance

\subsection{Primal-Dual Algorithm}

The variables $\bs{x}^{(t)}$, $\bs{v}_j^{(t)}$ and $\bs{u}_i^{(t)}$, $\forall i, j$, are guaranteed to converge to the solution of the PD problem (\ref{split-min-problem-gamma})-(\ref{split-min-dual-problem}) for a proper set of configuration parameters.
The convergence, defined given two general preconditioning matrices $\bm{U}$ and $\bm{W}$, requires \citep[Lemma 4.3]{Pesquet2014} that
\begin{equation}
	\| \bm{U}^{\sfrac{1}{2}} \bm{L} \bm{W}^{\sfrac{1}{2}} \|_{\rm{S}}^2 < 1,
	\label{convergence-req-pd}
\end{equation}
with the linear operator $\bm{L}$ being a concatenation of all the used operators, in our case a concatenation of both $\bm{\Psi}^\dagger$ and $\bm{\Phi}$.
By choosing diagonal preconditioning matrices, with the config parameters $\tau$, $\sigma_i=\sigma$ and $\varsigma_j=\varsigma$, $\forall i, j$, on the adequate diagonal locations, the conditions from (\ref{convergence-req-pd}) can be restated explicitly for Algorithm \ref{alg-primal-dual} as%
\begin{equation}%
		\left\| \begin{bmatrix}
			\sigma \bm{I} & \bm{0} \\
			\bm{0} & \varsigma \bm{I} \\
		\end{bmatrix}^{\sfrac{1}{2}}
		\begin{bmatrix}
			\bm{\Psi}^\dagger \\
			\bm{\Phi} \\
		\end{bmatrix}
		\begin{bmatrix}
			\tau \bm{I}
		\end{bmatrix}^{\sfrac{1}{2}} \right\|_{\rm{S}}^2 \!\!\!\!\leq
		 \tau \sigma \left \| \bm{\Psi}^\dagger \right \|_{\rm{S}}^2 + \tau \varsigma \left \| \bm{\Phi} \right \|_{\rm{S}}^2 \!<1,
	\label{convergence-req-explicit-pd}
\end{equation}
with the use of the triangle and Cauchy-Schwarz inequalities and with the diagonal matrices $\bm{I}$ of a proper dimension.
It should be noted that this formulation does not limit the use to only two parameters $\sigma$ and $\varsigma$.
However, having more independent update steps scales poorly due to the increasing difference between the resulting bound, computed similarly to (\ref{convergence-req-explicit-pd}), and the requirements (\ref{convergence-req-pd}).
This translates to having increasingly small values for the update steps, the more independent parameters we employ, with the convergence speed slowing down considerably in such situation.
It is also required that the relaxation parameter is chosen such that  $0 < \lambda \leq 1$.
The additional parameter $\gamma > 0$ imposes that $\kappa > 0$ as well.

For the randomised setup, the same parameters satisfying (\ref{convergence-req-pd}) suffice, granted that the probabilities of update $p_{\mc{P}_i}$ and  $p_{\mc{D}_j}$ are nonzero and the activated variables are drawn in an independent and identical manner along the iterations.

\bsp
\label{lastpage}
\end{document}